\documentclass[twocolumn]{aastex701}

\usepackage{hyperref}

\RequirePackage{color}

\newcommand{\mc}[1]{#1}

\newcommand{\kh}[1]{#1} 
 
\newcommand{\ld}[1]{#1} 
\newcommand{\acs}[1]{#1} 
 
\newcommand{\yh}[1]{#1}

\newcommand{\FeH}{\left[\mathrm{Fe}/\mathrm{H}\right]}
\newcommand{\onohiula}{`\=Onohi`ula}

\shortauthors{PFS-SSP GA Working group}
\shorttitle{PFS-SSP GA Science}

\begin{document}

\title{Galactic Archaeology with the Subaru \onohiula\ Prime Focus Spectrograph Strategic Program}

\collaboration{100}{The Subaru/PFS-SSP GA Collaboration}

\author[0000-0002-9053-860X]{Masashi Chiba}
\affiliation{Astronomical Institute, Tohoku University, 6-3 Aoba, Sendai, Japan}
\email{chiba@astr.tohoku.ac.jp}

\author[0000-0002-4013-1799]{Rosemary F. G. Wyse}
\affiliation{Johns Hopkins University, Department of Physics \& Astronomy, 3400 N Charles Street, Baltimore, MD 21218}
\email{wyse@jhu.edu}

\author[0000-0001-6196-5162]{Evan N. Kirby}
\affiliation{University of Notre Dame, Department of Physics \& Astronomy, 225 Nieuwland Science Hall, Notre Dame, IN 46556}
\email{ekirby@nd.edu}
\correspondingauthor{Evan N. Kirby}

\author[0000-0002-8039-4673]{Judith G. Cohen}
\affiliation{California Institute of Technology, Department of Astronomy \& Astrophysics, 1200 E California Boulevard, Pasadena, CA 91125}
\email{jlc@astro.caltech.edu}

\author[0000-0001-7679-9478]{L{\'a}szl{\'o} Dobos}
\affiliation{Johns Hopkins University, Department of Physics \& Astronomy, 3400 N Charles Street, Baltimore, MD 21218}
\affiliation{E{\"o}tv{\"o}s Lor{\'a}nd University, Department of Information Systems, P{\'a}zm{\'a}ny P{\'e}ter s{\'e}t{\'a}ny 1/c, Budapest, Hungary 1117}
\email{dobos@jhu.edu}

\author[0000-0003-0398-639X]{Roman Gerasimov}
\affiliation{University of Notre Dame, Department of Physics \& Astronomy, 225 Nieuwland Science Hall, Notre Dame, IN 46556}
\email{rgerasim@nd.edu}

\author[0000-0003-4656-0241]{Miho N. Ishigaki}
\affiliation{National Astronomical Observatory of Japan, 2-21-1 Osawa, Mitaka, Tokyo 181-8588, Japan}
\affiliation{Department of Astronomy, School of Science, SOKENDAI (The Graduate University for Advanced Studies), 2-21-1 Osawa, Mitaka, Tokyo 181-8588, Japan}
\affiliation{Kavli Institute for the Physics and Mathematics of the Universe (WPI), The University of Tokyo Institutes for Advanced Study, The University of Tokyo, Kashiwa, Chiba 277-8583, Japan}
\email{miho.ishigaki@nao.ac.jp}

\author[0000-0002-8758-8139]{Kohei Hayashi}
\affiliation{National Institute of Technology, Sendai College, 4-16-1 Ayashi-Chuo, Sendai, Japan}
\email{khayashi@sendai-nct.ac.jp}

\author[0000-0001-5522-5029]{Carrie Filion}
\affiliation{Center for Computational Astrophysics, Flatiron Institute, 162 Fifth Ave, New York, NY 10010, USA}
\email{cfilion@flatironinstitute.org}

\author[0000-0001-7214-3009]{Magda Arnaboldi}
\affiliation{European Southern Observatory - ESO, Karl-Schwarzschild-Strasse 2, 85748 Garching, Germany}
\email{marnabol@eso.org}
 
\author[0000-0003-4594-6943]{Souradeep Bhattacharya}
\affiliation{Centre for Astrophysics Research, Department of Physics, Astronomy and Mathematics, University of Hertfordshire, Hatfield AL10 9AB, UK}
\email{s.bhattacharya3@herts.ac.uk}

\author[0000-0002-5661-033X]{Yutaka Hirai}
\affiliation{Department of Community Service and Science, Tohoku University of Community Service and Science, 3-5-1 Iimoriyama, Sakata, Yamagata 998-8580, Japan}
\email{yutaka.hirai@koeki-u.ac.jp}
 
\author[0000-0002-4343-0487]{Chiaki Kobayashi}
\affiliation{Centre for Astrophysics Research, Department of Physics, Astronomy and Mathematics, University of Hertfordshire, Hatfield AL10 9AB, UK}
\email{c.kobayashi@herts.ac.uk}

\author[0000-0002-3852-6329]{Yutaka Komiyama}
\affiliation{Department of Advanced Sciences, Faculty of Science and Engineering, Hosei University, 3-7-2 Kajino-cho, Koganei, Tokyo 184-8584, Japan}
\email{komiyama@hosei.ac.jp}

\author[0000-0003-1980-8838]{Pete B. Kuzma}
\affiliation{Institute for Astronomy, University of Edinburgh, Royal Observatory, Blackford Hill, Edinburgh, EH9 3HJ, UK}
\email{pete.kuzma@ed.ac.uk}

\author[0000-0001-8239-4549]{Itsuki Ogami}
\affiliation{National Astronomical Observatory of Japan, 2-21-1 Osawa, Mitaka, Tokyo 181-8588, Japan}
\affiliation{The Institute of Statistical Mathematics, 10-3 Midoricho, Tachikawa, Tokyo 190-8562, Japan}
\affiliation{Department of Advanced Sciences, Faculty of Science and Engineering, Hosei University, 3-7-2 Kajino-cho, Koganei, Tokyo 184-8584, Japan}
\email{itsuki.ogami@nao.ac.jp}

\author[0000-0003-3220-0165]{Ana L. Chies-Santos}
\affiliation{Departamento de Astronomia, Instituto de Física, Universidade Federal do Rio Grande do Sul (UFRGS), Porto Alegre, R.S 90040-060, Brazil}
\email{ana.chies@ufrgs.br}

\author[0009-0009-3088-5886]{Nicole L. Klock-Miranda}
\affiliation{Departamento de Astronomia, Instituto de Física, Universidade Federal do Rio Grande do Sul (UFRGS), Porto Alegre, R.S 90040-060, Brazil}
\email{ana.chies@ufrgs.br}

\author[0000-0002-3182-3574]{Federico Sestito}
\affiliation{Centre for Astrophysics Research, Department of Physics, Astronomy and Mathematics, University of Hertfordshire, Hatfield AL10 9AB, UK}
\email{f.sestito@herts.ac.uk}
 
\author[0000-0002-7034-4621]{Tam{\'a}s Budav{\'a}ri}
\affiliation{Johns Hopkins University, Department of Applied Mathematics and Statistics, 3400 N Charles Street, Baltimore, MD 21218}
\affiliation{Johns Hopkins University, Department of Physics \& Astronomy, 3400 N Charles Street, Baltimore, MD 21218}
\email{budavari@jhu.edu}

\author[0000-0001-8274-158X]{Andrew P. Cooper}
\affiliation{Institute of Astronomy and Physics Department, National Tsing Hua University, 101 Section 2, Kuang Fu Road, Hsinchu 30013, Taiwan}
\email{apcooper@gapp.nthu.edu.tw}

\author[0000-0001-8470-1725]{Keyi Ding}
\affiliation{Department of Astronomy, University of Maryland, College Park, MD 20742, USA}
\email{kyding@umd.edu} 

\author[0000-0002-9933-9551]{Ivanna Escala}
\affiliation{Space Telescope Science Institute, 3700 San Martin Drive, Baltimore, MD 21218, USA}
\email{iescala@stsci.edu}

\author[0000-0002-5032-8368]{Elisa G. M. Ferreira}
\affiliation{Kavli Institute for the Physics and Mathematics of the Universe (WPI), The University of Tokyo Institutes for Advanced Study, The University of Tokyo, Kashiwa, Chiba 277-8583, Japan}
\email{elisa.ferreira@ipmu.jp}

\author[0000-0003-3333-0033]{Ortwin Gerhard}
\affiliation{Max-Planck-Institut für Extraterrestrische Physik, Giessenbachstrasse, 85741 Garching, Germany}
\email{gerhard@mpe.mpg.de}

\author[0000-0001-6207-4388]{Lauren Henderson}
\affiliation{University of Notre Dame, Department of Physics \& Astronomy, 225 Nieuwland Science Hall, Notre Dame, IN 46556}
\email{lhender6@nd.edu}

\author[0000-0002-2453-0853]{Jihye Hong}
\affiliation{University of Notre Dame, Department of Physics \& Astronomy, 225 Nieuwland Science Hall, Notre Dame, IN 46556}
\email{jhong5@nd.edu}

\author[0000-0001-5251-2284]{Shunichi Horigome}
\affiliation{Astronomical Institute, Tohoku University, 6-3 Aoba, Sendai, Japan}
\email{shunichi.horigome@astr.tohoku.ac.jp}

\author[0000-0002-2634-9169]{Ryota Ikeda}
\affiliation{National Astronomical Observatory of Japan, 2-21-1 Osawa, Mitaka, Tokyo 181-8588, Japan}
\affiliation{Department of Astronomy, School of Science, SOKENDAI (The Graduate University for Advanced Studies), 2-21-1 Osawa, Mitaka, Tokyo 181-8588, Japan}
\affiliation{Astronomical Institute, Tohoku University, 6-3 Aoba, Sendai, Japan}
\email{ryota195ikeda@gmail.com}

\author{Ryo Ishikawa}
\affiliation{Astronomical Institute, Tohoku University, 6-3 Aoba, Sendai, Japan}
\email{ryo.ishikawa@astr.tohoku.ac.jp}

\author[0000-0001-6503-8315]{Takanobu Kirihara}
\affiliation{Kitami Institute of Technology, 165, Koen-cho, Kitami, Hokkaido, 090-8507, Japan}
\email{tkirihara@mail.kitami-it.ac.jp}

\author[0000-0002-1126-9289]{Zhuohan Li}
\affiliation{National Astronomical Observatories, Chinese Academy of Sciences, Beijing 100101, PR China}
\email{zhli@nao.cas.cn}

\author[0000-0001-9178-3992]{Mohammad K.\ Mardini}
\affiliation{Department of Physics and Kavli Institute for Astrophysics and Space Research, Massachusetts Institute of Technology, Cambridge, MA 02139, USA}
\email{m.mardini@ipmu.jp}

\author[0000-0002-1349-202X]{Nicolas Martin}
\affiliation{Universit\'e de Strasbourg, CNRS, Observatoire astronomique de Strasbourg, UMR 7550, F-67000 Strasbourg, France}
\affiliation{Max-Planck-Institut f\"{u}r Astronomie, K\"{o}nigstuhl 17, D-69117 Heidelberg, Germany}
\email{nicolas.martin@astro.unistra.fr}

\author[0000-0001-9780-0220]{Yohei Miki}
\affiliation{Information Technology Center, The University of Tokyo,
6-2-3 Kashiwanoha, Kashiwa, Chiba 277-0882, Japan}
\email{ymiki@cc.u-tokyo.ac.jp}

\author{Rin Miyazaki}
\affiliation{Astronomical Institute, Tohoku University, 6-3 Aoba, Sendai, Japan}
\email{rin.miyazaki@astr.tohoku.ac.jp}

\author[0000-0002-7866-0514]{Sakurako Okamoto}
\affiliation{Subaru Telescope, National Astronomical Observatory of Japan, 650 North A’ohoku Place, Hilo, HI 96720, U.S.A.}
\affiliation{National Astronomical Observatory of Japan, 2-21-1 Osawa, Mitaka, Tokyo 181-8588, Japan}
\affiliation{The Graduate University for Advanced Studies (SOKENDAI), 2-21-1 Osawa, Mitaka, Tokyo 181-8588, Japan}
\email{sakurako.okamoto@nao.ac.jp}

\author[0000-0003-3835-9898]{Rohan Pattnaik}
\affiliation{Johns Hopkins University, Department of Physics \& Astronomy, 3400 N Charles Street, Baltimore, MD 21218}
\email{rpattna1@jh.edu}

\author[0009-0002-9382-7068]{Kyosuke Sato}
\affiliation{Department of Astronomy, School of Science, SOKENDAI (The Graduate University for Advanced Studies), 2-21-1 Osawa, Mitaka, Tokyo 181-8588, Japan}
\affiliation{National Astronomical Observatory of Japan, 2-21-1 Osawa, Mitaka, Tokyo 181-8588, Japan}
\email{kyosuke.sato@grad.nao.ac.jp}

\author[0009-0009-9769-534X]{Yoshihisa Suzuki}
\affiliation{National Astronomical Observatory of Japan, 2-21-1 Osawa, Mitaka, Tokyo 181-8588, Japan}
\affiliation{The Institute of Statistical Mathematics, 10-3 Midoricho, Tachikawa, Tokyo 190-8562, Japan}
\email{yoshihisa.suzuki@nao.ac.jp}

\author[0000-0002-4108-3282]{Alexander S. Szalay}
\affiliation{Johns Hopkins University, Department of Physics \& Astronomy, 3400 N Charles Street, Baltimore, MD 21218}
\affiliation{Johns Hopkins University, Department of Computer Science, 3400 N Charles Street, Baltimore, MD 21218}
\email{szalay@jhu.edu}

\author[0009-0004-1126-0286]{Dafa Wardana}
\affiliation{Astronomical Institute, Tohoku University, 6-3 Aoba, Sendai, Japan}
\email{dafaward@astr.tohoku.ac.jp}
 
\author[0009-0008-1002-2621]{Viska Wei}
\affiliation{Johns Hopkins University, Department of Applied Mathematics and Statistics, 3400 N Charles Street, Baltimore, MD 21218}
\affiliation{Johns Hopkins University, Department of Computer Science, 3400 N Charles Street, Baltimore, MD 21218}
\email{swei20@jhu.edu}

\author[0000-0002-3354-9492]{Wenbo Wu}
\affiliation{National Astronomical Observatories, Chinese Academy of Sciences, Beijing 100101, PR China}
\email{wbwu@nao.cas.cn}

\author[0009-0002-4998-0567]{Zhenyu Wu}
\affiliation{National Astronomical Observatories, Chinese Academy of Sciences, Beijing 100101, PR China}
\email{wuzhenyu@nao.cas.cn}

\author[orcid=0000-0002-9217-7051]{Xinfeng Xu}
\affiliation{Department of Physics and Astronomy, Northwestern University,
2145 Sheridan Road, Evanston, IL, 60208, USA}
\affiliation{Center for Interdisciplinary Exploration and Research in
Astrophysics (CIERA), 1800 Sherman Avenue,
Evanston, IL, 60201, USA}
\email{xinfeng.xu@outlook.com}

\author[0000-0002-5805-8112]{Xianhao Ye}
\affiliation{National Astronomical Observatories, Chinese Academy of Sciences, Beijing 100101, PR China}
\email{yexianhao@nao.cas.cn}

\author[0009-0003-8713-6946]{Ziqing Ye}
\affiliation{Johns Hopkins University, Department of Computer Science, 3400 N Charles Street, Baltimore, MD 21218}
\email{zye20@jhu.edu}

\author[0009-0000-7432-1390]{Xiangwei Zhang}
\affiliation{National Astronomical Observatories, Chinese Academy of Sciences, Beijing 100101, PR China}
\email{zhangxw@bao.ac.cn}

\author[0000-0002-8980-945X]{Gang Zhao}
\affiliation{National Astronomical Observatories, Chinese Academy of Sciences, Beijing 100101, PR China}
\email{gzhao@nao.cas.cn}

\author[0000-0003-2868-8276]{Jingkun Zhao}
\affiliation{National Astronomical Observatories, Chinese Academy of Sciences, Beijing 100101, PR China}
\email{Jingkunzhao123@gmail.com}

\author[0000-0002-8328-1447]{Xiaosheng Zhao}
\affiliation{Johns Hopkins University, Department of Physics \& Astronomy, 3400 N Charles Street, Baltimore, MD 21218}
\email{xzhao113@jh.edu}

\suppressAffiliations

\begin{abstract}
The recently commissioned Subaru \onohiula\ Prime Focus Spectrograph (PFS) will obtain spectra from nearly 2,400 fibers that cover 1.24 square degrees.  The 360 night Subaru Strategic Program  for PFS is dedicating approximately one-third of its allocation (130 nights) to study the structure and evolution of galaxies in the Local Group.  This Galactic Archaeological survey has three pillars.  (1) We will determine whether the mass density profiles of dwarf galaxies are consistent with cusps, as expected for cold dark matter, or cores, as expected from alternative dark matter theories or baryonic feedback.  We will deduce the density profiles as a function of radius from modeling of the full line-of-sight velocity and abundance distributions for six dwarf galaxies.  Our total sample will consist of 18,000 member stars to beyond the nominal tidal radius of each system.  (2) From measurements of the [$\alpha$/Fe] abundance ratio, we will learn the difference in assembly history of the two most massive galaxies in the Local Group: M31 and the Milky Way.  We will observe 30,000 member stars over 45 square degrees of M31's halo and outer disk.  (3) We will uncover how the most fragile (outer) part of the Milky Way responded to accretion events both in the distant past (such as Gaia--Sausage Enceladus) and in more recent history (such as the  Sagittarius dwarf spheroidal galaxy).  To support this study, PFS will provide velocities and metallicities---from which, in combination with photometry, we will deduce ages---for tens of thousands of main-sequence stars out to a Galactocentric distance of $\sim$30 kpc.
\end{abstract}

\section{Introduction}

\mc{Unraveling how large disk galaxies like the Milky Way (MW) and Andromeda (M31) formed and evolved in the expanding Universe is the ultimate science goal of Galactic Archaeology (GA; also known as Near-Field Cosmology). This is due to the fact that such nearby galaxies offer us the most detailed views of galactic structure and evolution through the properties of their resolved stars. In particular, ancient galactic components, such as extended thick disks and stellar halos, provide valuable information on the early chemodynamical evolution,  probing star formation during epochs more than 10 billion years ago. Indeed, much of the assembly history of a galaxy is written in the elemental abundances, kinematics and spatial structure of long-lived stars, with signatures of mergers and interactions persisting over many dynamical times \citep{Chiba2000,Belokurov2006,Frebel2015,DeSilva2015}. In particular, recent observational evidence, as provided by {\it Gaia} astrometry combined with ground-based spectroscopy,  suggests that the stellar halo interior to the solar circle (most of the stellar mass) is predominantly tidal debris from a single accreted system, dubbed ``Gaia-Sausage-Enceladus'' \citep{Belokurov2018,Helmi2018}. This accretion event likely occurred $\sim 10$~Gyr ago, and may be the last significant mass ratio merger experienced by the Milky Way, perhaps creating the thick disk \citep[cf.,][]{Wyse2001}. The earlier stages of this ancient merger should have created chemodynamic structure in the very outer halo.  

Gaia plus ground-based spectroscopy has also revealed the response of the local disk to the ongoing interaction/merger with the Sagittarius dwarf \citep[e.g.,][]{antoja_2018,Bland-Hawthorn2021}, debris from which dominates the outer halo \citep{Majewski2003}.  What of the many more subhaloes predicted to have merged with the Milky Way during the hierarchical growth of structure predicted in the currently popular $\Lambda$~Cold Dark Matter (CDM) paradigm? They too should leave evidence of their existence and subsequent demise in the multi-dimensional chemodynamic structure of the Milky Way. The stellar populations of M31 should similarly contain evidence of its assembly.  Thus, detailed studies of ancient stellar populations in the MW and nearby galaxies provide us with important clues to understanding galaxy formation in addition to constraining the nature of dark matter.}

The \onohiula\footnote{Hawaiian name that recognizes the instrument's embodiment of the idea of ``perceiving the realm of our origins,'' honoring  the people and land of Hawai`i and expressing our commitment to exploring the Universe together with the local community.} Subaru Prime Focus Spectrograph (PFS), a massively multiplexed spectrometer on the 8.2~m Subaru Telescope, will make the GA science described in this paper feasible. The instrument consists of four 3-arm spectrographs, equipped with a total of 2,394 reconfigurable fibers distributed across a 1.24~deg$^2$ hexagonal field-of-view. The three arms cover a wide wavelength range (380-1260~nm) in a single exposure \citep{Takada2014, Tamura2024}. The PFS GA survey in the Subaru Strategic Program (SSP) will use the blue arm (380--650~nm with $R \sim 1900$), the medium-resolution red arm 
(710--885~nm with $R \sim 5000$), and the near-infrared arm (970--1260, $R \sim 3500$). The  medium-resolution (MR) mode in the red arm was  specifically designed to allow measurements of multiple chemical abundances and precise radial velocities for our target stars.\footnote{The default red arm covers 630–970~nm with $R \sim 3000$.}

Scientific observations with PFS started in semester 2025A for both the SSP and open-use programs. The overall scientific thrust of the SSP is ``Cosmic  Evolution and the Dark Sector,''  and it will dedicate 360 nights over six years (2025--2030) to three avenues of investigation: Cosmology (100 nights), Galaxy Evolution \citep[130 nights,][]{Greene2022}, and Galactic Archaeology (130 nights).  \mc{We describe below the primary objectives of the PFS/GA science and the survey strategy for the PFS/GA-SSP.} 

\mc{
\subsection{Primary goals of the PFS/GA-SSP program}
}
We plan to determine the line-of-sight velocities and chemical abundances of large samples of stars in selected galaxies within the Local Group.  We will compare their derived properties with predictions from the concordance $\Lambda$CDM model of structure formation, plus alternative models, primarily those with lower power on small scales. 

 The three main thrusts of the PFS/GA-SSP survey are:

\begin{itemize}
\item {\bf Determination of the dark-matter density profiles and chemical evolution histories of dwarf spheroidal galaxies with a range of stellar mass and star-formation history.} As is well-established, these extremely dark-matter dominated systems are the best-suited targets for testing the robust prediction of $\Lambda$CDM that the density profile, in the absence of baryons, should be cusped \yh{\citep[e.g.,][]{Navarro1997}}. Time-dependent gravitational perturbations - due to either baryons or dark matter - can and will modify this profile \yh{\citep[e.g.,][]{ReadGilmore2005, Mashchenko2008, Pontzen2012}}, and we have designed our survey to enable quantification of the possible effects of strong bursts of star formation, as have been proposed to erase cusps in dark matter. We will apply multiple different analysis techniques, discussed in more detail below, to our sample of dwarf spheroidals (dSphs).  One approach is the essentially model-independent determination of the inner slope of the density law introduced by \citet{Walker2011}.  Another approach is a dynamical analysis of the line-of-sight velocity distribution across the face of each galaxy \yh{\citep[e.g.,][]{Hayashi2020, Hayashi2022, Wardana2025}}. The Walker \& Pe\~narrubia approach requires the robust identification of sub-populations based on chemical abundances, velocity dispersion profiles, and spatial extent, while the velocity distribution approach requires large samples of line-of-sight velocities for stars across the face of the system. Both require robust membership probabilities.  With the expectation that cored mass density profiles can result from repeated bursts of very active star formation, we will pair the kinematic/dynamical measurements with chemical estimates of star-formation histories, based on [Fe/H], [$\alpha$/Fe], plus detailed abundances of up to 18 individual elements.

\item {\bf Comparison of the stellar populations in M31 with those of the MW, through a large-scale spectroscopic survey of 30,000 individual stars across all structural components of our companion large disk galaxy.} The data will be used to infer the  major-merger history of M31 and compare that with the quiescent major-merger history since a redshift of $\sim 2$ inferred for the MW \citep[e.g.,][]{Wyse2001, Helmi2018, Belokurov2018,Pu2025}. Such a quiet past  is rare in $\Lambda$CDM, and the proposed  PFS dataset of kinematics and chemical abundances will allow us to establish whether the MW is indeed unusual. A key observable is the pattern of [$\alpha$/Fe] against [Fe/H] for the stars in the disk and disk/halo interface, at a large range of M31-centric distances. The MW shows two distinct sequences for thin and thick disks, albeit with different relative contributions as a function of location within the Galaxy \citep[e.g.,][]{Kordopatis2015, hayden15}, together with differences in mean stellar kinematics. These different abundance patterns and kinematics plausibly reflect two distinct formation mechanisms, namely major-merger induced heating for the thick disk, followed by gas cooling and a quiescent phase of thin-disk (re-)formation and growth \yh{\citep[e.g.,][]{Renaud2021}}. Establishing whether or not the stellar disk(s) of  M31 shows a bimodal chemical structure at locations across the extent of the disk is a strong test of the  major-merger history: a similar chemical dichotomy as the Milky Way disks would favor a similar quiescent history, while  establishing that the M31 disk(s) differ in chemical structure from the MW disk(s) would imply a qualitatively and quantitatively different merger history. Similarly, the chemical and kinematic structure and substructure in the halo of M31 will reflect the merger history, through tidal debris removed from satellite galaxies during earlier stages of accretion.  Such chemodynamic phase-space structure will be especially helpful in revealing whether the outer halo of M31 is dominated by stellar debris from a minor merger \citep[mass ratio 1:10 or smaller,][]{Fardal2007} or a major merger \citep[mass ratio 1:5 or larger,][]{Hammer2018,2025Tsakonas}, and constrain whether the mergers were ``wet'' or ``dry.''

\item {\bf Investigation of the response of the MW to the ongoing (minor) mergers with the Sagittarius dwarf and the LMC\@.} We will target faint main-sequence turn-off stars in selected lines-of-sight that probe the outermost regions of the disk and stellar halo, where dynamical times are longest. Main sequence turn-off stars are an unbiased sampling of the underlying population  and crucially allow for the determination of ages from isochrone fitting, in addition to kinematics and chemical abundances.  In the spirit of Galactoseismology \citep[e.g.,][]{Widrow2012}, we aim to decompose the derived disk kinematics into bending and breathing modes. We will also  characterize the coherent response of the stellar halo to gravitational perturbations, together with identifying kinematic phase space structure from the earliest stages of satellite accretion. We will target known stellar debris streams to distinguish those from star clusters and galaxies.  The data will also allow a refined determination of the edge of the stellar halo and of the thick-disk/halo interface in kinematic/chemical phase space. 
\end{itemize}

\begin{deluxetable}{lcccc}[pbt]
\tablecaption{PFS/GA-SSP survey plan\label{tab:survey_plan}}
\tablewidth{0pt}
\tablehead{ \colhead{Target} & \colhead{Mag} & \colhead{Exp.\ time} & \colhead{$N_{\rm pointings}$} & \colhead{$N_{\rm clear\  nights}$}  \\
            \colhead{  }     & \colhead{}      & \colhead{(hrs)}      & \colhead{  }                 & \colhead{ }    }
\startdata
MW dSphs      & $g<23$ & 3  & 63  &  24   \\
LG dIrr       & $g<22$ & 5  & 2   &   1   \\
\hline
MW halo       & $g<22$ & 3  & 45  &  17   \\
MW disk       & $g<22$ & 3  & 44  &  17   \\
\hline
M31 halo/disk      & $i<23$ & 5  & 44  &  28   \\
M33 halo      & $i<23$ & 5  &  7  &   4   \\
\hline
Total         &        &    &     &  91
\enddata
\tablecomments{Pointings for the MW dSphs and the LG dIrr include a second visit for those fields within the nominal tidal radius of each galaxy (outlined in red in Fig.~\ref{fig:dwarf_pointings} below). The total number of nights, 91, includes the assumption of a clear-sky fraction of 0.7, i.e., $130 \times 0.7 = 91$.}
\end{deluxetable}

\mc{The summary of our PFS/GA-SSP survey plan is given in Table~\ref{tab:survey_plan}.  The details of each survey component are described below.}

\section{Dwarf galaxies: Cusps, Cores, and Starbursts}
\label{sec:dwarfs}

\subsection{Dark matter content of dwarf galaxies}
\label{sec:darkmatter}

\begin{figure}[tpb]
\centering
\includegraphics[width=1.0\columnwidth]{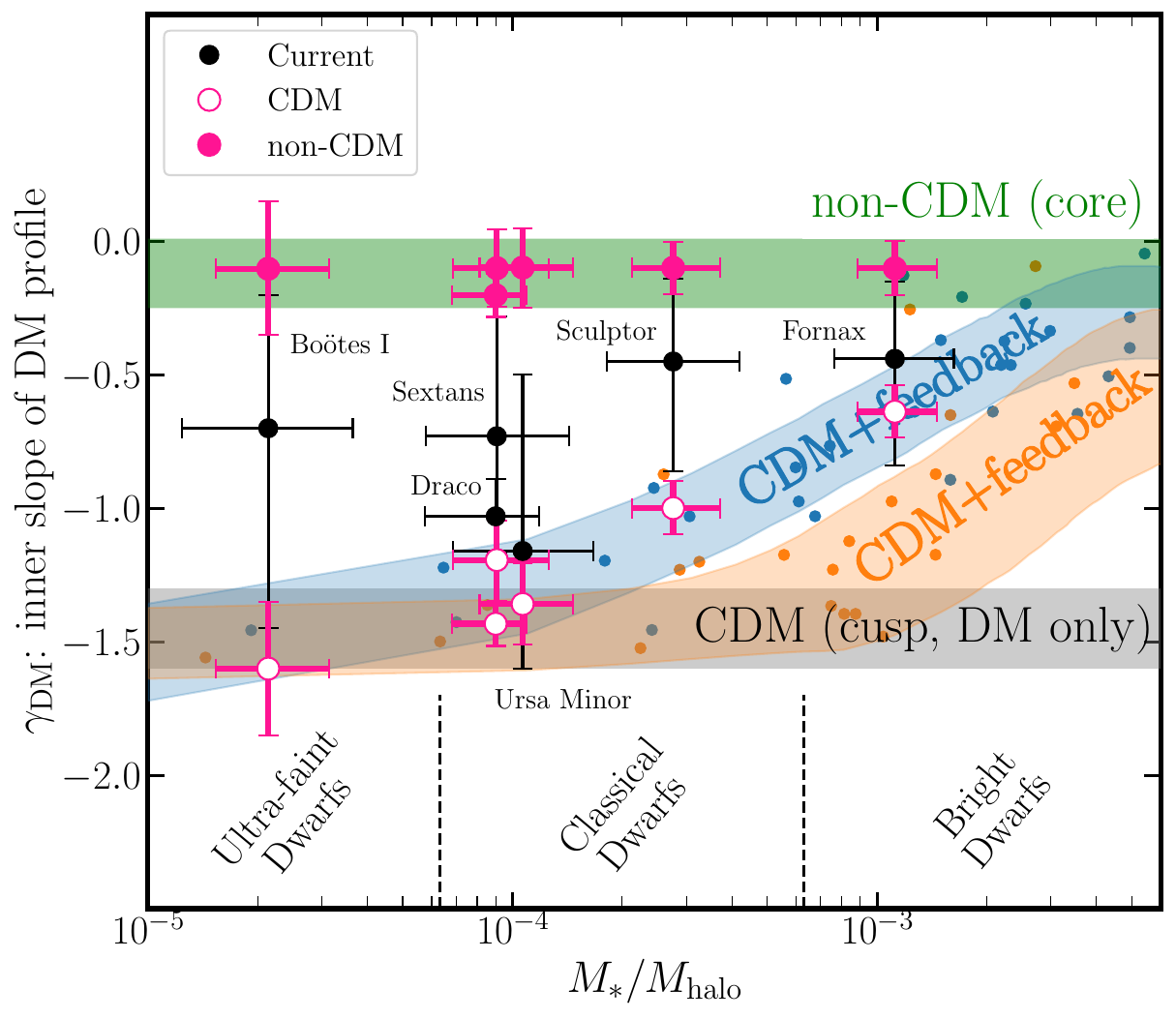}
\caption{(Figure after \citealt{Bullock2017}, their Figure 13.) The inner dark-matter density profile slope ($\gamma_{\rm DM}$) for our selected sample of dSphs, derived by axisymmetric Jeans analysis, plotted against their derived stellar-to-halo mass ratio~\citep[black points, ][]{Hayashi2020,Hayashi2022}.
Pure dark matter $\Lambda$CDM predicts NFW cusps $\gamma_{\rm DM} \lesssim -1$ ({\it gray shading}), while alternative dark-matter models predict cores $\gamma_{\rm DM} \sim 0$~ ({\it green shading}).  Hydrodynamical simulations~(\yh{{\it orange}, FIRE-2; \citealt{Lazar2020};  {\it blue}: NIHAO, \citealt{Tollet2016}}) predict that strong, episodic baryonic feedback can modify cusps into cores.  
The black points depict the current estimates of $\gamma_{\rm DM}$ in MW dSphs as reported by \citet{Hayashi2020}.
The open and filled magenta points, shown with reduced error bars, illustrate the expected constraints on $\gamma_{\rm DM}$ from the proposed PFS data, under the two prospective scenarios: CDM and cored non-CDM\@.
\label{fig:cuspcore}}
\end{figure}

\subsubsection{Scientific Motivation}
The existence of high-density ``cusps'' in the inner parts of CDM halos remains one of the most robust predictions of the theory. The orbital motions of gas and stars can in principle be analysed to determine the distribution of matter that produces the gravitational potential. The most DM-dominated galaxies require the least correction to the derived mass profile for the baryonic contribution, and thus dwarf galaxies - both gas-rich and gas-poor -  which tend to be more DM-dominated than larger systems, are favored targets.   
The ``core-cusp'' problem ~\citep[e.g.,][]{Moore1994, Moore1999,deBlok2001,Gilmore2007} arises because the most robust observational determinations of the inner DM density profiles for both gas-rich and gas-poor dwarfs favor a flattening towards the center (a ``core'') rather than a continuing steep rise~(the predicted ``cusp'').
However, the absence of a DM cusp does not necessarily imply the absence of CDM\@. 
Baryonic physics, such as  intense supernova feedback, can induce dynamical heating, which could transform a cusp into core \citep[e.g.][]{Pontzen2012, Pontzen2014}.
Dwarf galaxies with a larger ratio of baryonic mass to DM mass would be more susceptible to these processes.
The lowest mass (ultra-faint) dwarf galaxies are the most DM-dominated and therefore the least prone to the erosion of a CDM cusp. Higher mass dwarf galaxies, such as the classical dSphs, have increasingly complex SFHs and correspondingly more intense baryonic feedback.
The current derived values for the inner density slope, $\gamma_{\rm DM}$, in several dSphs are in tension even with the cored profiles predicted by $\Lambda$CDM based hydrodynamical simulations that include baryonic feedback (Figure~\ref{fig:cuspcore}).

These determinations of $\gamma_\mathrm{DM}$ could also be hints of dark matter other than the often-adopted WIMP extension to  the standard model.  Possible alternatives  include Fuzzy Dark Matter (FDM) or Self-Interacting Dark Matter (SIDM), which would induce a central core \citep{Hui2017,Spergel2000,Chan2022}.
Moreover, understanding the dark matter distributions within galaxies is of substantial importance for dark-matter searches through multiwavelength and multimessenger astrophysics. 
Our PFS-derived dark-matter density profiles will be essential to determine accurate astrophysical factors for such indirect searches and thus to set robust constraints on keV- to TeV-scale dark-matter particles \citep{Ando2022}.

\begin{deluxetable*}{lcccccrrrrr}
\tablecaption{Dwarf Galaxies Targeted by PFS\label{tab:dwarfs}}
\tablewidth{0pt}
\tablehead{ \colhead{Galaxy} & \colhead{Distance} & \colhead{$r_{\rm tidal}$} & \colhead{$M_*$             } & $\langle {\rm [Fe/H]} \rangle$ & \colhead{mean age} & $N_{\rm field\, centers}$ & $N_{\rm PFS}$  \\
            \colhead{  } & \colhead{(kpc)} & \colhead{($\arcmin$)} & \colhead{($10^6~M_{\odot}$) } & \colhead{(dex)} & \colhead{ } & \colhead{ } & \colhead{ }}
\startdata
Bo{\" o}tes I & \phn 66 & 38 & \phn 0.044              & $-2.6$ & ancient &  4 & 1000 \\
Draco         & \phn 76 &  48 & \phn 0.47\phn           & $-1.9$ & ancient &  4 & 1500 \\ 
Ursa Minor    & \phn 76 &  77 & \phn 0.66\phn           & $-2.1$ & ancient &  8 & 1650 \\ 
Sextans       & \phn 86 &  61 & \phn 0.70\phn           & $-1.9$ & ancient & 15 & 3000 \\ 
Sculptor      & \phn 86 &  74 & \phn 3.1\phn\phn        & $-1.7$ & ancient &  6 & 4800 \\ 
Fornax        &     147 & 69 &     25\phd\phn\phn\phn  & $-1.0$ & intermediate &  8 & 5100 \\ 
NGC 6822       &     459 & \nodata &     83\phd\phn\phn\phn  & $-1.0$ & young &  1 & 1000  \\
\enddata
\tablerefs{Distances and [Fe/H] are from \citet[][updated 2021]{McConnachie2012}.  Tidal radii are from \citet{Munoz2018}. Stellar mass is from the $V$-band magnitude given in \citet{McConnachie18} scaled by the mass-to-light ratio from \citet{Woo2008} and assuming $M_*/L_V = 2$ for Bo{\" o}tes ~I\@.  The number of field centers corresponds to the PFS pointings shown in Figure~\ref{fig:dwarf_pointings}.  The last column gives the approximate number of stars in each galaxy for which a radial velocity can be measured to better than 3~km~s$^{-1}$ \citep{Dobos2024}.}
\end{deluxetable*}

{\bf The PFS SSP is the spectroscopic survey that will provide the largest sample sizes of velocity measurements of the precision required to determine whether the density profiles of multiple dSphs are more consistent with $\Lambda$CDM or alternative dark matter models.}
PFS's ability to probe dark matter on small scales stems from several instrumental strengths: a wide field-of-view, the large 8.2~m Subaru mirror, which allows deeper observations down the red giant branch luminosity function than 4m-class telescopes, and a highly multiplexed spectrograph. 
These capabilities enable (1) sample sizes about five times larger than current samples, (2) velocity precision several times smaller than the velocity dispersion of a dSph, (3) chemical abundance measurements, and (4) synergy with Subaru/Hyper Suprime-Cam (HSC) pre-imaging with both broad-band and narrow gravity-sensitive filters, making PFS particularly well-suited for dSph studies.
\kh{While the Dark Energy Spectroscopic Instrument (DESI) has recently expanded the spectroscopic sample sizes of the classical dSphs, namely Draco, Ursa Minor, and Sextans, with a particular increase for Ursa Minor~\citep{2025arXiv250702284Y}, PFS is expected to surpass these numbers by obtaining substantially larger samples (Table~\ref{tab:dwarfs}).}

The PFS/GA-SSP will investigate several aspects of dark matter in dwarf galaxies.
First, we will measure the density profile, and we will definitively determine whether each of the six dSphs has a cusp or a core~(see the magenta points in Figure~\ref{fig:cuspcore}).
Second, we will measure individual stellar metallicities and [$\alpha$/Fe] ratios, and we will compare them to simulations to determine whether the inferred burstiness can explain the presence of a core.  PFS will be able to statistically distinguish populations with a mean offset as little as 0.15~dex in [$\alpha$/Fe]~\citep[][see also Figure~\ref{fig:dsph_dm}(c)]{Hirai2024}. The PFS/SSP data will extend well beyond the nominal tidal radii, allowing the effects of Galactic tides to be quantified and placing complementary limits on the dark-matter density profiles, relative to that of the Milky Way.

\subsubsection{Dynamical modeling of PFS data}
We will quantify the density distributions of the dwarf galaxies with several  independent techniques. 

Current datasets for luminous dSphs show that most of them contain more metal-rich stars in a more centrally concentrated distribution, with colder kinematics, whereas the more metal-poor stars are distributed over a larger area and have hotter kinematics \citep[e.g.,][]{Battaglia2008,Walker2011,Kordopatis2016}.
These kinematically different populations should be in equilibrium within the same dark matter potential well, and thus the coexistence of multiple populations enhances our ability to infer the inner structure of the dark matter halo.  We will use the elemental abundance data to identify distinct chemodynamical populations within a given galaxy, as independent kinematic tracers of the underlying mass distribution.

\citet[][hereafter WP11]{Walker2011} used the existence of two distinct chemodynamical populations to quantify $\gamma_{\rm DM}$ in each of the Fornax and Sculptor dSphs. 
Under the assumption of  spherical symmetry, the WP11 method determines the dynamical mass enclosed within the (different) half-light radius of each chemodynamical population, and thus infers the inner density slope without needing to assume a functional form for the dark matter profile or to measure the velocity anisotropy~\citep{Wolf2010}.

The ability to identify and characterize  chemodynamical populations, including their   distinct dispersion profiles, requires thousands of stars with velocity precision better than 3~km~s$^{-1}$ \cite[e.g.,][]{Amorisco2012,Kordopatis2016}. 
{\bf PFS will meet these criteria, as it will enable us to obtain stellar spectroscopy not only for the inner regions of dSphs but also for stars out to the tidal edge of each dSph, providing a large radial range over which to separate subpopulations.
Our ability with PFS to measure detailed elemental abundances (see Section~2.2) will allow us to determine the existence of multiple chemical populations not just in metallicity but also in the space of [$\alpha$/Fe] vs.\ [Fe/H]\@.}
Identifying distinct multiple stellar populations 
is crucially aided by the fact that we have obtained
HSC pre-imaging in a gravity-sensitive narrow band
to identify candidate member RGB stars from their lower gravity, compared to  foreground   main-sequence stars. 

While the WP11 technique is powerful, it is limited to estimations of enclosed mass at just a few radii. These radii are set by each sub-population (its half-mass radius), and they will not be uniform across the sample of dSphs, so that the slope is measured across different ranges of radius in each system.  
Moreover, the  assumption of spherical symmetry can lead to a systematic bias, and the value of the inner slope inferred by this method depends  on the viewing angle~\citep[e.g.,][]{Kowalczyk2013,Genina2018}.
Therefore, we will also infer density profiles by modeling the second-order velocity moment (i.e., velocity dispersion) based on a Jeans analysis. Although spherical Jeans analyses are widely used for constraining dark matter density profiles, such a symmetry is not a good assumption for $\Lambda$CDM halos. Furthermore, it is difficult to distinguish between an inner core or cusp for the dark matter distribution, even with a large dataset of line-of-sight velocities ~\citep[e.g.,][]{Chang2021,Splawska2026}, due to the well-known degeneracy between the dark halo parameters~(scale density, scale radius, and inner slope) and the anisotropy of the stellar velocity dispersion tensor. 
To address this difficulty, we will apply dynamical models that are more sophisticated than the spherical Jeans analysis. We will   (i)~relax the assumption of spherical symmetry and (ii)~use more information available in the shape of the stellar line-of-sight velocity distribution (LOSVD) than just the line-of-sight velocity dispersion.

\begin{figure*}[tpb]
\centering

\includegraphics[width=0.99\textwidth]{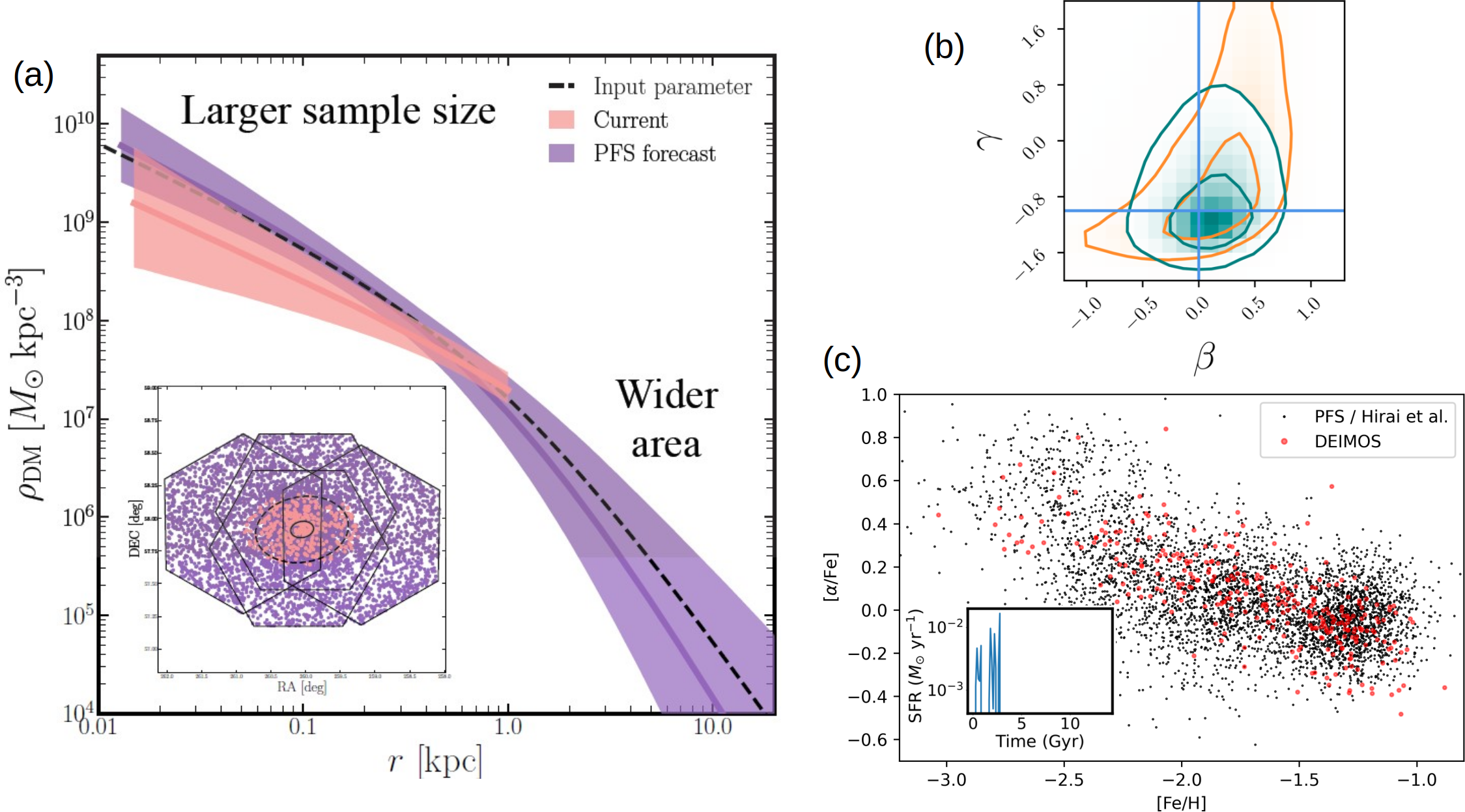}
\caption{(a)~DM density profiles derived by an axisymmetric, 2nd-velocity-moment Jeans analysis. The underlying model is shown as a dashed line. The shaded bands and colored curves correspond to the recovered density profiles and uncertainties obtained using line-of-sight velocities for stars that match ``Current'' ($N = 500$, orange) and ``PFS forecast'' ($N = 5,000$, purple) samples, distributed on the sky as shown in the inset. The ellipses correspond to the half-light and tidal radii. 
(b)~MCMC posteriors on the DM inner profile slope~$(\gamma)$ and velocity anisotropy $(\beta)$ from a spherically symmetric Jeans analysis using first only the 2nd-order velocity moment (orange) and then both the 2nd- and 4th-order velocity moments, thus allowing the inclusion of non-Gaussianity (cyan) \citep{Wardana2025}. 
The vertical and horizontal lines depict the input values of this mock analysis, for the case of a cusp $(\gamma= -1)$ and isotropy~$(\beta=0)$.
(c)~[$\alpha$/Fe] vs. [Fe/H] for uniformly analyzed spectroscopic data from DEIMOS for the Sculptor dSph \yh{\citep[red,][]{Kirby2011a}} plus anticipated results from PFS data (black), assuming bursty star-formation histories taken from the cosmological simulation of \citet{Hirai2024}. The repeated starbursts, indicated in the inset, are sufficient to modify the DM profile and result in clustering of the black points in elemental abundance space. 
\label{fig:dsph_dm}}
\end{figure*}

Non-spherical Jeans modeling can alleviate the effect of parameter degeneracies (especially between the dark matter density profile and velocity anisotropy), even using only the projected positions and line-of-sight velocity dispersions of the stars~\citep{Hayashi2020}.
Figure~\ref{fig:dsph_dm}(a) demonstrates such a Jeans analysis applied to mock PFS data for a Draco-like galaxy, adopting  axisymmetric stellar and dark matter components, generated from the \texttt{AGAMA} public code~\citep{Vasiliev2019}.
As indicated by the purple shading in the figure, the PFS mock data, which provide a large sample over a wide area, can recover the input density profile over the full range of radii probed by the spectroscopic sample.
This Jeans modelling can also be applied to each of the distinct  stellar populations within any one system~\citep{Zhu2016,Hayashi2018}.
We will carry out such an analysis with PFS data. 

The shape of the LOSVD that can be characterized by higher-order velocity moments is  sensitive to the underlying velocity ellipsoid and thus a higher-order Jeans analysis should offer a powerful technique for mitigating the degeneracy between dark matter density and velocity anisotropy.
The mock analysis by \citet{Read2021} indicates that the sample sizes we intend with PFS will suffice to break the mass--anisotropy degeneracy,  using {\sc GravSphere} \citep{Read2017}, {\sc MAMPOSSt} \citep{Mamon2013}, and {\sc AGAMA} \citep{Vasiliev2019}. 
\kh{{\sc GravSphere} is a dynamical modeling method that uses the second- and fourth-order moments of the spherical Jeans equation to simultaneously constrain mass distribution and velocity anisotropy. By incorporating higher-order velocity moments, it helps break the mass–anisotropy degeneracy that plagues simpler models. The method employs Bayesian inference with MCMC to reconstruct the dark matter density profile in a largely non-parametric way, though it assumes spherical symmetry.
{\sc MAMPOSSt} directly fits galaxy position–velocity data within the framework of spherical Jeans models. It uses parametric forms for both mass and anisotropy profiles, allowing efficient inference without binning of the data. While computationally practical and widely used, its reliability depends on the assumed functional forms and spherical symmetry.
{\sc AGAMA} is a flexible dynamical modeling framework that uses action–angle variables to construct distribution functions consistent with a given gravitational potential. Unlike Jeans-based approaches, it can model spherical, axisymmetric, and even triaxial systems in a fully self-consistent way. This makes it powerful for galaxy-scale modeling, though it requires careful selection of distribution functions and is computationally more demanding.}
These techniques assume spherical symmetry, but they can treat some non-Gaussianity of the LOSVD\@.

Figure~\ref{fig:dsph_dm}(b) demonstrates the power of including deviations away from Gaussianity when considering the shape of the LOSVD, even when still assuming spherical symmetry \citep{Wardana2025}.\footnote{In some cases,  rotation also needs to be considered \citep[e.g., see][for an analysis of rotation in Bo{\"o}tes~I]{Sandford2026}.} As may be seen in the figure, the MCMC posteriors for the dark-matter density slope, $\gamma_\mathrm{DM}$, and for the velocity anisotropy parameter, $\beta$, that are obtained using only the 2nd-order velocity moment (i.e., velocity dispersion; orange coloring), shows a strong degeneracy, while incorporating additional information from the 4th-order velocity moments (i.e., velocity kurtosis; cyan coloring) significantly reduces the degeneracy.
We will extend this 4th-moment analysis technique to non-spherical mass distributions,  in order to place tighter constraints on the inner density slope and velocity anisotropy.

A sample size of order  a thousand stars with secure membership is needed to define the shape of the LOSVD\@. As noted in Table~\ref{tab:dwarfs}, our planned PFS SSP satisfies this for each of the five classical dSphs (the ultra faint galaxy Bo{\"o}tes~I\@ contains too few stars). 

There are additional sources of non-Gaussianity in the LOSVD other than the intrinsic stellar kinematics, and these must be addressed. Contamination of the sample by non-member stars must be minimized and their possible effects incorporated into the modeling (see Section~\ref{sec:HSCdwarftargetselection}). Orbital motions within binary systems will inflate the wings of the LOSVD and again the effects must be mitigated (see next section). 

The high-quality data we will acquire offer the possibility of carrying out several different dynamical analyses, allowing the investigation of possible systematics.    
Thus we will also apply the Schwarzschild method of orbit superposition \citep{Schwarzschild1979,vandenBosch2008} to the PFS data.
For a given gravitational potential, this approach  constructs a library of numerically integrated orbits, and then weights the orbits to reproduce the observed stellar density profile and line-of-sight velocity distribution.
This model does not make any assumptions about the velocity anisotropy and can use all kinematic information, including higher-order moments.
Further, this method  can also be applied to multiple distinct stellar populations~\citep{BreddelsHelmi2014,Kowalczyk2022}.

\subsubsection{Estimating Membership and Binarity}
Both non-member stars and binary stars that are members have an impact on  dynamical analyses, in particular by modifying  the shape of the LOSVD\@.  
Secure membership determination and contaminant/binary  identification  are crucial for constraining the dark matter density profile.

For membership estimation, we first use Gaia proper motions \citetext{\hyperlink{cite.Gaia2023}{Gaia Collaboration et al.\ 2023}} to identify MW foreground stars. (They will be removed from the dynamical analyses of the dSphs, but will be retained for the spectroscopic sample, as tracers of the distant field halo.)   However, the majority of our targets are too faint for Gaia proper motions to aid in membership.  We therefore use the dedicated HSC photometric data we obtained to select  candidate member stars, based on a Bayesian analysis of the color-magnitude diagram, together with constraints from the color-color diagram including photometry in the NB515 filter~(for all the classical dSphs; see Section~\ref{sec:targeting_strategy} for details). The spectroscopic targets selected this way will  still contain some contamination by stars in the outer limits of the MW stellar halo, as such stars have negligible proper motions and are not completely separable by photometric data  (and are interesting science targets in their own right).
 We will therefore construct models for the LOSVD that incorporate the expected velocity distributions of the contaminating MW stars, extending the approach of, e.g., \citet{Ichikawa2017} and \citet{Horigome2020}, who forward-modeled the velocity dispersion only.

Binary stars alter the velocity distribution~\citep{Spencer2018,Buttry2022}.
Through detailed simulations of binary orbits and of PFS observation strategies\footnote{The binary model is based on \citet{Spencer2018}.  Our implementation of binary simulations is available on Github: \url{https://github.com/Subaru-PFS-GA/ga_binaries}.}, we determined that we can mitigate  the effects of binaries simply by observing each dSph twice, separated by at least three months \citep{Ye2024}.
Rather than discarding binary stars, we will construct a Bayesian model to infer their center-of-mass velocities for input into the mass modeling.  This Bayesian technique does not require that we re-observe every star in every field within each dSph.  Rather, we will estimate properties of the binary population by re-observing a subset of the fields.  Eventually, the binarity, orbital parameters and dSph membership will be determined probabilistically in parallel with the determination of the density profile.

\subsubsection{Target dwarf galaxies}\label{sec:targetdwarf}
Table~\ref{tab:dwarfs} summarizes the observational properties of the dwarf galaxies targeted by the PFS SSP\@. 
They have been selected to  span a range in $M_*/M_{\rm halo}$, in mean metallicity, and star-formation history (SFH). As illustrated in Figure~\ref{fig:cuspcore}, they thus represent different expected effectiveness of baryonic feedback in modifying the dark-matter density distribution. 

At the high stellar-mass end, Fornax is expected to have been strongly affected by baryonic feedback~\citep{Amorisco2013}. At the low stellar-mass end, feedback in Bo{\"o}tes~I is expected to have been too weak to erode an NFW cusp into a core~\citep[e.g.,][]{Tollet2016,Lazar2020}. 
The SFHs \citep[as derived from Subaru and HST imaging;][]{Okamoto2017,Weisz2014,2025PASJ...77.1259S} span from nearly continuous (Fornax) to truncated $\sim 10$~Gyr ago (Sculptor and Bo{\"o}tes~I).  
The proposed PFS sample includes three galaxies---Draco, Sextans, and Ursa Minor---with very similar $M_*/M_{\rm halo}$ but with disparate chemical properties: 
Draco has a steep radial metallicity gradient, whereas Sextans and Ursa Minor do not;   Draco and Sextans have  metallicity distributions that are typical of dwarf galaxies, with metal-poor tails, whereas Ursa Minor has an unusual metal-rich tail \citep{Norris2010,Kirby2011a}.  Ursa Minor stars also show higher values of [$\alpha$/Fe],  relative to Draco and Sextans at the same iron abundance  
\citep[see Fig.~13 of][]{Kirby2011b}, as would be predicted if the onset of Type~Ia supernovae occurred at lower iron enrichment in Draco and Sextans than in Ursa Minor.  These properties suggest that the galaxies have different SFHs \citep[e.g., Fig.~1 of][]{Wyse1993} that could have influenced the density profiles differently.
Both the quantity and burstiness of baryonic feedback determine the degree of transformation of an NFW cusp into a core \citep{Pontzen2012}.  The [$\alpha$/Fe] abundance pattern reflects the burstiness \citep{Gilmore1991,Koch2008, Hirai2024} and thus is an independent probe of the possible effectiveness of feedback.

\begin{figure*}[tpb]
\centering
\includegraphics[width=2.3in]{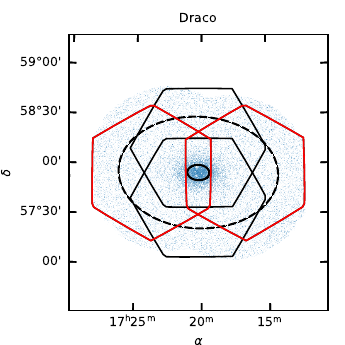}
\includegraphics[width=2.3in]{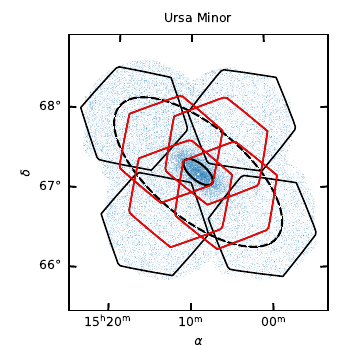}
\includegraphics[width=2.3in]{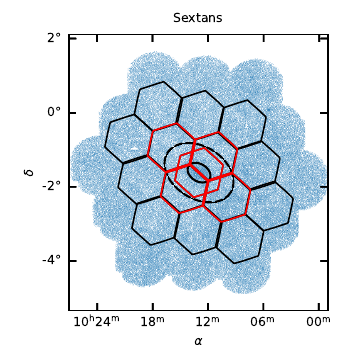}
\includegraphics[width=2.3in]{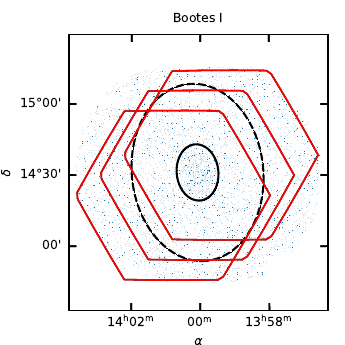}
\includegraphics[width=2.3in]{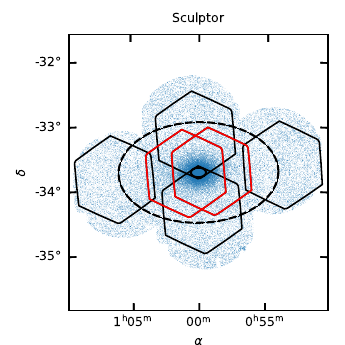}
\includegraphics[width=2.3in]{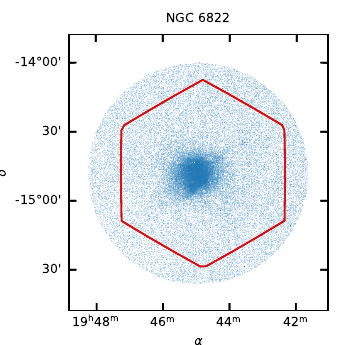}
\includegraphics[width=2.3in]{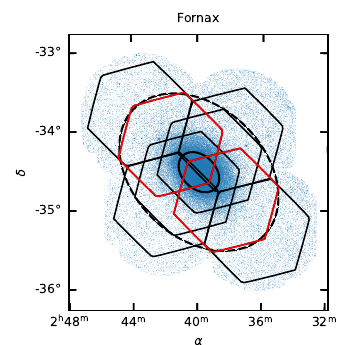}
\caption{The planned PFS pointings for the dwarf galaxies. Red hexagons indicate pointings that will be repeated to identify candidate binary systems.   The blue and gray dots in each panel are member and non-member star candidates selected by HSC photometry.
The solid and dashed ellipses show the core and nominal tidal radii that result from fits to King model profiles~\citep{Munoz2018}.}
\label{fig:dwarf_pointings}
\end{figure*}

Table~\ref{tab:dwarfs} gives the number of PFS pointings for dwarf galaxies in the SSP, plus the  estimated number of stellar spectra we will obtain. 
Figure~\ref{fig:dwarf_pointings} shows the planned PFS pointings for each of the dwarf galaxies.  The wide field of view of PFS allows us to target  stars well beyond the dSphs' nominal tidal radii.
Such \lq extra-tidal' stars enable study of the dynamical stability of the system, providing  additional constraints on the (relative) density profiles of the dwarf and Milky Way, together with  the chemodynamical properties of the outskirts of each dwarf galaxy~\citep[][]{FilionWyse2021,Longeard2022,Yang2022,Waller2023,Sestito2023a,Sestito2023b,2024MNRAS.527.4209J,2025ApJ...993L...7S}.

The pre-selection of candidate member red giant stars from photometric data is enabled through deep and precise Subaru/HSC $g$- and $i$-band images, which we obtained for most of the PFS target galaxies during open-use programs conducted in 2014--2016 [Sextans: S14A (PI:Chiba) \& S16A (Cohen), Ursa Minor: S15A (Okamoto), Fornax \& Sculptor: S15B \& S16B (Cohen), Draco: S16A (Cohen)]. The separation of (low surface gravity) RGB stars in these galaxies from foreground MW (high surface gravity) dwarf stars is achieved through HSC images we obtained with the narrow-band filter NB515, which isolates the gravity-sensitive Mg~b triplet \citep{Komiyama2018}. Broad-band HSC imaging data in ($g$, $r$) for Bo\"otes~I was obtained from the data archive, SMOKA (observations from 2014/7/2--3 and 2015/5/24). For NGC~6822, HSC images in four broad bands ($g$, $r$, $i$, $y$) and the narrow band, NB515 were taken during both the engineering (for installation of the new filter) and GTO runs. Section~\ref{sec:HSCdwarftargetselection} describes the photometric membership determination.

\subsection{Stellar elemental abundances in dwarf galaxies}
\label{sec:elements}

Along with kinematics, chemical abundances are a pillar of the spectroscopic study of galaxy evolution and its attendant baryon physics.  The chemical information can be categorized into various levels of complexity.  At the basic level, the chemical evolution of a galaxy can be characterized with simple metallicity (bulk metal content) measurements of stars.  The history of star formation, gas accretion, and gas removal shapes the stellar metallicity distribution function (MDF)\@.  The MDF can be interpreted to infer this history even with analytic, one-zone models of chemical evolution \yh{\citep[e.g.,][]{Kirby2013, kobayashi20, Kvasova2024}}.  An additional layer of complexity can be provided by examining the abundances of two or more elements simultaneously, exploiting the fact that different elements are created and ejected into the ISM on different timescales  \yh{\citep[e.g.,][]{Wallerstein1962,Tinsley79,Gilmore1991,Kobayashi2011,Nissen2010,Kirby2011b,Hirai2019,delosReyes2022}}. The most widely used combination pairs Fe with one or more $\alpha$ elements, such as Mg, resulting in the  Tinsley--Wallerstein diagram, which shows the evolution of [$\alpha$/Fe] with [Fe/H]\@.  The iron-rich ejecta from Type~Ia supernovae depresses [$\alpha$/Fe], after some delay time, unless the SFR increases with a corresponding burst of $\alpha$-rich core-collapse supernovae.  Numerical models of chemical evolution can recover the SFH of a galaxy from its Tinsley--Wallerstein diagram.  The most complex level of information is in analyzing several different elements to discern the contribution to the enrichment of the galaxy from multiple distinct nucleosynthetic sources: Type~Ia supernovae \citep{Kirby2019,delosReyes2020,Kobayashi2020_Ia}, core-collapse supernovae \citep{ishigaki18}, asymptotic giant branch winds \citep{Skuladottir2020}, and even rare events such as  magnetorotational supernovae or neutron star mergers \citep{Ji2016,Duggan2018}.

The PFS SSP will address outstanding questions in galactic chemical evolution at all of these levels of complexity.  In order to do so, the PFS GA survey will provide measurements of chemical abundances in unprecedentedly large samples.  The PFS-SSP will be the largest source of measurements of metallicity and detailed elemental abundances for {\it each} of the seven dwarf galaxies in our sample.  We will measure [Fe/H] to better than 0.2~dex for a total of approximately 17,900 member stars, [$\alpha$/Fe] to better than 0.2~dex for 8,900 member stars, and individual elemental abundances, like Mg, for 7,000 stars across all seven dwarf galaxies.  The precision for [Fe/H] and [$\alpha$/Fe] will be 0.1~dex for stars as faint as $r = 21.6$ (see Section~\ref{sec:abund} and Figure~\ref{fig:abun_precision}).  

Abundance surveys can be classified into high-resolution and low- (or medium-) resolution.  The Dwarf Abundances and Radial velocities Team \citep[DART,][]{tol06} observed four dSphs (Sculptor, Fornax, Sextans, and Carina) in both high- and low-resolution modes with the VLT/FLAMES multi-fiber spectrograph.  Among other successes, DART found distinct chemodynamical components in Sculptor \citep{tol04}, which is highly relevant to our science case on dark matter in dSphs (Section~\ref{sec:darkmatter}).  They also confirmed that Sculptor has very small scatter in detailed abundances at a given metallicity \citep{hil19}, which indicates efficient mixing at all times.  \citet{kir10} surveyed eight dSphs (Sculptor, Fornax, Leo~I, Sextans, Leo~II, Canes Venatici~I, Ursa Minor, and Draco) with the medium-resolution spectrograph Keck/DEIMOS\@.  They found that the MDFs of high-mass dSphs were more heavily shaped by gas accretion than low-mass dSphs \citep{Kirby2011a}, and they determined that sub-Chandrasekhar-mass Type~Ia supernovae were prevalent in dSphs \citep{kir19,del20}.  APOGEE has recently expanded the study of dSph chemical abundances into the infrared \citep{Hasselquist2013,Hasselquist2021}, for example, quantifying the intrinsic dispersion of elements to quantify the efficiency of mixing \citep{Mead2024}.  DESI has recently added hundreds of new spectroscopic members of dSph satellites of the Milky Way \citep{Yang2025,Ding2025,Sestito2025}.

Although we have learned a great deal about chemical evolution in dwarf galaxies, no survey has yet breached a sample size of 1,000 stars with detailed abundances in a single dwarf galaxy.  Reaching this threshold  provides multiple avenues for breakthroughs on understanding chemical evolution and nucleosynthesis.

Large samples will drive us to a major advance in the chemical evolution of dwarf galaxies: \textbf{multi-zone} GCE models.  Multi-zone models will introduce spatial variation.  Most importantly, they will allow the timescale for GCE to vary in different parts of the galaxy.  In general one expects evolution to proceed quickly in the center of the galaxy, where the gas density and star formation rate were high \citep{deBoer2012a} and dynamical times are shorter.  The lower-density outskirts of the galaxy experience more protracted evolution.  Furthermore, the outskirts are likely to receive enriched gas ejected from the inner parts of the galaxy \citep{Emerick2019}.  Only a multi-zone model can incorporate intra-galaxy enrichment of different populations, thus relaxing the instantaneous mixing approximation.  Multi-zone models have been applied to the MW \citep[e.g.,][]{Chiappini2001,Spitoni2021}, but demand for such models in dwarf galaxies has been limited due to modest sample sizes, predominantly in the inner regions.  

Large sample sizes will also capture a feature of dwarf galaxies that has been long predicted but poorly observed: \textbf{bursty star formation} \citep{Hodge1989,Gilmore1991}.  As established in Section~\ref{sec:darkmatter}, starbursts are predicted to be very important for shaping the dark matter distribution of dwarf galaxies (Figure~\ref{fig:dsph_dm}(c)).  However, a history of burstiness in Local Group ancient dwarf galaxies, like Sculptor and Ursa Minor, has not yet been confirmed.  \citet{Ting2024} found hints of 300~Myr bursts in the abundance pattern of the Sculptor dSph, but current spectroscopic sample sizes limit the significance of that detection.  Furthermore, the duty cycle and amplitude of burstiness even in simulations is subject to debate.  \citet{Zhang2024} found that the strength of burstiness decreases with increasing numerical resolution. 

\begin{figure*}
    \includegraphics[width=\linewidth]{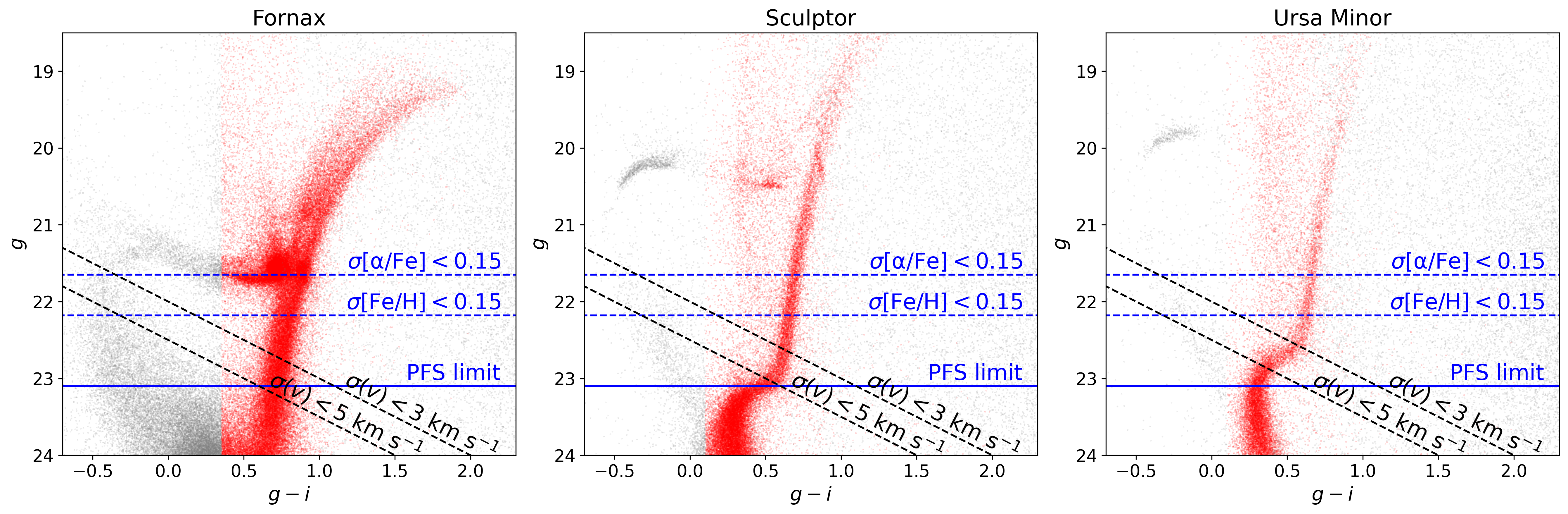}
    \caption{The HSC color--magnitude diagrams for three dSphs in order of decreasing distance.  Red points show stars that pass the narrow-band selection for giants.  The approximate $g$ magnitude limit for PFS spectroscopy is shown as a solid blue line.  The approximate $g$ magnitude at which [Fe/H] and [$\alpha$/Fe] uncertainties are less than 0.15 are shown as dashed blue lines (constants in $g$ magnitude).  The limits for 3 and 5~km~s$^{-1}$ uncertainties on line-of-sight velocity are shown as dashed black lines (constants in $i$ magnitude).\label{fig:fnx_scl_umi_cmd}}
\end{figure*}

The PFS-SSP improves upon or complements each of the existing 
spectrographs in a variety of ways.  Most importantly, the PFS/GA-SSP survey will exploit the 8.2~m diameter of the Subaru Telescope in order to reach down to the entire RGB for five of the six dSphs in the survey, as well as most of the RGB of Fornax and NGC~6822.  Figure~\ref{fig:fnx_scl_umi_cmd} shows the HSC CMDs of three dSphs along with the PFS magnitude limit.  The steeply increasing luminosity function of the RGB with fainter magnitudes permits us to observe member stars at large projected distances from the centers of the dwarf galaxies, where only those fainter magnitudes have sufficient surface density to fill a significant fraction of the PFS fibers.  
Therefore, PFS is particularly suited to observing member stars out to and even beyond the tidal radii of the dwarf galaxies.  

One of the major strengths of the PFS GA survey will be the large sample sizes.  Whereas previous surveys of Fornax have yielded [Fe/H] measurements in fewer than 700 stars, PFS will provide such measurements for 7,500 stars down to $r=21$.  The power of such a sample is not simply a more precise treatment of a one-zone chemical evolution model.  The PFS samples for each dwarf galaxy will be large enough to split into sub-samples based on a variety of properties.  For example, the MDF could be divided into ten radial bins to determine how the timescales of chemical evolution varied with radius and to investigate the origins of radial metallicity gradients \citep[e.g.,][]{Kirby2011a, Hirai2024}.  

The MDF alone is powerful enough to test for other aspects of galaxy evolution, including the interaction with other galaxies.  For example, ram pressure stripping can truncate the SFH, which would cause the high-metallicity end of the MDF to deviate from simpler models \citep{Kirby2013,Kvasova2024}.  This signature has mostly proved elusive in existing samples, but PFS will expand the sample sizes for all of the dwarf galaxies by roughly an order of magnitude.  The MDF could also encode information about the number and type of mergers a dwarf galaxy has experienced \citep[e.g.,][]{Deason2014}.  A dry, major merger would yield a fairly ordinary MDF but cause the remnant galaxy to lie off the mass--metallicity relation (possessing a high mass for its metallicity).  A wet merger would result in a strong burst of star formation, possibly over a narrow range of metallicity.  Minor mergers cause minor disturbances to the MDF that would be detectable only with the sample sizes that PFS will achieve.  Furthermore, radial mixing induced, for example by mergers \citep{Querci2025}, can move stars to large radius and flatten a metallicity gradient.

Dwarf galaxies are archaeological probes of the early universe.  Essentially all dwarf galaxies have ancient stars, and some of them, like Bo{\"o}tes~I, Sculptor, and Ursa Minor, have exclusively ancient stellar populations.  Their SFHs could contain clues to the timing and duration of reionization.  Traditionally, SFHs have been measured from CMDs \citep[e.g.,][]{Weisz2014}.  However, CMDs have logarithmic accuracy in determining stellar ages, which means they are at best precise to 1~Gyr at ages greater than 10~Gyr. Further, they generally require spectroscopic metallicities to break the age-metallicity degeneracy of
optical colors.  Chemical SFHs using the Tinsley--Wallerstein diagram can provide higher precision for ancient populations \citep{delosReyes2022}.

The large sample sizes for the dSphs targeted by the PFS-SSP  mean that their SFHs  can be measured with unprecedented precision \yh{\citep[Figure \ref{fig:dsph_dm}(c);][]{Hirai2024}}.  The ability  to measure [$\alpha$/Fe] as precisely as $\pm 0.1$~dex for stars well below the tip of the RGB will result in well-populated Tinsley--Wallerstein diagrams.  For most of the dSphs, the sample sizes will be large enough to construct Tinsley--Wallerstein diagrams for sub-samples (selected, for example, by projected radial position,  or by kinematics).   For example, the strong radial metallicity gradient in Sculptor has been attributed to outside-in formation \citep{Kirby2011a}.  PFS will test this hypothesis by measuring chemical SFHs in radial bins.  If the hypothesis is correct, then the SFH of the inner regions will be more extended that the SFH of the outer regions.

Low- to medium-resolution spectroscopy, such as obtained with the PFS blue and red optical arms, is even able to measure detailed abundances of individual elements, rather than an overall $\alpha$-abundance, given sufficient S/N\@.  The quality of abundances of individual elements, such as  Mg, measured from low-resolution spectroscopy of dSph stars compares favorably to those abundances measured from high-resolution spectroscopy \citep{Hill2019}.  Among other elements, we expect to measure C, Mg, Si, Ca, Ti, Cr, Mn, Fe, Co, Ni, Y, Ba, and Eu, all of which have been measured and validated from low-resolution Keck/DEIMOS spectroscopy \citep{Duggan2018,Kirby2018,delosReyes2020,Henderson2025a}.

There are many applications of these detailed abundance measurements.  For example, C and Ba abundances can be used to investigate stellar evolution, including binary mass transfer \citep[e.g.,][]{Arentsen2019}, and the influence of AGB stars on galactic chemical evolution \citep{Kirby2015}.  Mn and Ni abundances can be used to distinguish between sub-Chandrasekhar and Chandrasekhar-mass Type~Ia supernovae \citep{Seitenzahl2013}. Ba and Eu abundances can be used to estimate the delay times of the $s$- and the $r$-process.

Detailed abundances can also be used to discover exceptionally rare stars.  Our PFS-SSP is not only the best survey to find the most metal-poor stars in dSphs \citep[e.g.,][]{Frebel2010}, but it can also find evidence for rare nucleosynthesis events.  Neutron-star mergers happen infrequently, but they engender highly $r$-process-enhanced stars \yh{\citep[e.g.,][]{Ji2016, Roederer2016, Hirai2022}}. Either in the very early stages of evolution, or as a galaxy runs out of gas near the end of its life, individual stars can be susceptible to single nucleosynthetic events \citep[e.g.,][]{Audouze1995}.  Eu-enhanced stars have been identified in Sculptor from medium-resolution Keck/DEIMOS spectra; PFS is capable of identifying many more such stars, especially on the outskirts of the dSphs \citep{Henderson2025a}. A nearly mono-enriched star also appears in Ursa Minor, formed from gas enriched by a sub-Chandrasekar-mass Type~Ia supernova \citep{McWilliam2018}.

Section~\ref{sec:abund} establishes the feasibility of this science case.  It explains our planned methodology for measuring stellar parameters and abundances.  It also contains our estimates of abundance precision as a function of S/N\@.

\subsection{The Dwarf Irregular Galaxy NGC~6822}

The Local Group contains two major types of dwarf galaxy: dSph and dIrr.  DIrrs differ from dSphs in their morphology, gas content, SFH, current SFR, and local environment.  Whereas the surface brightness distributions of dSphs appear nearly featureless, dIrrs appear disordered and pockmarked by star-forming regions.  Unlike dSphs, dIrrs usually have copious amounts of gas and ongoing star formation.  Some---but not all---dIrrs have prominent ancient populations, and they all have young populations.  Except for the LMC and SMC, dIrrs are found far from the MW and M31.  The dividing line in distance from the larger `host' galaxy between dwarf galaxies with gas (dIrr) and without gas (dSph) lies around 200--300~kpc \citep{Grcevich2009,Spekkens2014}.

The PFS/GA-SSP sample includes one dIrr, NGC~6822, in order to contrast the properties of a dIrr with six dSphs.  NGC~6822 is the closest dIrr to the MW (other than the LMC and SMC), at a distance of 460~kpc.  It is not known if it orbits the Local Group barycenter or the MW\@.  The SFHs of most dSphs were truncated many Gyr ago.  In contrast, NGC~6822 has copious recent and ongoing star formation, and it established a complex dynamical structure \citep{Valenzuela2007}.  Previous spectroscopic samples have been limited to within 10~arcmin of the center of the galaxy or to sparse, single-slit spectroscopy \citep{Kirby2013,Swan2016,Belland2020,Ness2024}.  PFS/GA-SSP will measure the line-of-sight velocities and metallicities for 1,000 stars over the entire face of the galaxy (up to 35~arcmin or 4.7~kpc from the center).  This dataset will hold the key to understanding the unusual red-giant population that is misaligned with the H$\,${\sc i} disk, in addition to providing insight into the cause of  the apparent dynamical instability in the outskirts of that disk \citep{deBlok2000}.

The kinematic distribution of RGB stars in the center of NGC~6822 is particularly puzzling. Towards the center of the galaxy, the velocity distribution indicates that the RGB stars are rotating about their apparent major axis \citep{Belland2020}. Rotation of a structure about its major axis is uncommon, indicating the possibility of an event in NGC~6822's history that may have perturbed the RGB stars into this shape. Prolate rotation has been used as evidence of a merger in the dSph Andromeda~II \citep{Ebrova2015} and in the ``transition'' dwarf galaxy Phoenix \citep{Kacharov2017}. The tidal stirring model has not been able to explain prolate rotation, as demonstrated for Andromeda~II by \citet{Lokas2014}. Interaction with another galaxy was invoked by \citet{deBlok2000} to explain separate anomalous features such as the extended H{\sc i} cloud in the northwest of NGC~6822 and the H{\sc i} hole to the southeast.  However, \citeauthor{deBlok2000}'s merger hypothesis predicts that NGC~6822 still has an interacting companion galaxy.  No such companion has been found, despite an investigation of stellar overdensities in the vicinity of NGC~6822 \citep{Cannon2012}. It is still possible that NGC~6822 did merge with another galaxy that is now entirely disrupted.  Its impact on the dynamics of NGC 6822 should be measurable.

While the central RGB stars in NGC~6822 appear to show prolate rotation, the rotation axis is aligned with the galaxy's young H{\sc i} disk. Furthermore, this H{\sc i} gas disk is rotation-dominated ($v_{\rm rot}/\sigma_v \approx 10$, \citealt{Weldrake2003}), but the old, prolate-rotating RGB stars in its center are more pressure-supported with $v_{\rm rot}/\sigma_v \approx 0.4$. It may be possible that the oblong appearance of the central RGB stars could be a bar that exists in dynamical equilibrium with the H{\sc i} disk \citep{Friedli1994,Perez2009,Zhuang2019}. In fact, NGC~6822 is classified morphologically as an irregular barred Magellanic type galaxy.  One hypothesis \citep{Belland2020} is that we are viewing the bar nearly along its axis.  A rotating bar almost aligned with the line of sight will have a velocity distribution with a less Gaussian shape than a dispersion-supported spheroid (i.e., the result of the merger hypothesis).

PFS/SSP-GA will address the many open questions about this galaxy.  Is there a companion galaxy---one that generated the northwest H{\sc i} cloud and the southeast H{\sc i} hole---waiting to be discovered?  Is there dynamical evidence for a bar that can explain the prolate rotation of the red giants?  How do the dynamics of the stellar population respond to the H{\sc i} disk beyond the center of the galaxy?  

NGC~6822 also has a substantial legacy for studying planetary nebulae\footnote{PNe are the late phases of intermediate (8 $M_\odot$) to low mass  (0.8 $M_\odot$) stars (see review by \citealt{Kwitter22}), i.e., $\sim 95\% $ of all stars will end their life as PNe, as they evolve from the AGB to the white dwarf stage.} (PNe hereafter) populations for direct chemical abundance\footnote{Abundances determined directly from flux measurements of temperature sensitive auroral lines, such as [\ion{O}{3}] 4363~\AA, and other associated optical lines.} determination and for constraining AGB nucleosynthesis processes in low-metallicity environments \citep{hmp09b,Rojas16}. PNe can be identified in broadband and narrow-band imaging because they have strong [\ion{O}{3}] 5007~\AA\ line emission and faint continuum while also being point sources at extragalactic distances \citep[e.g.,][]{Arnaboldi02,longobardi13, Bhattacharya19}. There are 33 PNe identified in NGC~6822 from CFHT/Megacam [\ion{O}{3}] narrow-band and broad $g$-band imaging (Bhattacharya et al., in prep.). A few additional PNe are known at its center from \citet{hmp09}. 
These PNe lie within the PFS/GA-SSP survey field and do not suffer from contamination by MW sources, as MW halo PNe would be brighter and likely resolved, or from background emission-line galaxies (primarily Lyman-$\alpha$ emitters, which are fainter; \citealt{Hartke17}). Inclusion of these PNe in our PFS observations are primarily expected to constrain their chemical enrichment history through O and Ar abundance determination (similar to the case for PNe in M31, discussed in Section~\ref{sec:pn}; \citealt{Arnaboldi22,Kobayashi23}), thus complementing equivalent abundance measurements for RGBs.

\section{M31: Assembly of Luminous and Dark Halos}
\label{sec:ga_assembly_light_and_dark_halos}

\subsection{Motivation for PFS spectroscopy}
\label{sec:m31pfs}

\mc{
Faint stellar halos in giant disk galaxies, such as the MW and M31 galaxies, hold much of the fossil record on the formation of such galaxies through hierarchical assembly and accretion events in the past. In the MW,}
{\it Gaia} and recent spectroscopic surveys, such as APOGEE, have
discovered that the inner ($R \lesssim 10$~kpc) stellar halo is dominated by debris from a single massive accretion event 8--10~Gyr ago  \citep[Gaia--Sausage Enceladus, GSE;][]{Belokurov2018,Helmi2018}. There has long been evidence from the age distribution and kinematics of stars in the thick disk and stellar halo that the MW had a quiescient  accretion history since that event, which is very unusual in $\Lambda$CDM \citep[see, for example,][]{Wyse2001}. Complementary information, such as the steep density profile of the MW's stellar halo---especially in contrast to M31---suggests that our Galaxy escaped the violent accretion history predicted for most $L^*$ galaxies, reflecting the hierarchical assembly that is generic in $\Lambda$CDM \citep{Pillepich2014}.  Therefore, it is imperative to test the lessons learned from the MW by carrying out a  spectroscopic survey of the resolved stellar populations of another $L^*$ galaxy.

\mc{
M31, the target of this program, is an excellent test-bed for probing the galaxy formation process through the studies of the resolved stars in its halo and outer disk.  It provides an external perspective of the nearest large galaxy similar to our own, yet it is close enough for individual stars to be studied in detail. This is largely motivated by several galaxy formation models that suggest that each disk galaxy developed through a different formation and evolutionary path.  In other words, the epoch of collapse, the star formation history, and the accretion rate of small subsystems are not universal.  Therefore, each stellar halo and disk are  expected to display  different morphology, dynamics, and chemical abundance patterns \citep[e.g.,][]{Bullock2005,Cooper2010}.}

\begin{figure*}[tpb]
\centering
\includegraphics[width=0.75\textwidth]{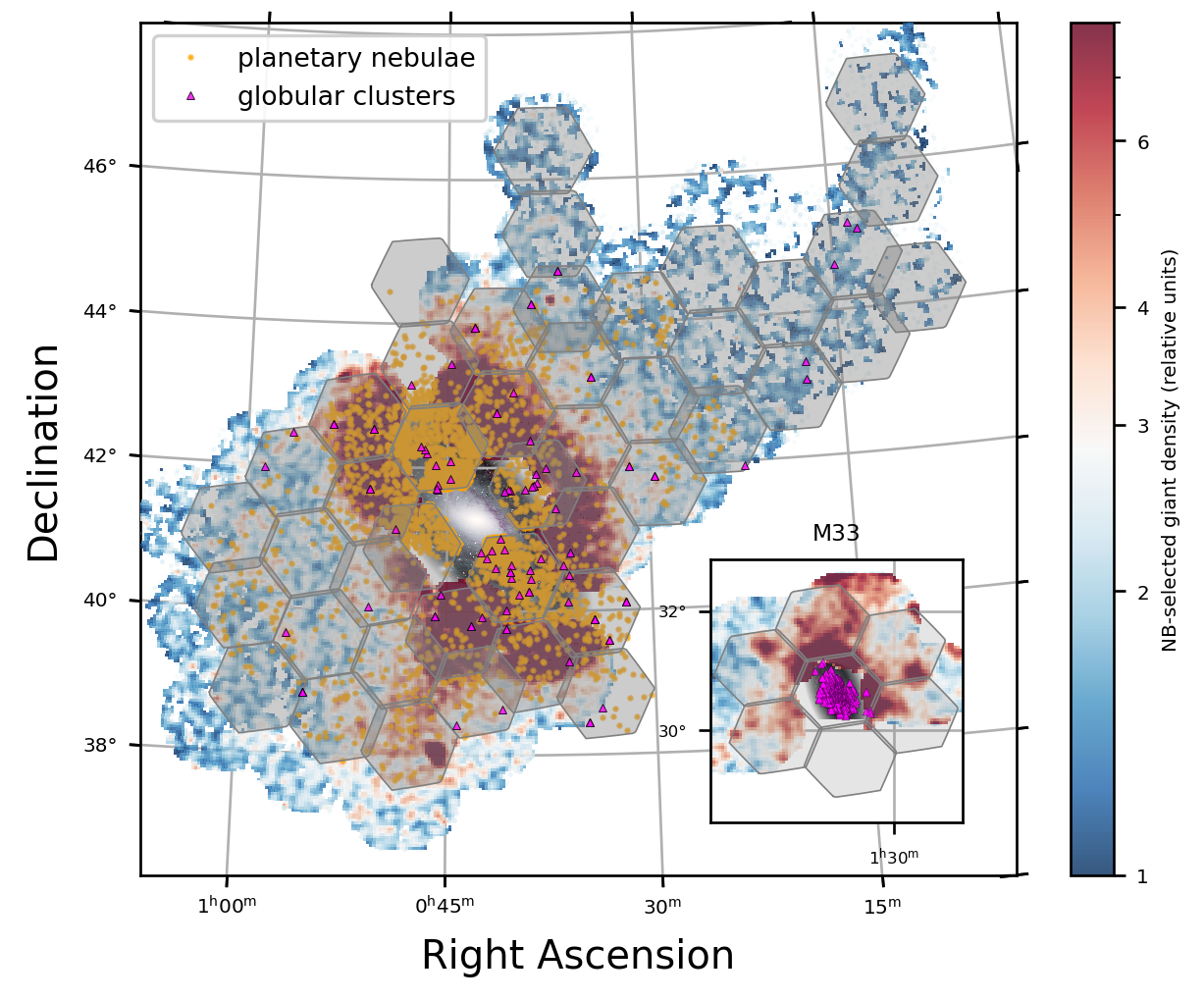}
\caption{Proposed PFS pointings ({\it gray hexagons}) in M31 and M33 ({\it inset}).  The color map shows the surface  density of candidate member stars selected through the combination of HSC broadband ($g$, $i$) and narrowband (NB515) imaging \citep{Ogami2025}.  We anticipate observing about 30,000 red giants in M31.  Also shown are planetary nebulae (\S\ref{sec:pn}) and candidate globular clusters (\S\ref{sec:gcs}).
}
\label{fig:m31}
\end{figure*}

\mc{
Past observational studies of M31's halo through large photometric and spectroscopic surveys of bright RGB/AGB stars (e.g., PAndAS survey using CFHT/MegaCam, SPLASH survey using Keck/DEIMOS, Subaru/SuprimeCam and Subaru/HSC) have revealed several characteristic features of M31's stellar halo in its both global and local spatial distributions. The similarities to the MW stellar halo appear in the outer parts, showing a power-law surface brightness profile, low stellar density and metal-poor stars \citep{Tanaka2010,Gilbert2014}. On the other hand, systematic differences are seen in the inner parts of the M31 halo, which reveal metal-rich and intermediate-age populations, in contrast to exclusively metal-poor and old halo stars in the MW\@. Also like the MW, the M31 halo contains many stellar streams or substructures, such as the Giant Southern Stream (GSS), the Northwestern Stream, Stream C, Stream D and more (see Figure~\ref{fig:m31}).  These streams are witnesses to past merging events with several dwarf galaxies or a massive gas-rich galaxy \citep{Fardal2007,Kirihara2017a,Kirihara2017,Hammer2018,DSouza18,2025Tsakonas}. 
}

Indeed, formation scenarios of GSS and other substructures have been put forward based on numerical simulations of either minor mergers of dwarf-type galaxies with $ M_{\rm total} \sim 10^9 M_{\odot}$ \citep{Mori2008,Kirihara2017} or a major merger, with mass ratio $\sim$ 1:5 \citep{Hammer2018,2025Tsakonas}. Both scenarios have succeeded in reproducing the  projected spatial distribution of stream and shell structures identified photometrically.  
On the other hand, the major-merger scenario is consistent with the observed age-velocity dispersion of the M31 disk(s) \citep{Dorman15, Bhattacharya19b}, its radial abundance gradient \citep{Bhattacharya22}, as well as the kinematics of the G1-clump substructure \citep{Bhattacharya23}. The major-merger scenario also reproduces the composite structure of the GSS along the LOS to the MW \citep{2016MNRAS.458.3282C,2025Tsakonas}  and the very large metallicity spread in the GSS \citep{2016MNRAS.458.3282C,2025Tsakonas}, NE and Western shelves \citep{Escala2022,Ogami2025,2025Tsakonas}. Further kinematic and chemical information for our targeted stars in M31 is essential to test any scenarios of accretion and merging histories of M31. In particular, their phase-space distribution, defined in line-of-sight velocities $v_{\rm los}$ vs.\ projected distances $R$, for given chemical abundances provides us with insight into several key dynamical parameters, e.g., the infall times of progenitor galaxies for producing observed streams/shells, apocentric pile-ups of debris from a merged galaxy, the weakness of shell features, and more \citep{Rocha2012,Genina2023}. This wealth of information will enable us to distinguish different models for halo and disk formation in M31, as well as constraining the current morphology of cannibalized satellites \citep[e.g.,][]{Kirihara2017, 2025Tsakonas}.

{\bf PFS will conduct a large-scale spectroscopic survey of the internal kinematics and chemistry of a spiral galaxy other than the MW\@.}  PFS will measure the velocities (precise to 3~km~s$^{-1}$; see Section~\ref{sec:rv}) and metallicities (precise to 0.2~dex; see Section~\ref{sec:abund}) for 30,000 member stars and [$\alpha$/Fe] ratios (precise to 0.15~dex) of 15,400 member stars, sampling all of the stellar components of M31.  This sample size is orders of magnitude larger than previous comparable quality spectroscopic
\mc{measurements \citep[e.g., fewer than 1,000 stars analyzed by ][]{Escala2022, Wojno2023}. Figure~\ref{fig:m31} shows the wide
coverage of the M31 halo and outer disk, plus M33,  to be targeted.}

\mc{
\citet{Dey23} demonstrated how massively multiplexed spectroscopy of red giant stars obtained with DESI on the 4m Mayall telescope can illuminate the accretion history of M31, by revealing the kinematic signature of a past merger event. Our survey with PFS will further revolutionize the study of M31, not only by measuring [$\alpha$/Fe], but also by obtaining such chemical data for a sample covering a much larger surface area and reaching up to one magnitude deeper than DESI \citep[for bluer RGB stars; we will reach 0.5~mag deeper for the red RGB stars that were the primary targets of][]{Dey23}. We also aim to target red giants over the full width of effective temperature found in M31, so that the MDF is unbiased and chemodynamic phase space is sampled fully.
}

A galaxy's accretion history imprints itself on the [$\alpha$/Fe] abundance pattern of its stars.  For example, the MW's quiet accretion history means that the GSE stands out clearly in the [$\alpha$/Fe] vs.\ [Fe/H] diagram of the stellar halo.  Furthermore, the GSE merger could have caused the bimodality present in the thin, alpha-poor and thick, low-alpha disks of the MW \citep{hayden15} by rearranging the disk $\sim 10$~Gyr ago \citep{Lu2024}.

PFS will study M31 in the Tinsley--Wallerstein diagram ([$\alpha$/Fe] vs.\ [Fe/H]) to distinguish between different formation scenarios.  For example, if M31 had a more violent recent accretion history than the MW, it would have a different [$\alpha$/Fe] distribution.  
\mc{Indeed, more frequent or more massive mergers of luminous or dark satellites cause more vertical and radial heating 
of the stellar disk \citep{Hayashi2006}, which can displace older, $\alpha$-enhanced stars.}  Such events would make for intriguing age structure at the interface between M31's thick disk and its halo, which PFS can test for.

\begin{figure}[tpb]
\centering
\includegraphics[width=\columnwidth]{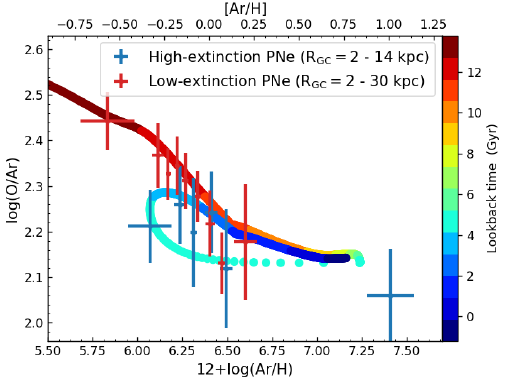}
\caption{Mean log(O/Ar) values of older ($>4.5$ Gyr) low-extinction PNe (red) over the $2-30$ kpc M31-galactocentric radial range, the younger ($\sim2.5$ Gyr) high-extinction PNe (blue) within 
$R_{\rm M31}\leq 14$ kpc, and the two-infall fiducial chemical evolution model for the M31 disk(s), colored by predicted lookback time (see \citealt{Arnaboldi22} for details).}
\label{fig:M31OArchem}
\end{figure}

The existing resolved spectroscopy of M31's disk (within a projected galactocentric radius of 2--30~kpc) provides initial insights, which indicate different radial ranges for
the disk bimodality in M31 with respect to that observed in the MW\@.  
Using PNe,  \citet{Arnaboldi22} showed evidence for a bimodal distribution of [O/Ar] at fixed [Ar/H]\footnote{Analogous to [$\alpha$/Fe] vs.\ [Fe/H] for RGBs, as also illustrated by \citet{Kobayashi23}.  For ensembles of star-forming galaxies, also see \citet{Bhattacharya2025a,Bhattacharya2025b}.} for the disk regions within $R_{\rm M31} \sim 14$~kpc (see Figure~\ref{fig:M31OArchem}).  These PNe consist of both younger ($\sim$2.5 Gyr old) thin disk and older ($>$4.5 Gyr old) thick disk populations. They also showed an apparently unimodal distribution for disk distances $>18$~kpc where these two PN populations overlapped. JWST/NIRSpec spectroscopy of RGBs \citep{Nidever24} in a small field at $18$~kpc\footnote{The disk region in M31, which is at the equivalent location as the solar neighborhood in the MW in unit of disk scale length is at 18 kpc distance from the center of M31 \citep{Yin09}.} also shows a unimodal distribution but in the appropriate stellar [$\alpha$/Fe] vs.\ [Fe/H] plane \citep{Kobayashi23}. This interesting  difference between M31 and the MW --- bimodality then single-track in [$\alpha$/Fe] vs.\ [Fe/H] in M31 as function of (increasing) radial distance and vice versa in the MW --- clearly calls for further investigation.

\begin{figure*}[tpb]
\centering
\includegraphics[width=0.95\textwidth]{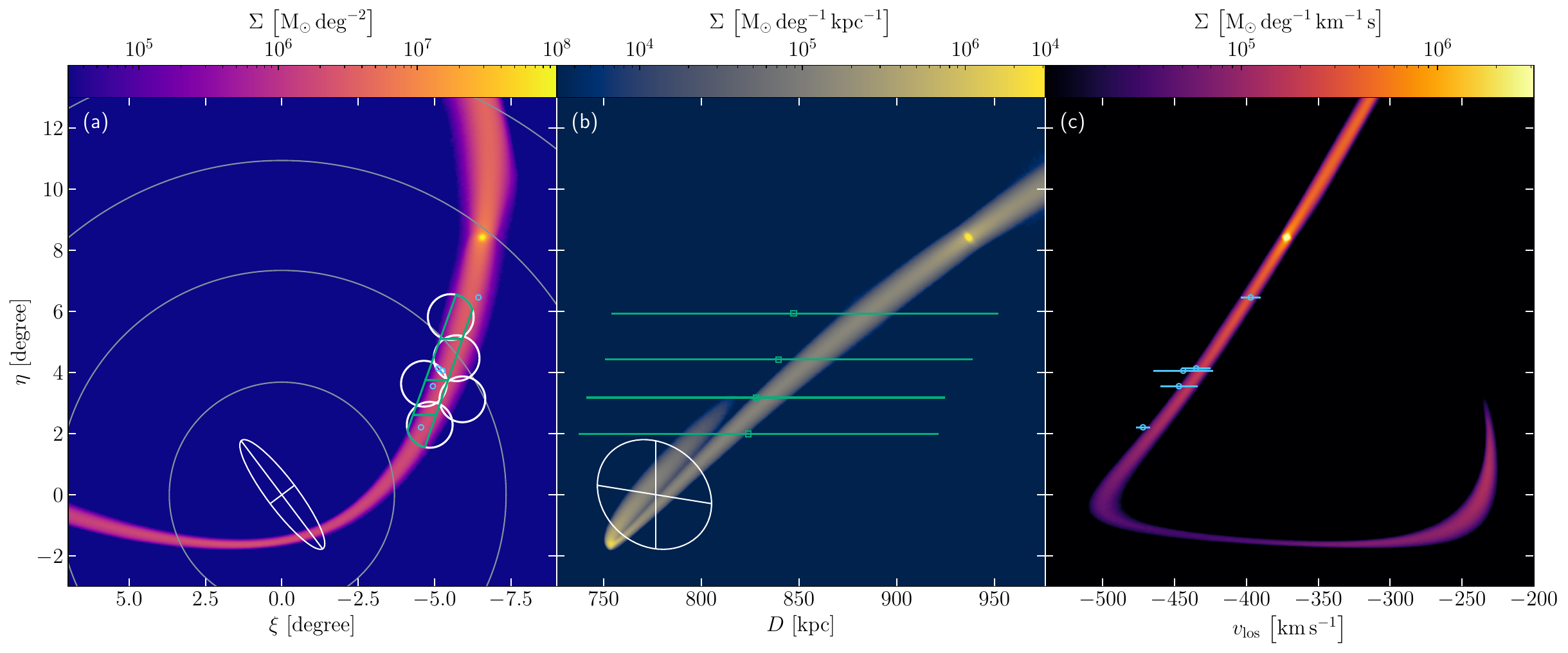}
\caption{
$N$-body simulation that reproduces the morphology and kinematics of the Northwestern Stream in M31 (Miki \& Kirihara, private communication): (a) The projected spatial densities of simulated stellar particles, where the main stellar disk of M31 is represented by an ellipse and the five HSC pointings for the stream \citep{Komiyama2018} are shown by the circles. These fields are included in the planned PFS pointings in Figure~\ref{fig:m31}. (b) The heliocentric distance distribution of stellar particles, where those of the observed stream fields are shown by the small open squares with error bars (teal). (c) The line-of-sight velocity distribution of stellar particles, where the small open circles with error bars (blue) are the observed velocities of the apparently associated globular clusters \citep{Veljanoski2014}.
\label{fig:m31_NWstream}}
\end{figure*}

\textbf{Our primary sample (detailed in Section~\ref{sec:M31_preimage}) consists of  a large number of RGB stars selected from Subaru/HSC photometry \citep[][see Figure~\ref{fig:m31}]{Ogami2025}, as well as an additional sample of $\sim$5,000 PNe discussed in Section~\ref{sec:pn}.} The line-of-sight velocities will allow us to distinguish between high angular momentum (disk) and dispersion-supported (halo) components of M31.  In addition to quantifying the accretion history through the Tinsley--Wallerstein diagram as a function of position and kinematics,   we will also model the dark matter profile out to 60~kpc by measuring the velocity dispersion of the halo as a function of projected radius.  The sample will inevitably contain a great deal of kinematic substructure in the form of stellar streams. The fact that our PFS fields  will contiguously cover the entire GSS, in contrast to the   pencil-beam approach of targeting  strategic locations \citep{Chapman2006,Gilbert2019}, will greatly facilitate the identification and treatment of substructure.    \mc{The PFS/SSP footprint also includes the Northwestern Stream, Stream C, and Stream D, for which only limited spectroscopic information is available at present \citep[e.g.,][]{Preston2024}.  The measurements of velocities and alpha-abundances will allow us to determine the orbits, stellar masses, and SFHs of the progenitor galaxies of these structures. Figure~\ref{fig:m31_NWstream} shows an example of an $N$-body simulation (Miki \& Kirihara, private communication) that reproduces many aspects of the observed  Northwestern Stream, such as the line-of-sight velocities of globular clusters that trace the over-density along the stream \citep{Veljanoski2014}. Such a cold stream \citep[the dispersion in line-of-sight velocity measured for the five associated globular clusters is consistent with zero;][]{Veljanoski2014} is also a dynamical probe of dark-matter subhalos should they interact with the stream. Further detailed comparisons between the available global/local chemo-kinematics of stars in M31's halo, as well as the MW's halo, and the predictions of high-resolution numerical simulations of the Local Group, such as the HESTIA suite of simulations \citep{Libeskind2020, Khoperskov2023}, will enable us to derive the accretion/merging history of M31. 
}

\subsection{Planetary nebulae}
\label{sec:pn}

Study of the PNe populations in M31 has a legacy of revealing its dynamical properties  \citep{Nolthenius1986, Nolthenius1987, merrett03,merrett06} as well as enabling direct abundance measurements, revealing radial gradients \citep{san12, Kwitter12, Pena19}. 
M31 was the target of an [\ion{O}{3}] narrow band imaging survey with Megacam@CFHT, identifying $5265$ PNe candidates over a total area of $\sim56$ sq. deg. centered on M31, covering the disk and inner halo, out to $R_{\rm M31} \sim 50$~kpc \citep{Bhattacharya19, Bhattacharya21} and falling within the PFS/GA-SSP survey area (yellow full dots in Figure~\ref{fig:m31}). Of these PNe candidates, $1251$ have LOS velocity measurements, while $205$ have measured O and Ar abundances from Hectospec@MMT MOS spectroscopy \citep{Bhattacharya19b,Bhattacharya22}. These well-studied PNe  are predominantly in the M31 disk; the  halo PNe are ideal supplementary targets for the PFS/SSP.  As in NGC~6822, PNe in M31 \citep{merrett06,Bhattacharya19,Bhattacharya21} do not suffer from contamination by either foreground MW stars or background galaxies.  

The PFS/GA-SSP survey will allow the measurements of the line-of-sight velocities for the entire PN MegacamCFHT imaging sample, including those located in the inner halo substructures (the GSS, the North-East and Western Shelves) and the more distant substructures (Stream D and Northern Clump), as well as for the smooth halo. Knowledge of the kinematics and metallicity distributions of the inner halo substructures will  constrain the mass  ratios of the past mergers that created the substructure. 
The PFS/GA-SSP plans to obtain chemical abundance measurements of PNe that are  much fainter than those previously studied in the M31 disk \citep{Bhattacharya22}, reaching down to a limiting magnitude of $m_{5007} = 25$ and a flux in the [\ion{O}{3}] 5007~\AA\ emission line of $F_{5007} \simeq 3.14 \times 10^{-16}$~erg~cm$^{-2}$~s$^{-1}$. Furthermore, O and Ar abundances can be determined in both the inner halo substructures of M31 and the smooth stellar component of the halo.  The resulting [Ar/H] vs.\ [O/Ar] plane will complement the equivalent [Fe/H] vs.\ [$\alpha$/Fe] determination from RGB stars.

\subsection{Pre-imaging with Hyper Suprime-Cam and sample selection}\label{sec:M31_preimage}

As with the MW dwarf galaxies (Section~\ref{sec:dwarfs}), we obtained HSC $g$- and $i$-band images for large areas of the M31 halo and outer disk during 9 nights in 2014 and 6 nights in 2015 (PI: Chiba), including most of the PFS/SSP M31 fields shown in Figure~\ref{fig:m31}.  To distinguish the intended target RGB stars in M31 from  foreground MW dwarf stars (which are  high surface gravity), we also obtained HSC images for many fields using the narrow-band filter NB515. This significant investment of HSC time is a strong point of synergy between wide-field photometry and spectroscopy at the Subaru prime focus.

\begin{figure}[tpb]
\centering
\includegraphics[width=\columnwidth]{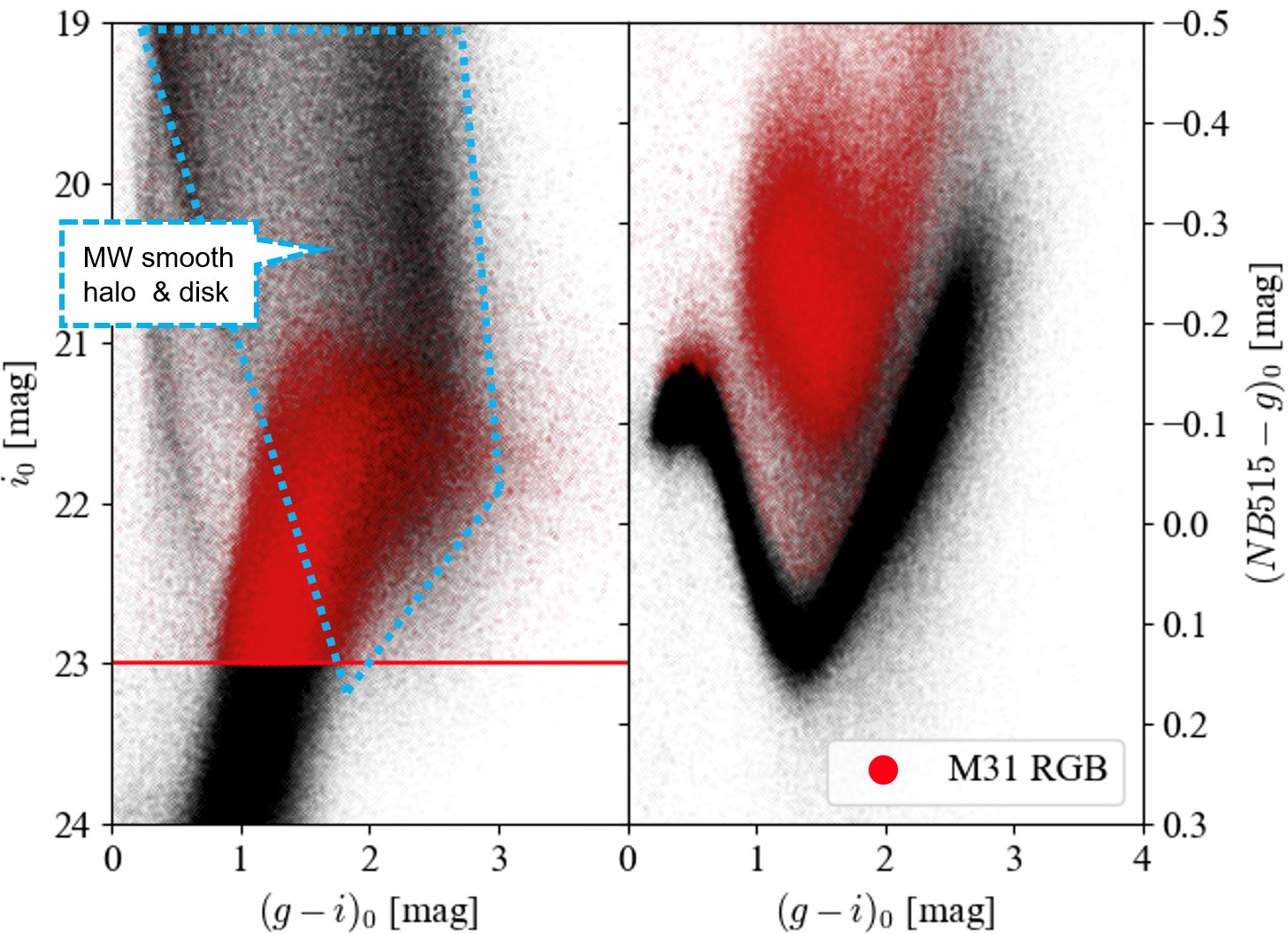}
\caption{
HSC color--magnitude ($(g-i)_0$ vs. $i_0$) and color--color ($({\rm NB515} - g)_0$ vs. $(g-i)_0 $) diagrams for stars in the  fields of the Northwestern Stream of M31, shown in Figure~5 \citep{Ogami2025}.  The NB515 filter discriminates M31 RGB stars from contamination by foreground MW dwarf stars (primarily located within the blue dotted polygon in the left panel).  Candidate red giant members ({\it red points}) are identified through the two-color diagram in the right panel. The redder, fainter RGB and dwarf stars will be classified using machine-learning techniques (not shown here; see \citet{ding_2025}).
\label{fig:m31_cmd}}
\end{figure}

An example of the extinction-corrected color-magnitude diagram (CMD) with $i_0$ vs.\ $(g-i)_0$ is shown in the left panel of Figure~\ref{fig:m31_cmd}, derived from the deep exposures of fields toward the Northwestern Stream \citep{Komiyama2018,Ogami2025}. The CMD consists of several stellar components: M dwarfs in the MW disk at $(g-i)_0 \sim 2$--3; the MSTO of the MW halo, extending from $(g-i)_0 \simeq 0.3$ and $i_0 \simeq 19$ to redder and fainter domains; the MW halo substructures \citep{Martin2014} located at $\sim 1$ mag fainter than the smooth halo; and the M31 halo including the RGB stars.\footnote{Fainter, red clump and horizontal-branch stars in M31, which are not included in Figure~\ref{fig:m31_cmd}, are also detected at high S/N in the deep imaging with HSC \citep{Komiyama2018}} The right panel shows a 2-color diagram, $({\rm NB515} - g)_0$ vs. $(g-i)_0 $, for brighter  stars with $19 < i_0 < 22$, where the thick (black) check mark-shaped sequence corresponds to the locus of MW dwarf stars, while the red points occupy the locus of low-gravity  stars and are selected RGB candidates in M31. These NB-selected halo stars are confirmed to show the same spatial distribution along the Northwestern Stream as that obtained from the selected RC stars in the M31 halo \citep{Komiyama2018}, thereby supporting our selection of RGB stars using NB515. 

\mc{The above selection method is insufficient to distinguish very metal-rich RGBs in M31 from the foreground dwarf stars for redder stars with $(g-i)_0 > 3$. For these stars, we developed  machine-learning techniques \citep{ding_2025} to select likely metal-rich RGBs in M31.} We first trained neural network models on synthetic colors derived from large empirical stellar spectral libraries, ensuring broad coverage of stellar parameters while optimizing the separation of M~dwarfs from M~giants. When applied to HSC data for fields in the inner halo of M31, the models demonstrate strong consistency with DESI spectroscopic membership selection \citep{Dey23}, validating the robustness of our membership classification.

\mc{
\subsection{Detecting the edge of the foreground stellar halo in the Milky Way}
Among the RGB stars selected using HSC/NB515 imaging, those brighter than the tip of RGB (TRGB) in M31, i.e.,
$i_0 < 21$~mag, and even some RGB stars occupying the locus of the broad red-giant branch of M31, can include some distant MW RGB stars along the line of sight to M31, including those at the edge of the MW stellar halo. 
According to cosmological hydrodynamical simulations for the formation of a Local Group analog \citep{Deason2020, Genina2023}, the edges of stellar halos can be defined as minima of both the logarithmic slope of the stellar density profile ($d \log{\rho} / d \log{r}$) and of the radial velocity profiles ($d V_r / d \log{r}$). This location, if identified, corresponds to the second caustic of the dark matter profile, estimated as $\sim 0.6r_{200m}$ \citep{Deason2020}, where $r_{200m}$ represents the radius within which the mean matter density of the galaxy equals 200 times the mean cosmic value, i.e., $200 \times \Omega_m \sim 60$ today, where $\Omega_m$ is the mass density parameter of the universe. Thus, the location of the estimated stellar halo edge can pin down the total dark halo mass of the MW, which remains a subject of much debate (e.g., \citealt{Magnus2022}). For example, a MW-mass halo of $\sim 1 \times 10^{12} M_{\odot}$ is expected to have $r_{200m} \sim 350$~kpc, suggesting that the RGB sample with $i_0 > 19$~mag can be used to probe these predicted properties at the halo edge.
These stars are so rare that only a survey as wide and as deep as the PFS/SSP survey in the direction of M31 can find them, albeit serendipitously.
}

\subsection{The M33 halo}

\begin{figure}[tpb]
\centering
\includegraphics[width=\columnwidth]{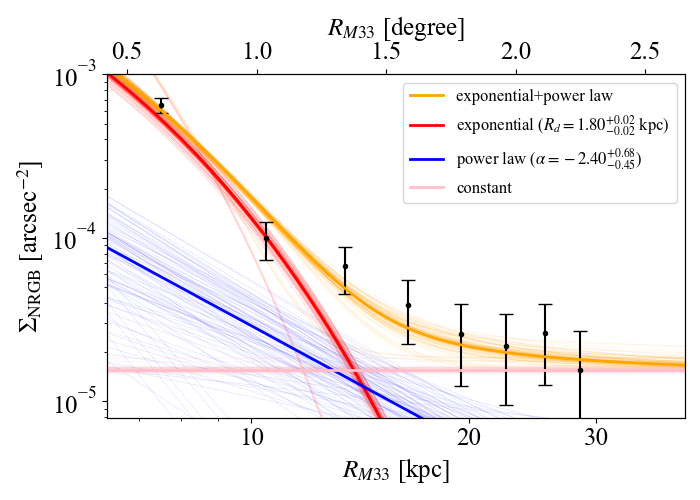}
\caption{Radial density profile of the HSC/NB515-selected RGB stars in the outer region of M33 \citep[reproduced by permission of the AAS from][]{Ogami2024a}. The black dots and their error bars indicate the number density of stars in each region. The red, blue, and pink lines are the derived individual profiles for the disk, halo, and contamination components, respectively, and the orange line shows the result of the fit to the disk plus halo components.
}
\label{fig:m33_RGB}
\end{figure}

M31 is distinct from the MW in possessing a {\it{bona fide\/}} spiral galaxy satellite, M33.  The PFS/GA-SSP survey will also include seven pointings towards M33 (see Figure~\ref{fig:m31}), allowing us to examine the structure of a galaxy that is  intermediate in mass between MW/M31 and the dwarf galaxies of Section~\ref{sec:dwarfs}.  Detecting the stellar halo of M33 has proved elusive, both photometrically \citep{McMonigal2016} and spectroscopically \citep{Gilbert2022}. \mc{\citet{Ogami2024a} detected an extended stellar component beyond the projected distance of $R_{\rm M33}=15$~kpc,  using Subaru/HSC multi-color imaging (Figure~\ref{fig:m33_RGB}). Interestingly, its radial density profile based on the photometrically identified RGB and red clump stars is shallow, with a fitted power-law index of $\alpha > -3$, shallower than that of the inferred inner halo population, which has been found to have $\alpha \sim -3$ at $R_{\rm M33}<5$~kpc, from Hubble Space Telescope (HST) multicolor imaging \citep[the Panchromatic Hubble Andromeda Treasury: Triangulum Extended Region survey or PHATTER survey,][]{Smercina2023}. This suggests that M33 may have a double-structured stellar halo, i.e., inner/outer halos, or a very extended low surface-brightess disk. Also, the slope of this extended component is shallower than those typically found for halos in large galaxies, implying intermediate-mass galaxies may have different formation mechanisms dominating their stellar halos (e.g., tidal interaction) from large spirals.}

PFS/SSP will observe 2,000 stars around M33 to determine once and for all whether M33 possesses a double-structured stellar halo.  If it exists, we will quantify its stellar mass by taking a census of its kinematic members and uncover its accretion history through elemental abundances. 

M33 also has 232 PNe identified from CFHT/Megacam narrow-band imaging (Bhattacharya et al., in prep.) that lie within the PFS/GA-SSP survey fields, though a vast majority of these sources are in the M33 disk. Only 26 PNe in M33 have direct O and Ar abundances determined  \citep{Magrini09}. Again, as in the case for NGC~6822 and M31, the PFS/GA-SSP observations of PNe in M33 will help constrain its chemical enrichment history through the determination of O and Ar abundances, complementing Fe and $\alpha$ measurements for RGB stars.

\subsection{Globular clusters}
\label{sec:gcs}
 
\acs{
The globular cluster system of M31 bridges the gap between the MW and more distant galaxies.
Of the globular cluster candidates in M31 and M33 \citep{galleti04,ma2012,huxor2014,beasley2015,caldwell2016,mackey2019,wang2022,usher24}, 103 fall within the planned PFS footprint of M31, and another 505 fall within the central PFS pointing of M33. 
While radial velocities have been determined for most of these GCs, stellar population parameters, such as ages and metallicities, have been determined mostly through optical integrated-light spectra using various instruments and methods. Thus, stellar population studies would benefit enormously from homogeneous UV-optical-NIR spectroscopic coverage.
Previous work has already demonstrated that the M31 GC system is very different from the MW system.  For example, it has roughly twice the number of GCs despite having a comparable total luminosity. Even though it contains the most metal-deficient massive globular cluster found to date  \citep{larsen2020,larsen2021}, the distribution of metallicities of the M31 GC system is weighted to higher values than the MW\@. Also, the MDF for the GC system of M31 is not obviously bimodal \citep{caldwell11},   as is the case for the MW GC system \citep{Zinn1985}. 
Moreover, the outer halo of M31 contains far more GCs than the MW, with higher masses and associated with several substructures in surface brightness \citep{Veljanoski2014}, likely created through past minor merger events. 
\citet{usher24}
measured ages for a sample of outer-halo GCs of M31 and found that those associated with halo substructure extend to younger ages and higher metallicities than those associated with the smooth halo. 

The kinematics of M31 globular clusters suggests two major accretion events \citep{mackey2019b}.  At the same time, \citet{Valenzuela24} found that the age distribution of globular clusters in M31 does not show bimodality, suggesting there were no major late-stage wet mergers, driving globular cluster formation. 
Having precisely measured age and elemental abundance distribution functions for the GC system of M31 will be invaluable to the use of globular clusters to understand the formation and evolution of galaxies at greater distances.
}

M33 has its own set of globular clusters, numbering at least 80 \citep{beasley2015}.  The population consists mostly of intermediate to young GCs in the inner 10~kpc, while a small number of older GCs have been detected in the outer regions of M33.  Intriguingly, these outer GCs appear to follow the tidal features of M33 \citep{mackey2019}, which are directed towards, and are most likely caused by, M31 \citep{Huxor2009}. However, it is unclear whether these outer GCs are indeed being pulled into the outer halo of M31 through this interaction or are just ``normal''  GCs in M33. In fact, it is unknown if M33 is currently undergoing its first infall or a second, the latter of which is expected to begin stripping M33's GCs into the M31 halo \citep{Patel2018}. PFS line-of-sight velocities can explore the nature of these GCs.

The purple triangles in Figure~\ref{fig:m31} show the locations of globular cluster candidates that fall within the PFS pointings in M31 and M33.  These candidate clusters were selected from \citet{bolognaCatalogGCS}, \citet{pandascatalogGCS}, \citet{chencatalogGCS,lamostcatalogGCS}, \citet{wangcatalogGCS,shouchengcatalogGCS}.

\section{Milky Way Structure and Substructure}\label{sec:mw}

The accretion of both baryonic and dark matter into the MW is ongoing, and observations suggest that there are a number of satellite galaxies, spanning a range of masses, at different stages of merging/assimilation. Of particular note are the mergers with the Sagittarius dwarf galaxy (Sgr) and the Large Magellanic Cloud/Small Magellanic Cloud system  (LMC/SMC), as discussed in more detail in the following subsections. The detailed dynamics of the merging process remain ill-understood, especially for disk galaxies like the MW in which there are several internal and external gravitational perturbations to model. It is also unclear how the structural components of the MW, such as the disk(s), have been shaped by past mergers. The MW's ongoing mergers present an exciting opportunity to map the current dynamical repercussions of merging in exquisite detail, and the role of past mergers can be explored through analyses combining age information with the chemodynamics of stars. The depth and field-of-view accessible to PFS make it an excellent instrument for such efforts. The planned PFS/SSP-GA observations will probe the furthest reaches of the stellar halo and disk(s), where dynamical times are longest, which will enable the characterization of the response of the Galaxy to these  perturbations, through  the quantification of the structure of the outer disk(s) and halo of the MW in spatial-coordinate, chemical, and age space.

\subsection{The Disk in Disequilibrium}\label{sec:disk_diseq}

In the past two decades, a deluge of data from photometric, spectroscopic, and astrometric surveys has provided ample evidence that the MW thin disk is in a state of disequilibrium. Signs of disequilibrium can be seen in coordinate-space over-densities and asymmetries (e.g., the \lq Triangulum–Andromeda cloud', also called TriAnd, \citealt{rochapinto_2004}; the Galactic Anticenter Stellar Structure, \citealt{Slater2014}; and the Monoceros Ring, \citealt{Morganson2016}), as well as kinematic phase space substructures, e.g., the `\textit{Gaia} phase-spiral' \citep{antoja_2018}, seen as over-densities in $z-V_z$ space, in addition to other dynamical projections \citep{Hunt2025}. Indeed, the disk is reacting to both internal non-axisymmetric perturbations, such as the bar and spiral arms, and external perturbations, such as the ongoing merger with the Sagittarius dwarf spheroidal \citep[Sgr,][]{Hunt2022}. Simulations have indicated that each of these perturbations can cause a variety of responses in the disk \citep[e.g.,][]{quinn_1993,sellwood_1993,JBH2021,Hunt2022}. Detailed study of the overall dynamical state of the disk, using both high-resolution dynamical simulations and chemodynamical observational data, is required to disentangle the signals of internal and external perturbations. 

In the outer disk (beyond the Solar radius), dynamical timescales are long and the restoring force of self-gravity is minimized. The combination of these properties cause the amplitude of the response to perturbations to be large, and the signs of past perturbations to be long-lasting. At large Galactocentric radii, it is possible that the kinematic phase-space contains information about \textit{multiple} previous passages of Sgr (e.g., \citealt{laporte_2019}), beyond just its most recent passage through the outer disk less than a billion years ago \citep{Ibata1997}. Thanks to these distinct properties, the outer disk is home to a wealth of substructure in both spatial/coordinate density \citep[e.g.][]{Xu2015} and kinematic phase spaces \citep[e.g.,][]{Ibata2021}. Some of these relatively nearby structures ($d \lesssim 15$~kpc) were misidentified as stellar streams in the halo, taken as evidence for accretion of dwarf galaxies, until more detailed distance, kinematic, and chemical abundance information revealed them to be connected high-amplitude corrugation responses of the perturbed thin disk (see e.g., \citealt{Li2017}). The chemodynamical and age information that PFS/SSP-GA will provide for stars in the outer disk (see Section \ref{sec:MW_targeting}) and will help to discriminate between some of the formation scenarios proposed for the observed outer disk substructures (e.g., \citealt{Laporte2020,laporte2022}). 

\begin{figure*}[tpb]
\centering
\includegraphics[width=2.0\columnwidth]{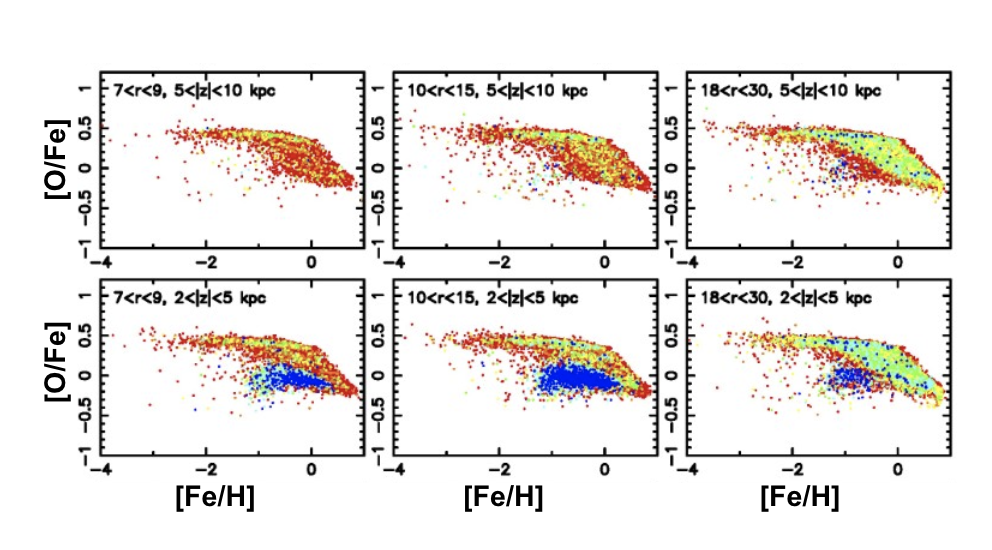}
\caption{The predicted abundance-ratio distribution of disk stars at different Galactocentric radial and height ranges, from the chemodynamical simulation model by \citet{Vincenzo2020}.  The color coding indicates the degree of rotation support: $v_{\rm rot}/\sigma < 3$ (blue), $2.5$ (cyan), 2 (green), 1.5 (yellow), 1 (orange), and 0.5 (red).}
\label{fig:disk_chem}
\end{figure*}

Star formation and chemical evolution in the outer disk may be distinct from the inner disk \citep[as suggested in, e.g.,][]{ciuca_2021,renaud_outer_disk_2021}. The added dimension of age, derivable from PFS/SSP data plus Pan-STARRS photometry,  will enable analyses of the chemical enrichment and star formation history of the outer disk, providing key insight into any potential radial gradients. These data can be compared to expectations from simulations, such as those shown in Figure~\ref{fig:disk_chem}, which   presents the predicted chemical structure at a range of Galactocentric radii and heights within the thin disk, from a state-of-the-art, fully cosmologically consistent hydrodynamic simulation of a MW-like galaxy, with chemical evolution incorporated \citep{Vincenzo2020}. The color-coding indicates the degree of rotation vs.\ dispersion support.  The patterns of elemental abundance ratios encode star-formation and merger/gas inflow histories and there are clear differences between the outer disk (right panels) and `Solar neighborhood' (left panels), at a level that is detectable with PFS spectra.

\subsection{Disk-Halo Interface}

The thin disk, thick disk, and stellar halo of the MW occupy overlapping regions in coordinate, kinematic, and chemical space. The overlap between these structural components of the Galaxy hint at the possibility that they are a continuum of co-evolving and inter-related populations. Indeed, there remains debate over how these components formed, and to what extent they are independent of one another. One possibility is that a kinematically cold, $\alpha$-rich proto-disk formed early on in the evolution of the Galaxy, and some portion this disk was then heated into (at least some fraction of) the in-situ halo via an ancient merger, while the remainder of the heated proto-disk constitutes what is now known as the thick disk. The current thin disk formed subsequently, possibly incorporating both gas from the outer disk and new gas brought in by the merger \citep[e.g.,][for discussion of this and similar theories in the \textit{Gaia} era]{belokurov_2020,bonaca_2020,xiang_rix_2020,conroy2022,chandra_2024}. Other theories, however, do not require a distinct past Galactic merger to form thin and thick disks \citep[see, e.g., the various scenarios presented in][] {schonrich_2009, Bird2021}. Further detailed analyses of the age, chemical abundance, and kinematic information of stars at the disk-halo interface are needed to better constrain the formation history and evolution of the Galaxy's structural components. For example, the planned PFS/SSP observations can better map the age, alpha-abundance, and metallicity distributions as a function of position and kinematics, to define the transitions in key physical parameters. 

\subsection{The Field Halo}\label{sec:halo}
The assembly history of the MW is further imprinted in the field halo stars in the form of chemodynamical phase-space substructures and/or halo streams.  Ancient merging and accretion processes---such as the likely major merger event of GSE $\sim$ 10~Gyr ago ($z \sim 2$)---can be detected in both kinematic phase space and detailed chemical abundances.  The numerous, thin halo substructures or streams, detected as overdensities in star counts \citep[e.g.,][]{Belokurov2006, Suzuki2024}, reflect the more recent accretion/merging/disruption of low-mass stellar systems \citep[e.g.,][]{ibata24}.  Indeed, the growth of our Galaxy since $z \sim 2$ was most likely dominated by  minor mergers---an ongoing process given the assimilation of Sgr (whose tidal stream dominates the structure in star counts) and interactions with the LMC---and smooth accretion from the circumgalactic medium, providing the fuel for sustained star formation in the thin disk.  Much like the situation in the outer disk, structures in chemodynamical (and coordinate) phase space can persist in the halo, reflecting the longer dynamical times there, and thus the halo contains clues about the emergence of the present-day MW\@.

As an example, intrinsically bright, but rare, halo tracers such as BHB stars are accessible to large distances with moderate-aperture telescopes, such as the SDSS 2.5~m.  BHB stars, plus local main-sequence stars, have been demonstrated to be on orbits with mean apocenters of about $r= 20$~kpc, coincident with a break in the halo stellar density profile  \citep{Deason2018}, interpreted by those authors as  apocenter turnarounds of the now-disrupted GSE progenitor galaxy. Recent analysis of distant red giants has revealed a doubly-broken power law with breaking radii at $r=12$ kpc and 28 kpc \citep[][using the H3 survey data]{Han2022}, again possibly indicative of the apocenter pile-up of stellar orbits. The PFS/SSP will provide spectra for distant MSTO stars, at radii covering these putative pile-ups, allowing for a more detailed characterization and dissection of the merger that created the tidal debris.

The debris from a disrupted satellite creates a ``chevron'' pattern in the phase-space of $(r,V_r)$ \citep{Merrifield1998}, as seen in $N$-body and cosmological numerical simulations for galaxy formation, such as FIRE \citep{Donlon2024}. This reflects wrapping of the debris’ phase-space density and has recently been discovered for the local \textit{Gaia} sample \citep{Belokurov2023, wu_2023}. The PFS/SSP data derived from the proposed deep, multi-pencil-beam survey in the MW halo will enable the quantification of this and other phase-space structure out to the probable furthest apocenter passage of the parent dwarf galaxy, thus revealing the early evolution of the merger, and containing information on the gradient of the gravitational potential at the shell/turnaround radius \citep{Merrifield1998}. In addition, the phase-space distribution of halo stars in the MW will be different from that of the M31 halo, and PFS is uniquely situated to facilitate the comparison of these structures. According to cosmological, $N$-body hydrodynamical simulations for the formation of Local Group analogs (APOSTLE: \citealt{Genina2023}), the galaxies with ``quiet" accretion histories, namely those that were assembled early, are more structured in phase space, with clear shells of particles moving on similar orbits. The ``active" galaxies, however, with fast recent growth exhibit less clear structure, with only some shells visible. The former and latter are MW and M31 analogs, respectively. 

The most significant ongoing satellite perturbation to the MW halo is likely due to the LMC, which is now estimated to be significantly more massive than previously thought ($M_{\rm tot} \sim 10^{11} M_{\odot}$, e.g., \citealt{Erkal2019}). The LMC may be on its first infall \citep{Besla2012} or on its secondary passage \citep[][see also \citealt{Diaz2012} for other models reproducing the Magellanic Stream]{Vasiliev2024}, and with either orbit, its high mass induces a density and kinematic wake in both the dark and stellar halos of the MW \citep[e.g.,][]{Garavito-Camargo2019, Chandra2025}. The kinematic signature is a systematic difference---as large as a few tens of km~s$^{-1}$---in the line-of-sight velocities of the distant halo ($\gtrsim 30$~kpc) compared to the inner halo \citep[e.g.,][]{petersen_2021}. This systematic effect in the velocity field will be evident in the PFS observations, facilitating further analysis of the gravitational interactions between the LMC and the Milky Way. 

\subsection{Stellar Streams}

Kinematically cold stellar streams in the MW halo provide tight constraints on the total mass and shape of the dark matter halo and the mass function of its dark substructures, especially when several streams are analyzed together  \citep[e.g.,][]{yoon11,Carlberg2012,Erkal2017, bonaca18, ibata24, Carlberg2024}. Obtaining robust kinematics of cold stellar streams in the outer halo is difficult for most surveys due to the extremely low stellar surface density of bright stream member (giant) stars. The PFS/SSP  observations toward the lines of sight of stellar streams located at heliocentric distances beyond 10 kpc will allow for kinematically and chemically identifying fainter, main-sequence member stars. This will provide much larger samples of member stars and enable stronger constraints on the shapes of the velocity distributions within the streams, potentially a signature of the type of dark matter \citep{Carlberg2024} and on the orbits of the stream progenitors. The line-of-sight velocity information toward the halo streams will also provide the sign of the vertical angular momenta of the streams. Recent analysis has suggested that there is a paucity of retrograde streams in the Southern sky \citep{Li2021}, and the PFS/SSP measurements will provide complementary information on the angular momentum distribution of accreted stellar systems in the Northern sky.

\subsection{Stellar ages}\label{sec:age}

The difficulty of estimating the age of a given field star is a long-standing problem in astronomy (e.g., \citealt{soderblom_2010}). Isochrone-based age estimation methods are perhaps the most straightforward to apply to large numbers of stars, however there is little separation between isochrones of different ages for both the red giant phase and the (cooler) main sequence phase, making estimates difficult and uncertain. Luckily, however, MSTO and subgiant stars are the exception to this statement.  Isochrone-based age estimates work best in these stages of stellar evolution. As shown by \cite{kordopatis2023}, it is possible to provide accurate ages for MSTO stars from isochrone grids with only effective temperature, surface gravity, and metallicity estimates, while additional information is typically needed to accurately constrain the age of giant and lower main sequence stars. PFS will provide the requisite spectroscopic information for MSTO and subgiant stars \citep[cf.,][]{Nataf2024}, and these data along with isochrone grids will be used to provide age estimates for the stars in these phases in the outer MW\@.

\subsection{Sample Selection}\label{sec:MW_targeting}

\begin{figure*}[tpb]
\centering

\includegraphics[trim={5mm 0 10.5cm 1cm},clip,width=1\textwidth]{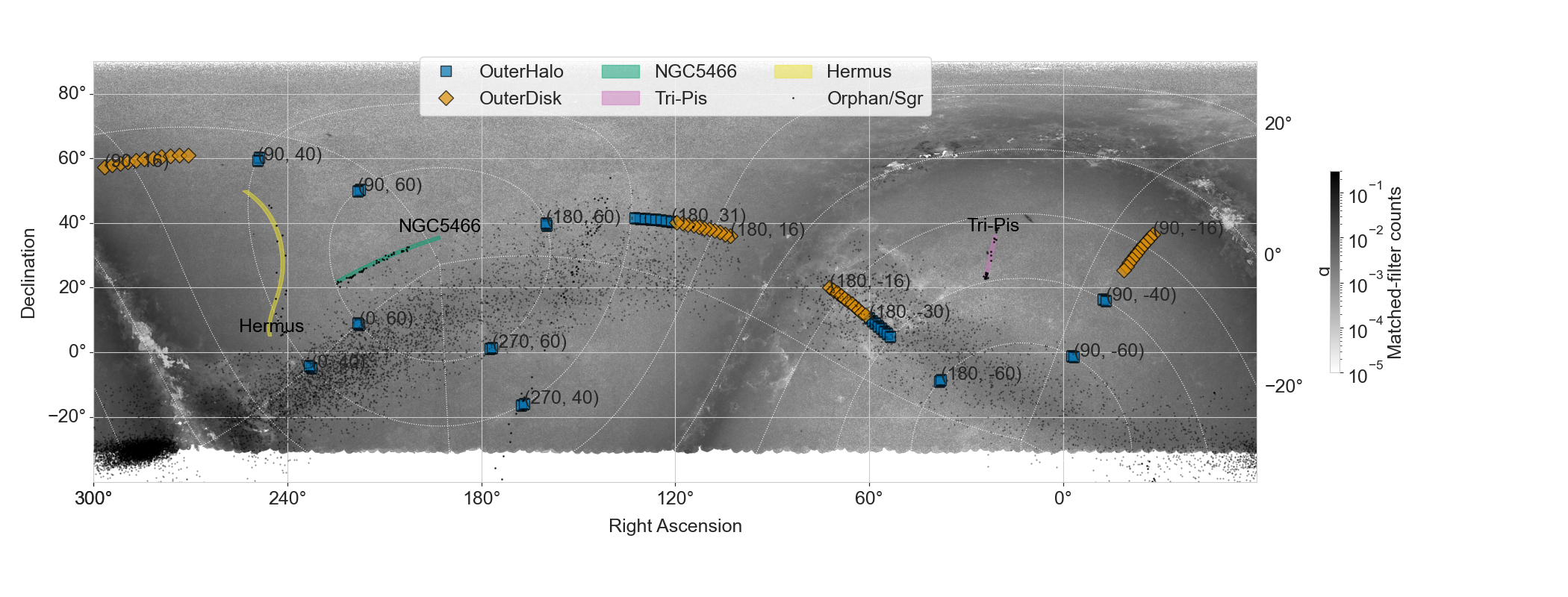}
\caption{
Planned PFS outer Milky Way fields for low and high latitude regions. The low-latitude fields target the outer disk and consists of 
44 contiguous pointings at $l=180^\circ$ and $l=90^\circ$ over $15^\circ < |b| < 30^\circ$. The high-latitude fields target the Galactic halo and include 45 pointings over $30^\circ < |b| < 60^\circ$, including those covering the three halo streams Triangulum--Pisces, NGC5466, and Hermus.  Known member stars of these streams, as well as the Orphan and Sagittarius streams, are shown as black dots \citep{martin18,koposov19,ramos20,mateu23,bonaca25}.
The background gray-scale image displays a matched-filter map optimized for old, metal-poor stellar populations at distances greater than 25 kpc (data adopted from \citet{bernard16}).
}
\label{fig:disk_halo}
\end{figure*}

The PFS/SSP data will enable quantification of the structure of the MW outer disk and halo in position, chemical, and age space.  Fig.~\ref{fig:disk_halo} shows the planned MW pointings. The primary science targets in the outer disk and halo are color-selected FGK-type stars, with  Pan-STARRS photometry \citep{PanSTARRS} within the apparent magnitude range $18 \lesssim g \lesssim 22$, and colors bluer than $(g-r)_0 = 0.8$. FGK-type stars are unbiased tracers of the underlying population, and the PFS/SSP data will allow for the determination of line-of-sight velocities, elemental abundances, and spectrophotometric distance estimates for these stars. The color selection is intended to include stars encompassing a range of phases of stellar evolution, including the BHB, the MSTO, the sub-giant branch, and red giants, while also removing red, M-type stars. While this cut excludes M-giants, so that metal-rich components of substructure, such in as the Sagittarius streams \citep{Majewski2003}, cannot be studied, it serves to remove the numerous nearby, faint M-dwarfs, thereby optimizing the range of distances probed by the sample.

Each of the phases of stellar evolution that are included within the color--magnitude selection can be used as tracers of different properties of the underlying stellar population. For example, precise isochrone-based age estimates will be derived for MSTO and sub-giant stars using PFS elemental abundance information and spectrophotometric distances (see Sections~\ref{sec:age} and \ref{sec:abund}). Age information adds an additional dimension with which to quantify the (sub)structure of the outer disk and halo, as discussed throughout this section. K giants probe a larger volume than would be possible with the less-luminous MSTO and sub-giants alone. Though fainter than K giants, BHB stars are well-calibrated distance indicators, which make them valuable as tracers of MW structure \citep[see][and Section~\ref{sec:bhb}]{xue_2008, fukushima19}. The outer MW fields will also include some extremely metal-poor (EMP, $[\rm{Fe}/\rm{H}] < -3$) stars. We describe the photometric selection of the EMP stars in Section~\ref{sec:emp}, where we further detail the science case motivating the inclusion of these stars. These stars are intrinsically rare, and there will be few (if any) in a given field-of-view in the outer disk and halo.

\subsubsection{Outer Disk Fields}

Given the current lack of sufficiently deep proper motion data (which will be added to the analysis as it becomes available), the PFS fields were chosen to maximize the dynamical information that can be obtained from line-of-sight velocities alone \citep[cf.][]{Kordopatis2017}.  PFS will target low to intermediate latitudes above and below the plane in the key lines of sight\footnote{These two cardinal directions alone are optimal for accessibility of the outer disk both above and below the plane in the Northern Hemisphere.} towards Galactic rotation ($l = 90^\circ$) and the anti-center ($l = 180^\circ$).   In the $l = 180^\circ$ direction, the line-of-sight velocity is dominated by contributions from $V_R$ at all distances, while in the $l = 90^\circ$ direction, the contributions from $V_\phi$ dominate at distances $\lesssim 15$~kpc, albeit with  position- and distance-dependent contributions from the other components of the 3D space velocity (e.g., \citealt{Morrison_1990}, \citealt{Kordopatis2017}).

Given this  mapping of line-of-sight velocity to $V_R$ and $V_\phi$ in the $l \sim 90^\circ$ and $l \sim 180^\circ$ directions, respectively, the PFS/SSP  will be able to engage in high-impact science with line-of-sight velocities alone. For example, the second data release of \textit{Gaia} illustrated the presence of over-dense ridges in the distribution of $V_\phi$ as a function of Galactocentric cylindrical radius in the MW (e.g., \citealt{antoja_2018}), and later studies revealed that some of these ridges correlate with chemical information \citep[e.g.,][]{khanna_2019,gaia_cartography}. In the absence of non-axisymmetric perturbations, the distribution of $R$ versus $V_\phi$ should be smooth, with no ridges, and numerous analyses have worked to link the observed ridges to their possible dynamical origins \citep[e.g.,][]{antoja_2018,khanna_2019,fragkoudi_2019, bernet_2024, jurado_2026}. In the $l \sim 90^\circ$ direction, where $V_\phi$ is the dominant contributor to the line-of-sight velocity, the SSP  will map $R$ vs.\ $V_\phi$  and also  provide the data required to analyze any observed structure as a function of chemical abundances and age. In the $l \sim 180^\circ$ direction, recent studies 
\citep[such as][]{gaia_anticenter,mcmillan_2022} have explored the kinematic substructures at low latitude ($|b|<10^\circ$) using proper motions alone, under the  assumption of an equilibrium state, so that the contributions from the line-of-sight velocities ($V_{\rm{los}} \sim V_R$) are negligible. However, as shown by, e.g., \citet{ding_2021}, \citet{drimmel_2022}, and \citet{gaia_disk}, there exist groups of stars with bulk inwards and outwards motion (i.e., positive and negative mean $V_R$) even within a few kpc of the Sun. PFS will quantify these Galactocentric radial motions ($V_R$) in the anti-center direction as a function of distance.

\subsubsection{Halo Fields}

In order to probe the distant field halo of the MW, our PFS/SSP will target 10 nominal halo field centers, allocating 2 PFS pointings per field, at $(l,b)=(0^\circ, 40^\circ)$,  $(0^\circ, 60^\circ)$, $(90^\circ, \pm 40^\circ)$, $(90^\circ, \pm 60^\circ)$, $(180^\circ, \pm 60^\circ)$ and $(270^\circ,40^\circ)$, $(270^\circ, 60^\circ)$. There will also be 16 halo fields toward $l = 180^\circ$ that will connect to the outer disk fields, with 8 contiguous pointings from $(l, |b|)$ of $(180^\circ, 30^\circ)$ to $(180^\circ, 40^\circ)$ both above and below the plane. 
Cold stellar streams in the outer halo are selected based on the midpoint distances defined by \citet{mateu23}, restricting the sample to those located beyond 15 kpc. To ensure an accurate characterization of the underlying Galactic gravitational potential, we exclude streams exhibiting morphological features suggestive of a dwarf-galaxy origin, and prioritize those with angular widths smaller than $\sim 1.0$ degrees \citep{riley20}. We plan to observe a subset of the outer Milky Way halo fields along the lines of sight toward Triangulum-Pisces \citep{bonaca12,martin13,martin22a,yuan22}, NGC5466 \citep{jensen21,ogami26}, and Hermus or Hyllus \citep{grillmair14,martin18} stellar streams.  Each stream will be observed with three PFS pointings, with an exposure time of three hours per pointing. 

The average PFS pointing will include $N = 200 \sim 400$ spectra for  stars (main sequence and red giants) belonging to the smooth component of the halo, including faint main sequence stars more distant than 10~kpc from the Sun, down to $g\sim 22$, while the total numbers per FoV including foreground disk/halo stars are more than 1,000.  
The systematic motion in each direction will be as precise as $\sigma / \sqrt{N} \sim 3$ km~s$^{-1}$, where $\sigma \sim 150$ km~s$^{-1}$ is the typical velocity dispersion of halo stars. The kinematics plus information from [Fe/H] and [$\alpha$/Fe] ratios will disentangle the stars from the GSE and other known substructures \citep[][]{Naidu2020}. The targeted  field halo observations, in combination with the data for foreground halo stars along the lines of sight to M31 and the 7 dwarf galaxies discussed above, will also capture the kinematic wake induced by the LMC and the distant halo debris associated with the GSE, as discussed above.

\section{Additional Science Cases}
\label{sec:ancillary}

The core GA science  presented above does not fill all of the available fibers of every pointing, leaving some available for ``ancillary'' science cases.  This section presents three examples of ancillary targets  relevant to Galactic Archaeology.

\subsection{Extremely Metal-Poor Stars}\label{sec:emp}

\subsubsection{Background}
Chemically pristine stars in our Galaxy 
hold the key to understand the earliest formation history of the MW \citep[reviewed by][]{Frebel_Norris_2015}. In particular, Extremely Metal-Poor Stars ([Fe/H] $<-3.0$; EMP) provide unique constraints on the yet-elusive nature of the first, metal-free (Population\,III, Pop\,III) stars to form in the universe
\citep[][and references therein]{bromm04,karlsson13,klessen23}. 

Extensive efforts have been devoted to discovering the most metal-poor stars over the last several decades \citep[e.g.,][]{beers&christlieb05,christlieb08, starkenburg17, dacosta19,chiti21}. The resulting photometric and spectroscopic estimates of stellar chemical abundances have been used to obtain the metallicity distribution function (MDF) down to the lowest metallicities \citep[e.g.,][]{schorck09,youakim20,bonifacio21}, to identify the most chemically pristine stars \citep{caffau11,keller14,nordlander19}, and to put constraints on the masses and explosion energies of the first metal-enriching Pop\, III stars \citep{tominaga14,placco15,ishigaki18}. These studies also confirmed that the fraction of carbon-enhanced stars increases at lower iron abundances \citep[cf.][]{NorrisC2013}, suggesting that the carbon enhancement in the lowest metallicity stars is due to nucleosynthesis yields of sources unique to environments in the early stages of chemical enrichemnt, such as the high-redshift universe \citep{placco14,li22}. Recent dynamical studies of metal-poor stars revealed that a certain fraction of very metal poor (VMP) and EMP stars have orbits confined to the Galactic disk, which has  significant implications for star formation during the earliest phase of the Galactic disk formation \citep{sestito19,sestito20,mardini22}. 

The current sample of confirmed VMP or EMP stars is limited in size and cannot fully address the questions of enrichment during the earliest phases of MW formation. In particular, the majority of previously identified VMP/EMP stars are located within only a several kpc of the Sun, where such stars are known to be very rare \citep{youakim17,Frebel2018}. Given that a large fraction of the outer part of the MW volume remains unexplored, the next crucial step is to spectroscopically confirm a large sample of VMP and EMP stars out to large distances,   located well beyond the Solar neighbourhood, and to measure chemical abundances of key elements. 

\subsubsection{Observing strategy}

A number of VMP and EMP stars will, inevitably, be observed serendipitously in the PFS/SSP  GA fields. In addition to these serendipitous detections, we selected candidate VMP and EMP bright stars that could be  filler targets for empty fibers in high-latitude fields---including fields that are part of the PFS/SSP Cosmology survey---using  publicly available photometric catalogs that include data in metallicity-sensitive filters (Ca H+K, or $u$-band). These catalogs include, for example, those from the Pristine survey \citep{starkenburg17}, which is a photometric survey that obtains metallicity estimates for halo stars based on the Ca H\&K regions and has been shown to be reliable down to the EMP regime. The Pristine survey DR1 catalogues \citep{martin_2024} provide photometric metallicities over the whole sky based on synthetic photometry from low-resolution \textit{Gaia} BP/RP spectra and higher quality photometric metallicities for stars in common ($G<17.6$). The latter will soon be superseded by Pristine DR2 that will provide photometric metallicities down to $G\sim20.5$ over $\sim8,500$ deg$^2$ (Yuan et al., submitted) and that we will use for target selection when it becomes available. Other ongoing (and future) photometric surveys (SAGES/J-PLUS survey, HSC-ZERO survey, UNIONS survey, etc.) will also be used to select up to $100$ candidate metal-poor stars (${\rm [Fe/H]} < -2$) per PFS  field-of-view down to $g\sim22$.

\subsection{Blue Horizontal Branch Stars in the Milky Way halo}\label{sec:bhb}

Candidate BHB stars, selected based on HSC ({\it g,r,i,z}) photometry in the HSC-SSP Wide layer footprint over $\sim 1,200$ deg$^2$ \citep{fukushima19,Fukushima2024}, are another category of rare, scientifically interesting stars that can be ancilary targets. These stars, important tracers of the old stellar populations in the Galaxy, have the advantage of well-calibrated absolute magnitudes ($M_g \simeq 0.45$~mag), so that accurate distances to individual stars better than 10\% are available \citep{Chen2009}. They are also luminous enough to serve as excellent dynamical tracers of the outer part of the stellar halo, out to and beyond $r \simeq 50$~kpc, extending to the halo's  outer edge, which is  possibly associated with the  ``splashback radius'' that separates infalling from bound material \citep[e.g.,][]{ONeil2021}.

Our scientific goals with PFS spectroscopy of these candidate BHBs 
are summarized as follows:
\begin{itemize}
\item We will spectroscopically identify {\it bona fide}\/ BHB stars, together with Blue Stragglers (BSs) having the same A-type color but higher gravity than BHB (and with $M_g = 2 \sim 3$~mag).  Both are  important tracers in the inner part of the halo within $r \sim 60$~kpc. For $g \leq 21$~mag, there exist 440 known BHBs and 1464 likely BSs the HSC-SSP footprint.
\item With this refined sampling of BHBs (for which heliocentric distances are about 130, 80 and 50~kpc at $g=21$, 20 and 19~mag, respectively), we will derive the density profile of the outer part of the stellar halo beyond $r \sim 50$~kpc to assess if it shows indeed a shallower slope than the inner, {\it in situ.} halo as predicted in numerical simulations of galaxy formation \citep{Rodriguez-Gomez2016}. 
\item Based on the estimation of the success rate for the selection of BHBs from HSC photometry, we will constrain the statistical significance of the presence of the splashback radius, which has been  identified from the spatial distribution of  the photometry-selected faint BHBs (with $g = 22.5 \sim 23$~mag) at around $r = 300$~kpc \citep{Fukushima2024}.
\item We will derive line-of-sight velocities of these halo tracers to determine the 3-dimensional velocity distribution of the outer part of the halo, by combining these data with existing Gaia proper motions, This will yield the mean rotational velocity in a given line-of-sight, $\langle V_{\phi}(r) \rangle$, velocity dispersions ($\sigma_r,\sigma_\theta,\sigma_\phi$), and velocity anisotropies, $\beta(r) = 1 -(\sigma_\theta^2+\sigma_\phi^2) / (2\sigma_r^2)$,  for different ranges of metallicities. 
Furthermore, this velocity information will also be an important dynamical probe of substructure, including a new feature recently identified at $r \sim 60$~kpc from photometry in the HSC-SSP footprint \citep{Suzuki2024}.
\end{itemize}

\subsection{Cool Subdwarfs in the Galactic Halo}\label{sec:ucd}

Over $75\%$ of the stellar census is comprised of late-type dwarfs, those  with $T_\mathrm{eff}<4000\ \mathrm{K}$ \citep{local_census}. This temperature range comprises mostly stellar spectral types M1\,$-$\,M6 (\textit{cool dwarfs}, or CDs), together with mixed stellar and brown dwarf spectral types $\geq$M7 (referred to as \textit{ultracool dwarfs}, UCDs). (U)CDs are prime targets for Galactic archaeology, as the dominant molecular opacity in such cool atmospheres makes their spectra particularly sensitive to elemental abundances. Furthermore, (U)CDs undergo minimal nuclear processing (due to their long evolutionary timescales), thus avoiding the need to correct their inferred stellar parameters for evolutionary effects, unlike the case for  higher-mass stars \citep{placco14,red_giant_abundances_bias_2}. For brown dwarfs in particular, the lack of nuclear fusion results in long-term cooling, thereby providing insight into the age of the parent population \citep{BD_kinematic_ages,SANDee}.

In addition to the utility of cool stars as chemical tracers, (U)CDs are especially suited to address a number of unresolved problems in galactic evolution and cosmology. For example, trends of (U)CD kinematics as a function of spectral type allow one to differentiate among proposed disk heating mechanisms that influence the structural evolution of galaxies \citep{UCD_disk_heating_1,UCD_disk_heating_2}. The (U)CD mass function serves as a key diagnostic of the extent of scale invariance in star formation processes \citep{20_pc_sample}. Brown dwarfs with masses below $\sim70\ \mathrm{M}_\mathrm{J}$ do not undergo lithium fusion \citep{lithium_test} and are therefore prime targets for studies of astrophysical solutions to the cosmological lithium problem \citep{Spite1982,cosmological_lithium_problem}.

Metal-poor (U)CDs are particularly valuable in galactic archaeological surveys as they plausibly represent the earliest stellar generations in the Galaxy. The faint luminosities of cool stars bias the known sample towards nearby objects that predominantly belong to the metal-rich thin disk \citep{thin_disk_to_halo_ratio}. Despite the large number of available empirical spectral libraries, the metal-poor (U)CD population remains vastly underrepresented. For example, the large \textit{MaStar} \citep{mastar} library of fluxed stellar spectra contains $\sim 24,000$ unique stars in total and explicitly endeavored to include cool, metal-poor stars, but the final catalog contains only a handful of low-metallicity CDs.
For yet intrinsically fainter and cooler UCDs, only $\sim 100$ subdwarf candidates have been confirmed spectroscopically \citep{BONES}.

Photometric filtering and the inclusion of astrometric information (e.g., from the \textit{Gaia} satellite) make it possible to identify candidate metal-poor (U)CDs  by their large transverse velocities, characteristic of the thick disk and  halo \citep{velocity_dispersion}. A cross-match of \textit{PanSTARRS} \citep{PanSTARRS} photometry with \textit{Gaia} parallaxes and proper motions \citep{Gaia2023} in the HSC-SSP fields (see Fig.~\ref{fig:disk_halo}) yields a typical target density of $20\ \mathrm{deg}^{-2}$ for M-dwarfs (based on the color selection from \citealt{2016ApJ...833..281K}) with transverse velocities over $100\ \mathrm{km}\ \mathrm{s}^{-1}$. Of these targets, approximately $1.5\%$ ($\sim 300$ targets in total) have $T_\mathrm{eff}<3000\ \mathrm{K}$ (based on the color-$T_\mathrm{eff}$ relation from \texttt{SANDee} isochrones; \citealt{SANDee}), corresponding to the UCD regime. Therefore, even with a conservative assumed success rate of $\sim30\%$, PFS is expected to double the number of known ultracool subdwarfs, and to significantly expand the existing spectral libraries of metal-poor CDs.

The primary goal of our (U)CD survey is to identify the spectral types and measure the fundamental stellar parameters ($T_\mathrm{eff}$, $\log(g)$, $[\mathrm{Fe/H}]$) of the observed targets, using available low-temperature spectral models in the literature \citep{SAND,ElfOwl}. The new spectral library will comprise a homogeneously-observed sample that spans the entire M dwarf regime, and enable population-level studies of the initial mass function, kinematic ages and galactic scale heights as a function of metallicity.

\begin{figure*}[tpb]
\centering
\includegraphics[width=2.0\columnwidth]{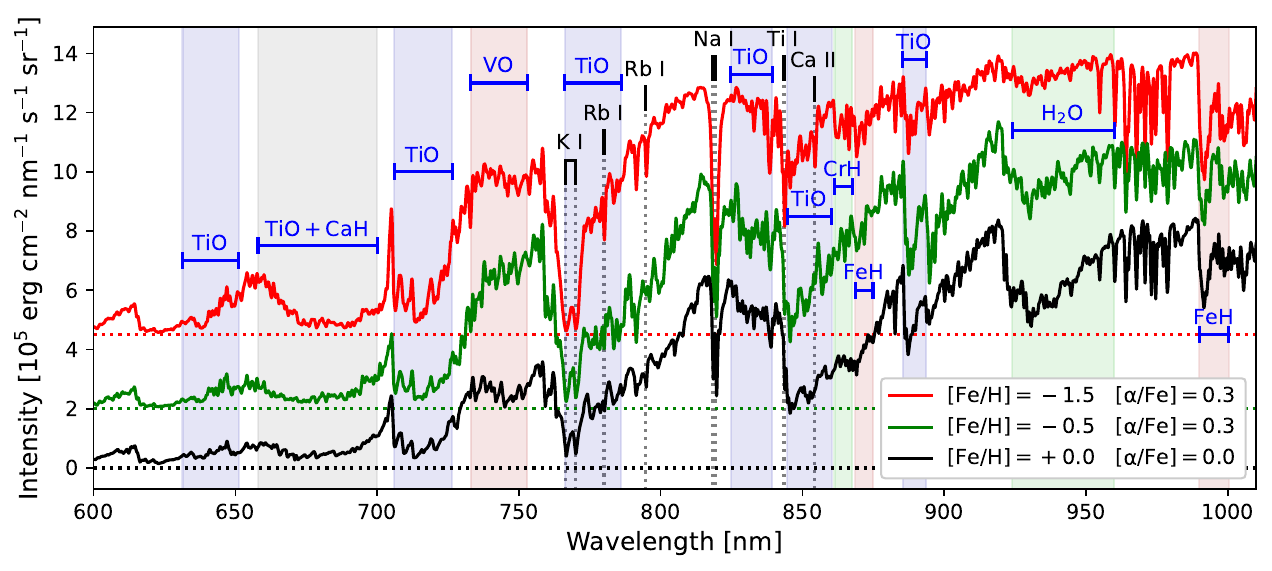}
\caption{Synthetic spectra of UCDs ($T_\mathrm{eff}=2700\ \mathrm{K}$) with assumed elemental abundances characteristic of the thin disk, thick disk and the Galactic halo \citep{APOGEE_zscale_alpha_distribution}, shown in \textit{black}, \textit{green} and \textit{red} respectively. Prominent molecular absorption bands and atomic lines are highlighted and labeled. Intensities have been offset for clarity, as indicated by dotted horizontal lines. Subsolar metallicity models were calculated using the \texttt{SAND} framework \citep{SAND}, while the solar metallicity model was adopted from \citet{BT-Settl}.}
\label{fig:ucd_spectra}
\end{figure*}

For a subset of brighter (U)CDs ($z<20$), a more detailed analysis will be carried out using custom grids of model atmospheres with varied abundances of individual elements, similar to the approach of \citet{speedyLdwarf}. The wavelength coverage of PFS allows for measurements of $[\mathrm{Ti/Fe}]$, $[\mathrm{V/Fe}]$, $[\mathrm{Ca/Fe}]$, $[\mathrm{Na/Fe}]$ and $[\mathrm{K/Fe}]$ from the strengths of molecular bands and pressure-broadened lines of alkali metals (see Fig.~\ref{fig:ucd_spectra}). In a few cases, constraints on magnetic activity may be derived from the intensity of chromospheric $\mathrm{H}\alpha$ emission \citep{optical_subdwarfs}.

\section{Targeting Strategy}
\label{sec:targeting_strategy}
\subsection{HSC Pre-Imaging}

We have obtained pre-imaging with HSC for the dwarf galaxies, M31, and M33, with the  details of the photometric observations given 
in Section~\ref{sec:targetdwarf} (for dwarf galaxies) 
and in Section~\ref{sec:M31_preimage} (for M31). 
The data were reduced with the standard HSC reduction 
pipeline  \citep[hscPipe][]{Bosch2018}. 
The hscPipe version 4.0.5 was used for the imaging data, 
with the exception of the narrow-band (NB515) data in M31, 
for which hscPipe 6.7 was used. 
We followed standard hscPipe procedures with the following modifications:
\begin{itemize}
\item 
For NB515 data, we first calibrated against Pan-STARRS (PS1) $g$-band photometry and adjusted the NB515 magnitude slightly 
to match the stellar locus of Galactic main-sequence stars 
in the two-color diagram. 
\item 
For dwarf galaxies, we used smaller patch sizes 
(1,000 pixels, instead of the standard 4,000 pixels) 
to avoid deblending problems that often occur
at the crowded galaxy centers. 
\end{itemize}

We chose those objects classified as stars (${\rm extendedness} = 0$) in 
either the $g$ or $i$ band ($g$ and $r$ for Bo{\"o}tes~I), and used them in the following analysis. 
Extendedness can separate stars and galaxies 
for the magnitude range relevant to the PFS survey 
\citep[$i < 23$,][]{Aihara2018}.
The foreground Galactic extinction along the line-of-sight to each star is calculated 
using the  estimates 
from \cite{Schlafly2011}, derived based on the dust
map of \cite{Schlegel1998}.

As shown in Figures~\ref{fig:m31_cmd} and \ref{fig:umi_hsc_cmd}, 
the photometry covers the RGB sequence of M31 
and of our target dwarf galaxies. 
The imaging data are 50\% complete down to 
$i \sim 25.5$ and deep enough to cover the 
HB in all the  dwarf spheroidal galaxies in our sample, 
and down to the MSTO of the more nearby dwarf galaxies (e.g., Ursa Minor, Draco, and Bo{\"o}tes~I).

\subsection{\ld{Photometry-based membership estimation for dSph targets}}
\label{sec:HSCdwarftargetselection}

\ld{Targeting of member stars of dwarf galaxies is based on HSC broadband and NB515 (except for Bo{\"o}tes~I and NGC~6822) photometry, combined with available parallax and proper motion measurements from Gaia DR3 (primarily to filter out foreground stars).   At the distances of the targeted dwarf galaxies, the survey $g$-band apparent spectroscopic magnitude limit of $g \lesssim 23$ corresponds to  the lower part of the RGB, and in some cases the MSTO\@. The steep nature of the RGB luminosity function means that significantly more potential targets are available at fainter magnitudes, and the member/foreground ratio decreases with brighter magnitudes. As a consequence, successful targeting of member stars toward the bright end of the RGB is the most challenging, especially when no narrow band photometry is available. }

\ld{The technique we outline here can give reliable estimates on the member/foreground probability for each star, but various targeting strategies are possible, depending on the objectives of the survey. Optimizing for success rate, i.e., targeting stars with the highest possible membership probability, might result in a sample too faint to measure important properties such as radial velocity or elemental abundances. Balancing targeting toward bright stars, on the other hand, has the risk of lowering the success rate, yet it might be a better strategy when a higher success rate of RV and abundance measurements is the goal. Overall, the targeting strategy has to be optimized for the scientific yield of the resulting data, which, in our case, is the best possible mapping of the chemodynamic stellar populations of the dSphs, required for characterization of the dark matter halo profile.}

\ld{To estimate the probability of membership of stars, we developed a stochastic model, based on the modeling of the stellar populations of each of the target satellite dwarf galaxy and the MW foreground, that generates synthetic color-magnitude diagrams from Dartmouth isochrones \citep{Dotter2008}. By generating millions of stars with randomly drawn physical parameters and photometric noise estimated from hscPipe, we can draw properly normalized probability maps that, when combined with the observed projected radial density profile of the satellite galaxy and the spatial density of foreground stars as a function of magnitude and color, can be used to estimate membership probability.}

\ld{We model the stellar populations of the dwarf galaxies by prescribing the probability density function of age and metallicity in the form of a strongly correlated multivariate normal distribution truncated at certain limiting values for each parameter. The values of the parameters of the distributions, such as mean age and its scatter and mean metallicity and its scatter, are taken from the literature, then adjusted to match the observed color-magnitude diagrams from the HSC data. 
A fiducial distance estimate for each dwarf galaxy target and a universal initial mass function \citep{Chabrier2003} are assumed. Parameter distributions for the different components of the MW foreground are taken from the Galaxia models \citep{Sharma2011}.} 

\ld{The simulation of CMDs containing large numbers of stars is a computationally expensive task, which  we achieved with the help of GPUs \citep[see][for details]{Dobos2023}.}

\ld{As noted above, the simulated color-magnitude diagrams presented here were made using the Dartmouth isochrone tables \citep{Dotter2008}, which do not follow the evolution of stars beyond the red giant phase. As a result, the model does not include the asymptotic giant branch (AGB) or the blue horizontal branch (HB)\@. The HB was excluded by a blue-end observed color cut, which was set at $g - i = 0.12$. HB stars are desirable targets and we include potential HB members in the observing lists by hand, through color and magnitude cuts, tuned for each dSph. Similarly, targeting AGB stars is accomplished by targeting stars bluer than the RGB with a lower priority than the RGB\@.}

\begin{figure}
    \centering
    \includegraphics{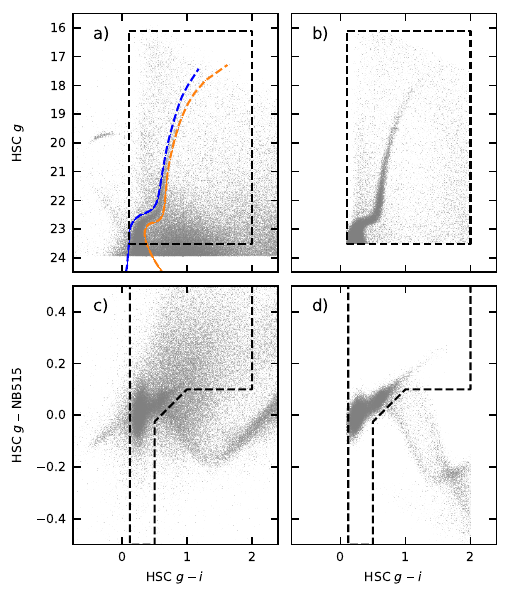}
    \caption{\ld{Color--magnitude and color--color diagrams, corrected for reddening and extinction, of the Ursa Minor dSph plotted from HSC observations (a and c) and our simulation (b and d). The broadband color cuts in (a) and (b) are $0.12 \le g - i \le 0.5$, and the magnitude cuts are $ 23.5 \le g \le 16$ in the HSC photometric system. The slanted line in the color--color selection was chosen to exclude the vast majority of main sequence stars. The two Dartmouth isochrones overplotted in panel a) assume $\FeH = -2.5, t_\mathrm{age} = 13.65$~Gyr (blue dashed curve) and $\FeH = -1.6, t_\mathrm{age} = 12.885$~Gyr (orange dashed curve). The width between the pair of isochrones is widened to simulate photometric error.}}
    \label{fig:umi_hsc_cmd}
\end{figure}

\ld{We also created synthetic two-color diagrams using the NB515 filter. The Dartmouth isochrone tables provide only broadband photometry, but give the atmospheric parameters and luminosity as a function of $\FeH$, age, and evolutionary phase. This allowed us to calculate the synthetic NB515 magnitudes from the BOSZ stellar spectrum grid \citep{Bohlin2017}. We achieved a good match between the synthetic data and the observed color--color distribution for RGB stars, but the synthetic narrow-band data for MSTO stars show much less variance than do the observed data. Further, the modeled color main sequence stars redder than $g-i > 0.5$ is incorrect, so we excluded this region entirely from the simulated CMD, as indicated in Figure~\ref{fig:umi_hsc_cmd}.}

\ld{The two top panels of Figure~\ref{fig:umi_hsc_cmd} show the observed (a) and simulated (b) HSC CMDs for the total area of the sky around the Ursa Minor dSph to be observed within the PFS/SSP, as shown in Figure~\ref{fig:dwarf_pointings}. The two overplotted isochrones in panel~(a) indicate the most extreme ages and metallicities consistent with those of  spectroscopically confirmed members of Ursa Minor published in the literature. While discrepancies between observed and simulated CMDs are visible, the simulations match the observations well for purposes of target selection.}

\ld{Forward modeling the CMD also allows for an assessment of the utility of narrow-band photometry for aiding the selection of bright stars near the tip of the RGB\@. The bottom two panels of Figure~\ref{fig:umi_hsc_cmd}, show the observed (c) and simulated (d) HSC color--color diagrams. Some key differences are readily observable: the dispersion of member stars in the simulated catalog along the $g - \mathrm{NB}515$ axis is smaller than observed, and the simulated RGB extends to redder colors than is observed. While the former probably reflects an issue with the stellar spectrum library we used to compute the synthetic magnitudes, the latter we attribute to there being stars of higher metallicity in the simulated data. As indicated by the selection boxes in panels (c) and (d), we completely excluded the lower right part of the color--color diagram from the analysis, due to the incorrect NB515 synthetic magnitudes of cool main sequence dwarfs.}

\ld{Simulations not only gave us the magnitudes for each star but also an estimation of their population membership. To determine the membership probability from simulated CMDs, we generated a 2-D histogram of the simulated stellar populations  in the $(g-i)$--$g$ color--magnitude space, treating  member stars and foreground stars separately, with a resolution of $0.01$ in $g-i$ and $0.065$ in $g$. In order to reduce the noise of the histograms, we generated 250,000 simulated stars for each population and smoothed the histograms with a $3\times3$ maximum filter.}

\begin{figure}
    \centering
    \includegraphics{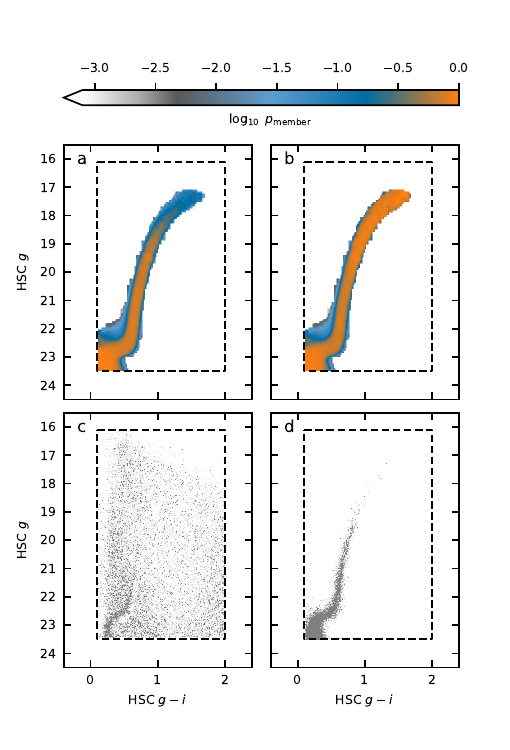}
    \caption{\ld{\textbf{Panels~a~and~b:} Membership probability of stars observed in the field of the Ursa Minor dwarf galaxy based on HSC broadband and narrow-band photometry from stellar population simulations. Panel~(a) shows the probabilities based on broadband colors only while panel~(b) is based on the broadband photometry plus the  NB515 narrow-band filter. 
    \textbf{Panel~c:} ``Ghost plot'' of the observed CMD with stars randomly removed according to their membership probability. The ghost plot is a useful visualization of the validity of the membership probability estimates: when the membership estimates are correct, the turn-off and the RGB of the dSph should disappear. \textbf{Panel~d:} The analogous ghost plot to that in panel~(c), now with the foreground stars removed randomly.}  All observed magnitudes and colors have been corrected for reddening and extinction.} 
    \label{fig:umi_hsc_pmap}
\end{figure}

\ld{The top two panels of Figure~\ref{fig:umi_hsc_pmap} show the probability maps derived from the CMD simulation of the field of Ursa Minor, with (a) and without (b) making cuts on the NB515 magnitude. While the membership probability is equally high in both cases for stars toward the base of the RGB, due to the very high number density contrast between the member stars and the foreground stars, the color--magnitude region toward the tip of the RGB is predicted to be dominated by foreground stars when only the broadband colors are considered. The NB515 filter very successfully excludes foreground main-sequence stars and increases the probability of successful targeting of actual UMi members to close to unity.}

\ld{Panels~(c)~and~(d) of Figure~\ref{fig:umi_hsc_pmap} show ``ghost plots.'' These CMDs were generated, for visualization purposes only, from the observed HSC data by removing stars according to their membership probability. In panel~(c), the remaining stars follow the distribution of the foreground stars, except for a few over- and under-densities where our membership model does not properly match the actual members. In panel~(d) we kept only observed stars with high  membership probability and successfully recovered the distribution of the end of the main sequence, the MSTO and the RGB of UMi\@.}

\subsection{Fiber allocation}

Fibers are allocated to stars based on their priorities, derived from the membership probabilities.  Each target is assigned a non-observation cost based on its priority.  The fiber allocation must satisfy certain constraints, such as the inclusion of a minimum number of sky fibers, a minimum number of flux-calibration standard stars, and avoidance of fiber collisions.  A flow network called \texttt{netflow}\footnote{\url{https://github.com/Subaru-PFS/ets_fiberalloc}} is run to find a solution that both satisfies the constraints and minimizes the non-observation costs.

\begin{figure}
    \includegraphics{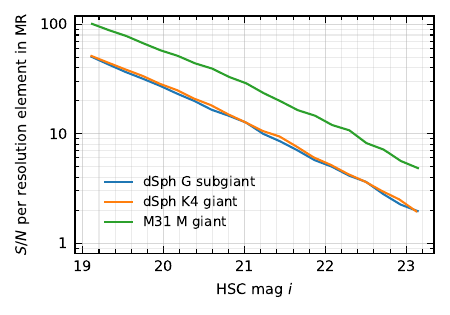}
    \caption{\ld{Expected signal-to-noise per resolution element in the medium resolution red arm, for three different types of targets. Targets in the fields of dSphs will have an exposure time of 3~hours,  while M31 targets will be observed for 5~hours.}}
    \label{fig:snr}
\end{figure}

Stars are observed for three hours or until their S/N reaches approximately 60, which should be achieved in 3 hours, under nominal conditions, for stars with $i \sim 19$ (see Figure~\ref{fig:snr}).  The maximum exposure time is scaled according to the brightness of the stars for stars with $i < 19$.  After a star has reached its maximum exposure time, the fiber can be reassigned, should  another star be available within the fiber's patrol region.  In this way, it is possible to observe more than 2,400 stars in one pointing (i.e., more than the number of fibers).

\begin{figure*}
    \centering
    \includegraphics{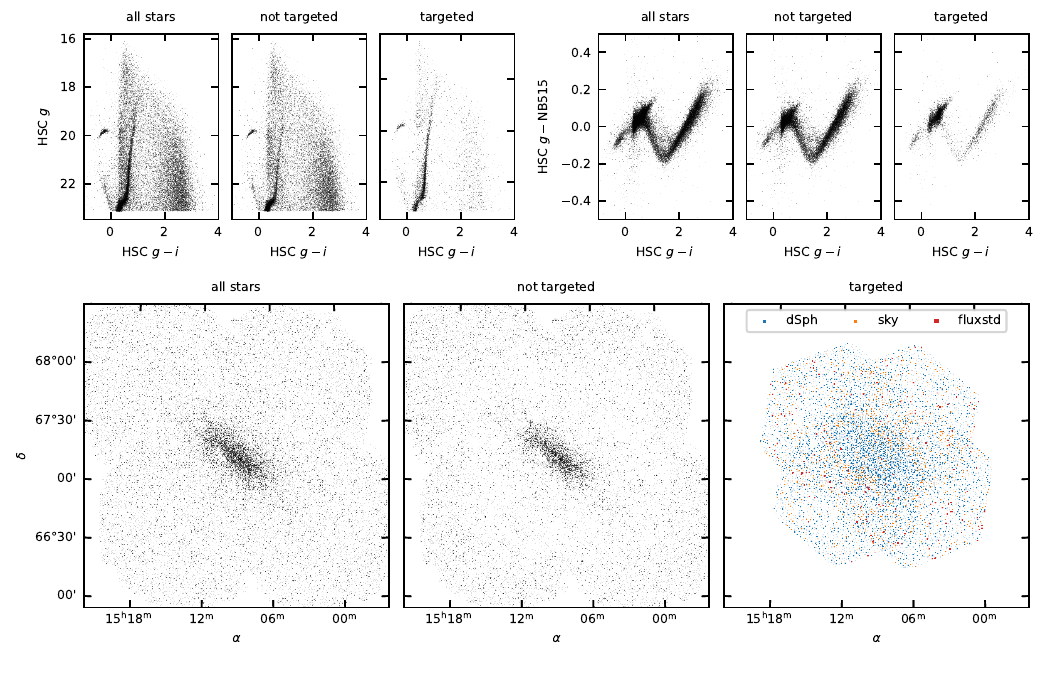}
    \caption{The color--magnitude (top left) and color--color (top right) diagrams, corrected for reddening and extinction, for the stars of the Ursa Minor dSph.  The innermost four pointings (of the planned eight pointings) are shown.  Stars with fibers assigned are shown in the right panels.  Untargeted stars are shown in the middle panels.  The spatial distribution of the stars is plotted in the bottom row. The bottom right panel shows the targeted stars in blue as well as sky fibers (orange) and flux calibration standard stars (red).}
    \label{fig:umi_target}
\end{figure*}

Figure~\ref{fig:umi_target} shows the stars selected for observation in the four central pointings of Ursa Minor (those outlined in red in Figure~\ref{fig:dwarf_pointings}).  Priorities were assigned according to the procedure in Section~\ref{sec:HSCdwarftargetselection}, which incorporates the probability map (Figure~\ref{fig:umi_hsc_pmap}) for RGB and MSTO stars.  The four pointings include 4,246 candidate member stars and an additional 3,233 non-member stars, some of which are ancillary targets (Section~\ref{sec:ancillary}).

\begin{figure}
    \centering
    \includegraphics[trim={0.3cm 0 0 0},clip]{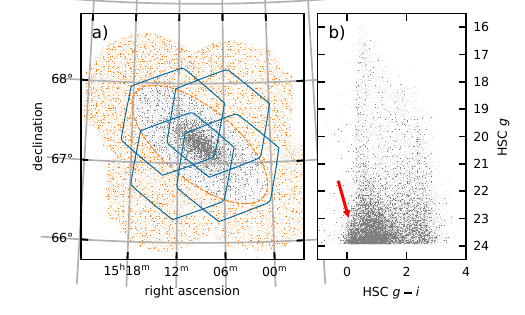}
    \caption{\ld{\textbf{Panel a:} Stars observed by HSC in four fields around the center of the Ursa Minor dSph. The dashed ellipse marks the nominal tidal radius, while the blue hexagons indicate these inner PFS pointings. \textbf{Panel b:} The color-magnitude diagram, corrected for reddening and extinction, of stars located outside the nominal tidal radius. Even though the sample is clearly dominated by likely foreground Milky Way halo and disk stars, the upper main sequence and MSTO of the Ursa Minor population are visible (red arrow).}}
    \label{fig:umi_hsc_tidal}
\end{figure}

One of the advantages of PFS is its wide field of view. For example, the four inner fields of the Ursa Minor dSph reach beyond the nominal tidal radius.  Several dSphs are known to have member stars beyond their nominal tidal radius \citep{FilionWyse2021,Longeard2022,Yang2022,Sestito2023a,Sestito2023b} and the ability to map ``extra-tidal'' stars provides important information about the equilibrium---or otherwise---of the system, and adds constraints on viable mass models. Indeed, Figure~\ref{fig:umi_hsc_tidal} shows that likely Ursa Minor member stars---especially those at the MSTO---are visible in the CMD of stars located beyond the nominal tidal radius.  Compare Figure~\ref{fig:umi_target} to Figure~\ref{fig:umi_hsc_tidal} to see how the PFS/SSP can map these extra-tidal stars in Ursa Minor.

\section{Planned Spectral Analysis}

We have developed methodologies and software to derive meaningful quantities from the stellar spectra obtained with PFS\@.  These quantities include radial velocities (RVs; Section~\ref{sec:rv}), as well as overall metallicity and selected elemental abundances (Section~\ref{sec:abund}). The broader PFS/SSP survey team  has developed a separate pipeline that is particularly attuned to measuring galaxy redshifts \citep{Greene2022}.  Here we describe the pipelines that will analyze the PFS/SSP GA survey of resolved stars and will be made available to the wider PFS community.

\subsection{Radial velocities}
\label{sec:rv}

The vast majority of surveys of resolved stars with massively multiplexed spectrographs have developed a dedicated pipeline to measure RVs and stellar parameters.  For example, the S$^5$ survey \citep{Li2019} uses maximum likelihood to find the best velocity and the best matching template spectrum (and therefore stellar parameters) for each observed spectrum. Their procedure is based on a least-squares fit between a library of synthetic spectra and the observed spectrum \citep{Koposov2011,Koposov2019}. Similar methodologies have been adopted by other surveys, including RAVE \citep{kordopatis2013}

\begin{figure}
    \includegraphics{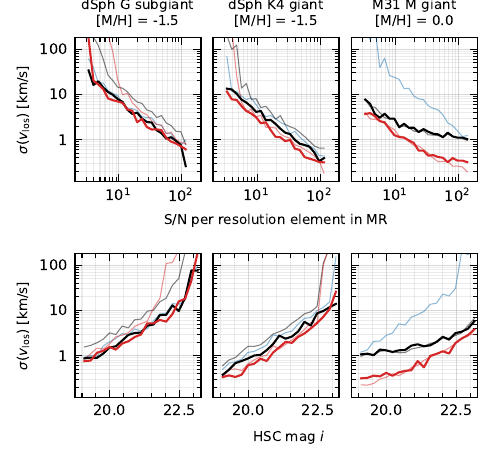}
    \caption{\ld{The uncertainty of RV measurements (random errors only) as a function of $S/N$ per resolution element, for the medium-resolution red arm (top row) and of HSC~$i$ magnitude (bottom row), for three different stellar types, similar to those that the PFS/SSP will target in dSphs ($t_{\rm exp} = 3$~hr) and in M31 ($t_{\rm exp} = 5$~hr). The colors indicate the spectrograph arms included in the simulations: blue -- blue arm, black -- low resolution red arm, red -- medium resolution red arm. The thin lines indicate that only one arm was included in the simulations, whereas the thick lines show the results for the simulation when the blue arm, as well as the low- or the medium-resolution red arm, was used. The thick red line (blue plus medium-resolution red) represents the default observational mode for PFS/SSP GA targets.}}
    \label{fig:rv_error}
\end{figure}

\ld{The PFS/GA-SSP program targets very faint stars in fields with low target density. This is particularly apparent in the outer regions of the dSph galaxies where the relevant target density can be lower than $1/\mathrm{fiber}$ at a magnitude cut of $m_i < 22.5$. This encourages us to target fainter stars with a maximum achievable signal-to-noise of ${\rm S/N} < 3$ per resolution element. Our simulations indicate that RV measurements, albeit with large random errors (ranging from 3 to 10~km~s$^{-1}$ depending on spectral type), can be obtained even at these high levels of noise \citep[][see Fig.~\ref{fig:rv_error}]{Dobos2024}. In order to make use of RV measurements in further analyses, such as modeling the radial mass profile of dSphs, understanding the distribution of the error in RV estimates is paramount \citep[e.g.,][]{Wardana2025}.}

\ld{Repeated simulations of RV fitting show that the asymptotic uncertainty estimators typically calculated around best-fit RV values, such as the Fisher information of the likelihood function, are reliable only down to  moderately high S/N, and fail to describe the error distribution at very low S/N, when the assumed Gaussianity conditions are no longer met. We found that, at low S/N, the variance of the error distribution is much larger than the asymptotic estimates of the variance, and higher moments of the error distribution are also significant.  As an alternative to such estimators, Monte Carlo sampling of the likelihood function, while significantly more computationally intensive, is capable of recovering the real distribution of RV error, even at lower S/N\@. }

\ld{The PFS/SSP will observe GA targets with a series of  15~minute exposures, for total exposure times of  $3$-$5$~hours. This results  in $12$--$20$ individual, low S/N spectra for a given target, each not necessarily observed with the same fibers. As a consequence, single-exposure spectra will be pixelized differently and will have slightly different S/N, in addition to different flux-calibration   systematics. In order to address these issues, we developed our own synthetic template-fitting algorithm that works from the per-pixel flux data of the individual exposures and relies on likelihood stacking instead of exposure stacking. Fluxing systematics, as well as the discrepancies of the continuum of synthetic templates, are treated with a wavelength-dependent multiplicative flux-correction function in the form of a low-order polynomial, in parallel with fitting RV and the fundamental stellar atmospheric parameters. 

}

\ld{To fit RV and the stellar atmospheric parameters, we start out with a very high resolution synthetic stellar spectrum grid, as described in Section~\ref{sec:abund}. During the fitting iteration, the model flux is interpolated to arbitrary values of the stellar parameters within the grid bounds, using multi-dimensional linear interpolation. Then, the templates are Doppler-shifted according to the required line-of-sight velocity and convolved with the line spread function and interpolated to the detector pixels. Even though the spectral resolution of the PFS instrument is low to medium, we work with templates of resolution at least $R=50,000$ in order to avoid any numerical errors originating from the convolution. To speed up the convolution step, we apply a trick based on Principal Component Analysis: We express the high resolution kernel function on a truncated basis of just a few (typically only five) eigenvectors and instead of evaluating the kernel at every wavelength, we calculate the convolution of the template spectrum with each eigenvector and interpolate the expansion coefficients of the kernel to each wavelength. This saves a lot of CPU time when the convolution is performed at high resolution and kernels can be as wide as hundreds of wavelength bins.}

\ld{When correcting for fluxing systematics, our template-fitting algorithm implements the maximum-significance method, which is similar to the maximum-likelihood method, but the parameter uncertainties are determined differently. This  approach was adapted from that of \citet{Kaiser2004}, who developed this technique  to identify very faint point sources in noisy 2-D images. The key idea of the method is that the parameters of the flux-correction function do not affect the significance of an RV measurement, but can increase its estimated uncertainty. Compared to the uncertainties derived from the second derivatives of the likelihood function, the difference is usually small, but becomes more pronounced at low S/N\@. We determine the uncertainty either from the Fisher information matrix by taking the numerical Hessian of the significance (likelihood) function at the maximum significance point, or run a full Monte Carlo sampling of the significance (likelihood) function. In addition, priors on the atmospheric parameters (based on photometric data), and a prior on RV (based on, for example, confirmed membership in a dSph) can be defined, if necessary. In the case of using priors, the Bayesian posterior probability density, which is the significance function multiplied by the priors, is either maximized using the Nelder--Mead simplex method, or sampled using adaptive Monte Carlo.}

\ld{Reliably estimating the likelihood (posterior probability density) function around its maximum has particularly high importance when the RV data are used to fit dynamical models, such as the determination of the radial mass profile of dSphs. Instead of excluding stars with large RV errors, they can be included in a detailed Bayesian model to infer the LOSVD and other parameters when the full error distribution is known.}

\ld{Using simulated observations, we found that template mismatch has minimal impact on the uncertainty of RV measurements in the signal-to-noise range of approximately $10 \lesssim S/N \lesssim 100$. Further, we found that  the bias of the RV estimates is negligible when a flux correction function is applied, in order to match the overall shape of the templates to those of the fluxed observations \citep{Dobos2024}. In the range of $5 \lesssim S/N \lesssim 10$, however, the bias of RV measurements increases to about $1~\text{km}~\text{s}^{-1}$, a value comparable to the systematics of the wavelength calibration,  which is expected to be in the $1$-$2~\text{km}~\text{s}^{-1}$ range but remains to be quantified. Above $S/N \sim 30 $, the expected bias remains below $0.5~\text{km}~\text{s}^{-1}$, independent of possibly  mismatched atmospheric parameters between template and star. However, the uncertainty of the RV measurement can double, with respect to the case of a perfectly matching template, when a template with the incorrect [Fe/H] or $\log g$ is used. Nevertheless, when the fundamental atmospheric parameters are fitted along with the RV, both the bias and uncertainty behave the same as in the case of  RV fitting  with a template that has matching values of the  fundamental parameters. Since RV fitting is not very sensitive to the choice of template, we use linear interpolation when interpolating the flux from a synthetic spectrum grid to an arbitrary set of fundamental parameters.}

\ld{In order to detect binary stars, we repeat some fields of the dSphs one year after the first observations and in some cases on a time-scale of a few months. Depending on how well we succeed in characterizing the systematics, and especially how stable they turn out to be over time, we will be able to measure velocity \textit{differences} with an uncertainty on the order of $1~\text{km}~\text{s}^{-1}$ or better. According to simulations, two observations one month apart will allow us to detect approximately 75~per~cent of the shortest period binary stars at $3\sigma$ confidence and about 50~per~cent of the binaries with periods shorter than a week. Our detection rate will drop significantly for binaries with periods longer than one month. If we can observe a field three times---one repeat on a time scale of one month and another repeat on a time scale of one year---binary detection rates will rise to as high as 75~per~cent for binaries with periods shorted than 20~days, and the detection rate only drops below 25~per~cent for binaries with periods longer than 0.5~year. In case of three observations, however, the detection rate can be as low as from a single repeat observation only in narrow windows of the time harmonics.  In practice, PFS will obtain three or more observations of those stars that lie in the overlap between two pointings.  Typically, we plan to observe each red-outlined pointing in Fig.~\ref{fig:dwarf_pointings} twice, separated by a few months.  Other pointings could be observed in subsequent years.  Hence, the overlap regions will have baselines on the order of months and years.}

\subsection{Element abundances}
\label{sec:abund}

\begin{figure}
    \includegraphics[width=1.0\columnwidth]{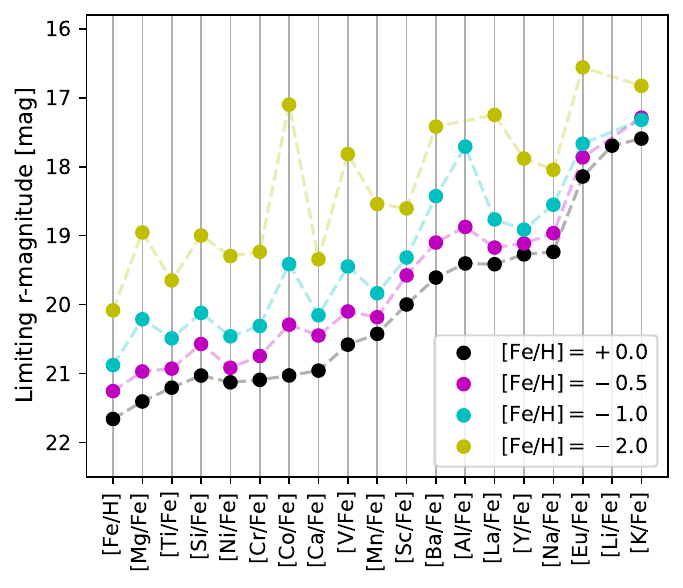}
    \caption{Expected limiting $r$-magnitudes for PFS measurements of chemical abundances for stars with different metallicities. Limiting magnitudes are defined such that the random error in the measurement does not exceed $0.1\ \mathrm{dex}$. The magnitudes were derived from simulated observations of a star with $T_\mathrm{eff}=5000\ \mathrm{K}$, $\log(g)=1.5$, and scaled solar abundances. Nominal 3-hour exposures and the MR mode in the red arm were assumed. The $\mathrm{Ca}\ \mathrm{II}$ triplet and the $\mathrm{Na}$ D line were excluded from the fit.}
    \label{fig:abun_precision}
\end{figure}

For all GA targets, we intend to estimate the metallicity ($[\mathrm{M/H}]$), $\alpha$-enhancement ($[\mathrm{\alpha/M}]$)\footnote{In this context, $[\mathrm{M/H}]$ and $[\mathrm{\alpha/M}]$ are defined by the best-fitting scaled-solar spectral model. In this model, all elements heavier than $\mathrm{He}$ are enhanced by $[\mathrm{M/H}]$ compared to standard solar chemistry, and the $\alpha$-elements ($\mathrm{O}$, $\mathrm{Ne}$, $\mathrm{Mg}$, $\mathrm{Si}$, $\mathrm{S}$, $\mathrm{Ar}$, $\mathrm{Ca}$, and $\mathrm{Ti}$) are also enhanced by $[\mathrm{\alpha/M}]$ on top of their $[\mathrm{M/H}]$-scaling.}, and the carbon-to-oxygen number-density ratio ($\mathrm{C/O}$)\@. For a subset of GA targets with sufficiently high S/N (quantified in Figure~\ref{fig:abun_precision}), we will also determine the chemical abundances of up to $18$ individual elements ($\mathrm{Li}$, $\mathrm{Na}$, $\mathrm{Mg}$, $\mathrm{Al}$, $\mathrm{Si}$, $\mathrm{K}$, $\mathrm{Ca}$, $\mathrm{Sc}$, $\mathrm{Ti}$, $\mathrm{V}$, $\mathrm{Cr}$, $\mathrm{Mn}$, $\mathrm{Co}$, $\mathrm{Ni}$, $\mathrm{Y}$, $\mathrm{Ba}$, $\mathrm{La}$, $\mathrm{Eu}$).

The abundances will be estimated in a two-stage procedure, similar to that described in \citet{Fisher_chemfit}. At the first stage, we determine the five fundamental stellar parameters ($T_\mathrm{eff}$, $\log(g)$, $[\mathrm{M/H}]$, $[\mathrm{\alpha/M}]$ and $\mathrm{C/O}$ ratio) by fitting a linearly interpolated 5D grid of fully self-consistent spectral models to both the observed spectrum (across all arms of the spectrograph) and photometric colors of the star. Depending on the science goal, this analysis stage can be carried out with gradient-based optimization methods or by sampling the parameter space with Markov chain Monte Carlo (MCMC).

At the second stage, we determine the abundances of individual elements by synthesizing their effects on the model spectrum during the fitting process using the \texttt{SYNTHE} and \texttt{BasicATLAS} codes \citep{SYNTHE,BasicATLAS}. To keep the computational cost manageable, we implement multiple variants of this stage that trade off accuracy against computational expense. The appropriate variant is selected for each star based on the scientific goals and the S/N of its spectrum. These variants differ in which first-stage parameters are held fixed versus re-determined, as well as in the simplifying approximations adopted in the spectral synthesis.

\begin{figure*}[tpb]
\centering
\includegraphics[width=2.0\columnwidth]{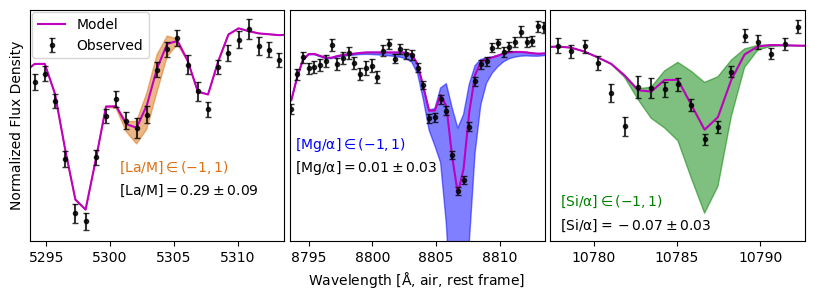}
\caption{Spectral model fit to a PFS SSP observation of a $r\approx16.8$ red giant member of Draco with estimated $T_\mathrm{eff}=4270\ \mathrm{K}$ and $[\mathrm{Fe/H}]=-1.7$. This spectrum was obtained with 3 hours of exposure. The three panels showcase small regions of the spectra obtained with the blue (\textit{left}), MR-mode red (\textit{center}) and infrared (\textit{right}) arms, centered around La {\sc ii} $\lambda \lambda 5302,5304$, Mg {\sc i} $\lambda 8807$, and Si {\sc i} $\lambda \lambda 10785,10787$, respectively. The shaded areas demonstrate the range of flux densities obtained by varying the abundances of $\mathrm{La}$, $\mathrm{Mg}$, and $\mathrm{Si}$ by $1\ \mathrm{dex}$ with respect to the best-fit metallicity and $\alpha$-enhancement of the star. The PFS/SSP pipeline  will measure chemical abundances not from any individual lines, but from the statistical ensemble of all lines within the wavelength range of the spectrograph, only $\sim0.8\%$ of which is shown in this figure.}
\label{fig:abun_spectra}
\end{figure*}

A detailed analysis of systematic errors associated with our fitting method will be presented in a future publication. Here, we provide a representative example, in which all stellar parameters are fixed to their stage one best-fit values except for $\log(g)$, which is re-optimized at the second stage. In this example, we adopt one of the most computationally efficient (and correspondingly least accurate) variants of the stage two analysis. In this variant, the atmospheric structure is assumed to be fully determined by the stage one parameters, and the effects of individual elements on the synthetic spectrum are treated as independent and linearly additive. For a representative PFS target ($T_\mathrm{eff}=4050\ \mathrm{K}$, $\log(g)=1.6$, $[\mathrm{M/H}]=-1.35$) with uniform-random enhancements of individual elements, the median systematic errors typically lie in the range $0.05$-$0.1\ \mathrm{dex}$, with a maximum error of $0.15\ \mathrm{dex}$ for $\mathrm{Eu}$. These errors are significantly reduced for stars with abundances closer to solar-scaled chemistry and become negligible when more computationally expensive variants of the stage two analysis are employed. In addition to fitting errors, physical limitations of the models (e.g., uncorrected effects of non-local thermodynamic equilibrium, inaccuracies in the line list, incomplete chemical reaction network, etc.)\@ introduce additional contributions to the error budget. These contributions were estimated to lie between $0.04\ \mathrm{dex}$ and $0.08\ \mathrm{dex}$, by analyzing the star-to-star scatter in abundances of selected elements from low-resolution spectra of the members of the globular cluster M13 \citep{evan_lithium}.

The expected random errors in the abundances of individual elements are shown in Figure~\ref{fig:abun_precision}, in the form of the value of the limiting $r$-magnitude at which the random error in the measurement begins to exceed $0.1\ \mathrm{dex}$. An example of a spectral fit to the PFS observation of a star in the Draco dwarf galaxy is given in Figure~\ref{fig:abun_spectra}.

\begin{figure}
    \includegraphics[width=1.0\columnwidth]{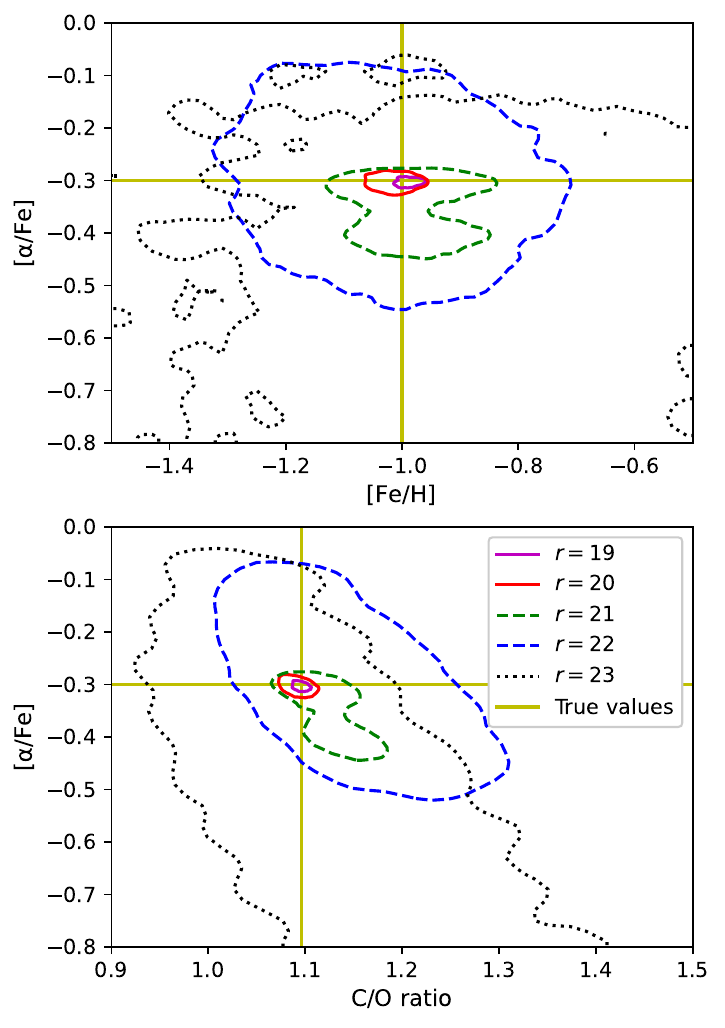}
    \caption{MCMC posterior distributions of chemical parameters inferred from simulated PFS observations of a star with $T_\mathrm{eff}=4000\ \mathrm{K}$, $\log(g)=1$ and $[\mathrm{Fe/H}]=-1$, assuming a range of apparent magnitudes as shown. The contours show $2$-sigma constraints on the best-fit parameters.}
    \label{fig:abun_mcmc}
\end{figure}

We expect that the S/N of faint ($r$-magnitude $\gtrsim 21$), metal-poor ($[\mathrm{Fe/H}]\lesssim -1$) giants will be insufficient to measure detailed chemical abundances, restricting our analysis for such stars to the 5 fundamental parameters: $T_\mathrm{eff}$, $\log(g)$, $[\mathrm{M/H}]$, $[\mathrm{\alpha/M}]$, and the $\mathrm{C/O}$ ratio. For these targets, MCMC sampling is necessary to capture the degeneracies and multi-modality of the parameter space. MCMC posteriors for simulated observations of a star with $[\mathrm{M/H}]=-1$ and $[\mathrm{\alpha/M}]=-0.3$ are shown in Figure~\ref{fig:abun_mcmc} for a range of assumed $r$-magnitudes.

\subsection{Standard stars}
\label{sec:standards}

During the commissioning observations conducted between March 2024 and January 2025, we observed several globular clusters and the Milky Way fields overlapping with other spectroscopic surveys to obtain spectra for objects that can be used as radial velocity and chemical abundance standards. The standard stars observed during this period span a wide range of effective temperatures, surface gravities, and metallicities, including very metal-poor stars with [Fe/H] $< -2$. This dataset, therefore, fully covers the stellar-parameter range relevant to our primary science sample. Several hundred stars from GALAH \citep{buder21}, APOGEE \citep{majewski17}, SEGUE \citep{lee08}, and DESI \citep{cooper23} were observed under moderately good conditions (seeing $<1.0''$ and transparency $>0.7$). In addition to these abundance and radial-velocity standards, several tens of fibers in each exposure were allocated to flux standard stars and are observed simultaneously with science targets (Tanaka et al. in prep). These flux standards were selected as candidate F-type stars based on Pan-STARRS1 and Gaia broad-band photometry, to ensure accurate flux calibration by the 2D data reduction pipeline. 

\section{Summary}

The PFS/SSP GA survey, as outlined here, will significantly improve our understanding of the formation, evolution, and structure of galaxies within the Local Group, by providing spectra for $\sim 100,000$ stars down to faint magnitudes. The planned observations include stars in galaxies that range over orders of magnitude in mass, from a low-mass, ultra-faint dwarf galaxy to the $L^*$ galaxies M31 and the MW\@. With these data, we will:

\begin{itemize}
    \item Constrain the mass-density profiles and chemical-evolution histories of a selection of nearby dwarf galaxies. The dwarf galaxies in the sample span a range of stellar masses and star-formation histories, including both the relatively high-mass Fornax dwarf galaxy and the ultra-faint dwarf galaxy Bo{\"o}tes I\@. As such, we will provide insight into the role of baryonic feedback, the mass profile of the dark-matter halo, and the interplay between star formation history and chemical evolution. In addition, the observations will reach across the furthest extents of the galaxies in the sample, allowing us to investigate their extended stellar envelopes, should such envelopes exist. The selection of candidate member stars in these systems is aided by deep, multi-band HSC imaging, representing synergy between photometry and spectroscopy on the Subaru telescope.
    \item Determine the assembly history of M31 and M33.  PFS will measure [Fe/H] and [$\alpha$/Fe] for 30,000 stars.  The distribution of abundances will inform theories of whether M31's halo and its most prominent tidal streams formed from a dominant major merger or a collection of minor mergers and when those mergers happened.  M31's qualitatively different accretion history will serve as a foil to the MW's comparatively quiescent past, with the last major accretion event having happened over 10~Gyr ago.  Furthermore, we will investigate the putative stellar halo of M33, which would probe an entirely different mass scale than the halos of the Milky Way and M31.
    \item Investigate the effect of recent mergers and interactions on the chemodynamical structure of the MW\@.  The PFS/SSP  survey will sample the velocities, metallicities, and ages of stars in the fragile outer disk, which has a long memory of gravitational perturbations from encounters with nearby satellite systems.  Fields in the MW halo will map structure in phase space so that we may further constrain the history of accretion by the Milky Way of small satellite galaxies.
\end{itemize}

The PFS-SSP commenced in February 2025 and will last for six years.  Data releases will include spectra of the stars observed during the GA survey, in addition to their derived data products, such as radial velocities and chemical abundances.

\begin{acknowledgements}

The PFS/SSP group at Johns Hopkins University thanks Schmidt Sciences for their support. 

The instrument \onohiula\ Prime Focus Spectrograph (PFS), including both hardware and software, was developed by the PFS collaboration to which over 25 institutes across multiple countries were committed. The technical activities were conducted by (in alphabetical order) Academia Sinica Institute of Astronomy and Astrophysics (Taiwan), California Institute of Technology, Johns Hopkins University, Kavli Institute for the Physics and Mathematics of the Universe in the University of Tokyo (Kavli IPMU), Laboratoire d'Astrophysique de Marseille, Laborat{\'o}rio Nacional de Astrof{\'i}sica (Brazil), Max-Planck-Institut f{\"u}r Astrophysik, Max-Planck-Institut f{\"u}r extraterrestrische Physik, NASA Jet Propulsion Laboratory, National Astronomical Observatory of Japan (NAOJ), Princeton University, and Universidade de S{\~a}o Paulo under the oversight by Project Office hosted by Kavli IPMU (later NAOJ). There were also essential commitments from academic and industrial partners such as Durham University (United Kingdom) and and Bertin Technologies (France).

The \onohiula\ PFS development work was supported by World Premier International Research Center Initiative (WPI), Ministry of Education, Culture, Sports, Science and Technology (MEXT), Japan. Kavli IPMU was established and supported by World Premier International Research Center Initiative (WPI), MEXT, Japan.
We gratefully acknowledge support from the Funding Program for World-Leading Innovative R\&D on Science and Technology (FIRST) program ``Subaru Measurements of Images and Redshifts (SuMIRe)'' by Council for Science and Technology Policy (CSTP), Japan. This work is supported by Japan Society for the Promotion of Science (JSPS) KAKENHI Grant Numbers JP15H05893, JP15K21733, JP15H05892, JP20H05850, JP20H05855, JP23H05438, JP23K13098, JP24K00669, JP25H00394, JP25H01553, and JP25KJ0017.  The work at Princeton University, Johns Hopkins University, and California Institute of Technology is supported in part by NSF Award 1636426. The work in ASIAA, Taiwan, is supported by the Academia Sinica of Taiwan. The work in Brazil is supported by grants from  CNPq (308994/2021-3) and FAPESP (2011/51680-6).  The work in France is supported by CNRS and Aix Marseille University.

\onohiula\ PFS makes use of the mechanical housing so-called POpt2 that accommodates the Prime Focus Instrument and integrates the Wide Field Corrector lens system generating a flat focal plane with good image qualities across the wide field of view at the Subaru's prime focus. We appreciate all efforts to make these crucial components of infrastructure operational in conjunction with the development of Hyper Suprime Cam (HSC). \onohiula\ PFS software components for instrument control and data processing utilize the platform developed and maintained for Vera C.\ Rubin Observatory. We appreciate their generosity of making it publicly available as an open source. We also appreciate the public catalogues from HSC-SSP PDR3, Gaia DR3, and the Pan-STARRS1 Surveys (PS1) which are exploited in PFS observations for field acquisition and auto-guiding of telescope pointing, characterization of sky spectra during exposures, and flux calibrations.

This work is based (in part) on data collected at the Subaru Telescope, which is operated by NAOJ\@. We are honored and grateful for the opportunity of observing the Universe from Maunakea, which has the cultural, historical, and natural significance in Hawaii.

We appreciate the development and operation of PFS Science Platform by Subaru Telescope and Astronomy Data Center at NAOJ which enables access to both PFS and HSC data and various analyses on the server side.

The HSC collaboration includes the astronomical communities of Japan and Taiwan, and Princeton University. The HSC instrumentation and software \citep{Miyazaki2018,Aihara2022} were developed by the National Astronomical Observatory of Japan (NAOJ), the Kavli Institute for the Physics and Mathematics of the Universe (Kavli IPMU), the University of Tokyo, the High Energy Accelerator Research Organization (KEK), the Academia Sinica Institute for Astronomy and Astrophysics in Taiwan (ASIAA), and Princeton University. Funding was contributed by the FIRST program from the Japanese Cabinet Office, the Ministry of Education, Culture, Sports, Science and Technology (MEXT), the Japan Society for the Promotion of Science (JSPS), Japan Science and Technology Agency (JST), the Toray Science Foundation, NAOJ, Kavli IPMU, KEK, ASIAA, and Princeton University.

This work has made use of data from the European Space Agency (ESA) mission Gaia \citep[][\url{https://www.cosmos.esa.int/gaia}]{Gaia2023}, processed by the Gaia Data Processing and Analysis Consortium (DPAC, \url{https://www.cosmos.esa.int/web/gaia/dpac/consortium}). Funding for the DPAC has been provided by national institutions, in particular the institutions participating in the Gaia Multilateral Agreement.

The Pan-STARRS1 Surveys \citep[PS1,][]{PanSTARRS,Flewelling2020} and the PS1 public science archive have been made possible through contributions by the Institute for Astronomy, the University of Hawaii, the Pan-STARRS Project Office, the Max Planck Society and its participating institutes, the Max Planck Institute for Astronomy, Heidelberg, and the Max Planck Institute for Extraterrestrial Physics, Garching, The Johns Hopkins University, Durham University, the University of Edinburgh, the Queen’s University Belfast, the Harvard-Smithsonian Center for Astrophysics, the Las Cumbres Observatory Global Telescope Network Incorporated, the National Central University of Taiwan, the Space Telescope Science Institute, the National Aeronautics and Space Administration under grant No.\ NNX08AR22G issued through the Planetary Science Division of the NASA Science Mission Directorate, the National Science Foundation grant No.\ AST-1238877, the University of Maryland, E{\"o}tv{\"o}s Lor{\'a}nd University (ELTE), the Los Alamos National Laboratory, and the Gordon and Betty Moore Foundation.
   
\end{acknowledgements}

\bibliographystyle{aasjournalv7}
\bibliography{pfs_ga_wp}

\allauthors

\end{document}